\documentclass[12pt,preprint]{aastex}

\shorttitle{Poynting--Robertson Cosmic Battery}
\shortauthors{Christodoulou et al.}

\begin{document}


\title{Simulations of the Poynting--Robertson Cosmic Battery \\ in Resistive Accretion Disks}


\author{
Dimitris M. Christodoulou,\altaffilmark{1}
Ioannis Contopoulos,\altaffilmark{2}
and
Demosthenes Kazanas\altaffilmark{3}
}


\altaffiltext{1}{Math Methods, 54 Middlesex Turnpike, Bedford, MA 01730.
            E-mail: dimitris@mathmethods.com}
\altaffiltext{2}{Research Center for Astronomy, Academy of Athens,
GR--11527 Athens, Greece. \\ Email: icontop@academyofathens.gr}
\altaffiltext{3}{NASA/GSFC, Code 663, Greenbelt, MD 20771.
            E-mail: kazanas@milkyway.gsfc.nasa.gov}


\def\gsim{\mathrel{\raise.5ex\hbox{$>$}\mkern-14mu
             \lower0.6ex\hbox{$\sim$}}}

\def\lsim{\mathrel{\raise.3ex\hbox{$<$}\mkern-14mu
             \lower0.6ex\hbox{$\sim$}}}

\begin{abstract}

We describe the results of numerical ``2.5--dimensional" MHD
simulations of an initially unmagnetized disk model orbiting a
central point--mass and responding to the continual generation of
poloidal magnetic field due to a secular source that emulates the
Poynting--Robertson (PR) drag on electrons in the vicinity of a
luminous stellar or compact accreting object. The fluid in the disk
and in the surrounding hotter atmosphere has finite electrical
conductivity and allows for the magnetic field to diffuse freely out
of the areas where it is generated, while at the same time, the
differential rotation of the disk twists the poloidal field and
quickly induces a substantial toroidal--field component. The secular
PR term has dual purpose in these simulations as the source of the
magnetic field and the trigger of a magnetorotational instability
(MRI) in the disk. The MRI is especially mild and does not destroy
the disk because a small amount of resistivity dampens the
instability efficiently. In simulations with moderate resistivities
(diffusion timescales up to $\sim$16 local dynamical times) and
after $\sim$100 orbits, the MRI has managed to transfer outward
substantial amounts of angular momentum and the inner edge of the
disk, along with azimuthal magnetic flux, has flowed toward the
central point--mass where a new, magnetized, nuclear disk has
formed. The toroidal field in this nuclear disk is amplified by
differential rotation and it cannot be contained; when it approaches
equipartition, it unwinds vertically and produces
episodic jet--like outflows. The poloidal field in the inner region
cannot diffuse back out if the characteristic diffusion time is of
the order of or larger than the dynamical time; it continues to grow
linearly in time undisturbed and without saturation, as the outer
sections of many poloidal loops are being drawn radially outward by
the outflowing matter of high specific angular momentum. On the
other hand, in simulations with low resistivities (diffusion
timescales larger than $\sim$16 local dynamical times), the
inflowing matter does not form a nuclear disk or jets and the linear
growth of the poloidal magnetic field is interrupted after about 20
orbits because of magnetic reconnection and asymmetric outflows.

\end{abstract}




\section{Introduction}

We revisit the theory of cosmic magnetic field generation via the
Poynting--Robertson (hereafter PR) battery in the vicinity of accretion--powered objects,
a process that has been proposed as a generic means of generating magnetic fields in a variety
of astrophysical sites. The principles of the PR battery have been enunciated
by \cite{ck} and then elaborated by \cite{ckc} (hereafter CK and CKC, respectively).
As was pointed out in CKC, the problem of the origin of cosmic magnetic fields
is largely of topological nature since the initial conditions of a homogeneous
and isotropic universe allow only for irrotational perturbations to exist.
It was also pointed out in the same reference that the vorticity ~{\boldmath $\omega$} =
${\bf \nabla \times v}$ ~of a velocity field ${\bf v}$ is subject to similar constraints and
obeys evolution equations
identical to those of the magnetic field \citep{KCOR97}. Since vorticity can only be
generated by the action of dissipative processes (Kelvin's theorem), it is not unreasonable to
also look for generation and growth of magnetic fields in sites that involve both
vorticity and dissipation, such as the accretion disks considered by CK and CKC.

The question of the source of cosmic magnetic fields has been vexing and it is believed to
still be open to investigation. It is generally thought that small fields (of the
order of $10^{-20}$~G) generated by the Biermann (1950) battery could be amplified
exponentially by dynamo processes to match the observed values of $\simeq 10^{-6}$~G of the
interstellar magnetic field in the Galaxy \citep{KCOR97}. But it has also been pointed out by
\cite{vr91} that turbulent dynamos can amplify fields to equipartition only on small scales,
while the large--scale field will remain well below equipartition, in apparent
disagreement with the observations (for simulations that support this point, see Fleming, Stone,
\& Hawley [2000] and references therein; but see also Igumenshchev, Narayan, \& Abramowicz [2003] for
recent developments concerning the evolution of the poloidal field).
These arguments rest on the assumptions that there are no
large--scale flows on top of the turbulent motions and that there are no other sources of magnetic field.
In contrast, the PR effect in accreting systems involves two qualitatively different
ingredients: a large--scale organized flow, namely the inflow that provides the accreting matter
onto the compact or stellar object; and a large--scale continuous source of azimuthal electric current
(or, equivalently, a toroidal electric field). It is our contention that these ingredients can lead
to magnetic--field evolution that is qualitatively very different than that envisioned in small--scale
turbulent environments and thus the PR--battery mechanism can be a better candidate for generating
and amplifying magnetic fields than the Biermann (1950) process and the turbulent dynamo criticized
by \cite{vr91}.

While it is apparent that an azimuthal current such as that of the PR battery would
lead to the generation of poloidal magnetic field, it is not immediately obvious that
such a field is necessarily of significance and some explanations are in order:
\begin{enumerate}
\item Although the PR current is quite robust, especially  in optically
thin accretion disks of the ADAF type \citep{ny}, in one dynamical time
it yields magnetic
fields well below equipartition values. But it was pointed out by CK
that the importance of this effect lies in its {\sl secular nature}:
the poloidal field will be amplified by the continual accumulation
of magnetic flux onto the central object due to the persistent
large--scale velocity field and its magnitude will eventually reach
dynamically significant levels.
\item The PR current generates naturally only closed poloidal loops. The
influx of an entire loop would amplify the central magnetic field
for just one typical accretion time to a value much too small to be
of consequence \citep{B-KB, B-Ketal}. But it was noted by CK and was
demonstrated explicitly by CKC that the possibility of accumulating
a nonzero net magnetic flux in the central region depends heavily on
the rate of inward--advected flux and the rate of outward diffusion.
Specifically, it was shown that a sufficiently high value of the
magnetic diffusivity could help the outer sections of complete
poloidal loops escape outward while the inner sections would still
be drawn inward by the inflow, leading to a secular net increase in
central magnetic flux.
\end{enumerate}
The work of CK and CKC shows that there exists a critical dimensionless parameter that specifies 
the boundary between different large--scale, long--term behaviors. This parameter is the 
inverse\footnote{CK defined the magnetic Prandtl number as $\eta /\nu$ (the ratio of magnetic diffusivity 
to turbulent kinematic viscosity), which is the inverse of what is commonly used in the literature.
Since most readers are familiar with the common definition ${\cal P}_m\equiv\nu /\eta$, confusion
can be avoided if we heretofore refer to the critical parameter of CK as $({\cal P}_m)^{-1}$. Field
amplification then occurs for $({\cal P}_m)^{-1}\gsim 2$ according to CK or, equivalently, for 
${\cal P}_m\lsim 0.5$.}
magnetic Prandtl number $({\cal P}_m)^{-1}$, i.e., the ratio ~$\eta /\nu$~ of magnetic diffusivity to 
turbulent kinematic 
viscosity, a number that is effectively determined by the ratio of the viscous to the diffusion timescale
$\tau_{\rm vis}/\tau_{\rm diff}$ in the disk's plasma. For $({\cal P}_m)^{-1} \gsim 2$
(or $\tau_{\rm vis} \gsim 2 \tau_{\rm diff}$), the outer section of each
loop escapes radially out as discussed above, and the magnetic flux in the vicinity of the central
object increases linearly in time and without bound. On the other hand, the central magnetic flux quickly
reaches saturation to dynamically insignificant levels in the opposite limit of 
$\tau_{\rm vis} / \tau_{\rm diff} < 2$.

The purpose of this work is to study numerically and in greater
detail the PR--battery mechanism and to explore whether the basic
premises outlined above can withstand the scrutiny of such a
treatment; and whether the scalings and behavior obtained in our
earlier simplified calculations carry over to detailed
multidimensional MHD simulations. To this end, we have decided to
dispense with the viscous prescription of the accretion disk and
opted instead to employ the magnetorotational instability (hereafter
MRI) \citep{bh1, bh2} as a means of establishing a global accretion
flow from a rotating disk toward the central object. Our global
accretion scenario begins with an initially unmagnetized,
geometrically thick torus embedded into a hotter, nonrotating
atmosphere and orbiting a gravitationally--dominant, central
point--mass. The magnetic field is produced continually by a secular
source term that emulates a toroidal electric field due to the PR drag
acting predominantly on the electrons. The fluid in the disk and in the
surrounding atmosphere is electrically conducting and allows for
magnetic dissipation to take place on a prescribed diffusion
timescale.

The effects that we are interested in are the large--scale response
of the fluid in the disk and the global evolution of the magnetic
field. These effects are nonperiodic and nonlocal and they cannot be
studied by using a local approximation such as the ``shearing box"
with field diffusion (e.g., Fleming, Stone, \& Hawley 2000; Fleming
\& Stone 2003). More specifically, the main questions that we wish
to investigate concern the fate of the gas and the magnetic field
that are driven from the accretion disk toward the central
point--mass by the MRI and by the process of magnetic diffusion. We
know from previous work on the applicable conservation laws in
magnetized fluids (Christodoulou, Contopoulos, \& Kazanas 1996,
2003) that azimuthal magnetic flux will also inflow from the
accretion disk as some of the angular momentum will be driven to
larger radii. This azimuthal flux along with gas depleted of its
angular momentum will have to accumulate in the vicinity of the
central point--mass. The gas that falls into the nuclear region becomes
effectively trapped deeper into the gravitational potential well of
the central object, but the toroidal magnetic field cannot be
trapped as easily, since it has at least two ways out of the
nucleus: it can dissipate if the resistivity of the fluid is high
enough; or it can unwind vertically and expel matter in an axial
(possibly bipolar) outflow if the nuclear azimuthal flux grows to a
sufficiently large value (Contopoulos~1995). In addition to the
study of the growth and evolution of magnetic fields in accretion
disks which constitutes our main goal, this work also provides some
insight about the influence of resistive effects to the global
dynamics of such disks and, in particular, about the possibilities
of MRI suppression and jet formation in the presence of anomalous
resistivity. To the best of our knowledge, the question of how the
anomalous resistivity influences the dynamics of the accreted matter
has not yet been explored to sufficient depth.

The remainder of the paper is organized as follows: In \S~2, we
present the MHD equations that we solve numerically. These equations
include resistive and dissipative effects in the magnetic field and
a source of poloidal field that emulates the effect of an azimuthal
electric current produced by the PR drag operating on electrons in
the area around a central object. In \S~3, we discuss the numerical
techniques that we use and we describe our initial disk model and
the surrounding isothermal atmosphere in equilibrium around the
central point--mass. In \S~4, we describe the most important
highlights from our model simulations. Finally, in \S~5, we discuss
the results of the simulations and their significance for
astrophysical accretion disks.

\section{The MHD Equations with Resistivity and PR Drag}

The MHD equations that we use are subject to the following conditions:
\begin{enumerate}
\item The fluid and the magnetic field extend in all three spatial dimensions which are
described in an inertial frame of reference in cylindrical coordinates ($R$, $\phi$, $Z$).

\item The fluid is perfect (it has no internal molecular viscosity) and ideal with adiabatic
exponent $\gamma$.

\item The fluid is electrically conducting and the electrical conductivity $\sigma$ can be
finite (fields with ohmic dissipation) or infinite (``frozen--in"
fields).

\item The fluid is not self--gravitating but it is influenced by the gravitational field
of a central point--mass $M$.

\item There exists a source of poloidal magnetic field ${\bf S} = (S_R, 0, S_Z)$ due to a
toroidal electric field produced by PR drag.

\item The displacement current is ignored in Ampr\`ere's law.
\end{enumerate}

The system of equations is parameterized by the adiabatic exponent $\gamma$, the resistivity $\eta$
(which is inversely proportional to  the electrical conductivity $\sigma$), and by the external sources
of the gravitational
field $\Phi$ and the poloidal magnetic field ${\bf S}$. The parameters $\gamma$ and $\eta$ are both
taken to be constant in space and time independent. Gaussian cgs units are chosen, in which case
$\eta = c^2/(4\pi\sigma)$, where $c$ is the speed of light.
The fundamental MHD variables are the mass density $\rho$, the entropy density
$\epsilon$, and the momentum density vector ${\bf\cal M}$ = (${\cal M}_R$, ${\cal M}_\phi$, ${\cal M}_Z$)
of the fluid, as well as the magnetic field vector ${\bf B}$ = ($B_R$, $B_\phi$, $B_Z$).
Additional derived variables are the components of the velocity vector
${\bf v}$ = ($v_R$, $v_\phi$, $v_Z$) = (${\cal M}_R /\rho$, ${\cal M}_\phi /(\rho R)$, ${\cal M}_Z /\rho$),
the angular velocity in the $Z$--direction $\Omega = v_\phi /R$ and the corresponding specific angular
momentum $L=R v_\phi$, the internal energy density $u = \epsilon^\gamma$, the internal fluid pressure
$P_{\rm fl} = (\gamma - 1)u$, and the electric current density vector
${\bf J} \equiv (c/4\pi){\bf \nabla\times B}$.

Under the above conditions, we can write the MHD equations in conservative form as follows:

The continuity equation expresses mass conservation:
\begin{equation}
\frac{\partial\rho}{\partial t} \ \ + \ \ {\bf\nabla\cdot}(\rho {\bf v}) \ \ = \ \ 0 \ ,
\label{continuity}
\end{equation}
where $t$ denotes time and ${\bf\nabla}$ is the usual del operator.

The components of the momentum equation describe the transfer of momentum within the fluid:
\begin{equation}
\frac{\partial {\cal M}_R}{\partial t} \ \ + \ \ {\bf\nabla\cdot}({\cal M}_R {\bf v}) \ \ = \ \
-\frac{\partial}{\partial R}\left(P_{\rm fl} + \frac{{\bf B}^2}{8\pi}\right) \ \ - \ \ \rho\frac{\partial\Phi}{\partial R} \ \
+ \ \ \frac{\rho v_\phi^2}{R} \ \ + \ \ \frac{1}{4\pi}({\bf B\cdot\nabla}) B_R \ ,
\label{rmomentum}
\end{equation}

\begin{equation}
\frac{\partial {\cal M}_\phi}{\partial t} \ \ + \ \ {\bf\nabla\cdot}({\cal M}_\phi {\bf v}) \ \ = \ \
-\frac{\partial}{\partial\phi}\left(P_{\rm fl} + \frac{{\bf B}^2}{8\pi}\right) \ \ - \ \ \rho\frac{\partial\Phi}{\partial\phi} \ \
+ \ \ \frac{R}{4\pi}({\bf B\cdot\nabla}) B_\phi \ ,
\label{phimomentum}
\end{equation}

\begin{equation}
\frac{\partial {\cal M}_Z}{\partial t} \ \ + \ \ {\bf\nabla\cdot}({\cal M}_Z {\bf v}) \ \ = \ \
-\frac{\partial}{\partial Z}\left(P_{\rm fl} + \frac{{\bf B}^2}{8\pi}\right) \ \ - \ \ \rho\frac{\partial\Phi}{\partial Z} \ \
+ \ \ \frac{1}{4\pi}({\bf B\cdot\nabla}) B_Z \ .
\label{zmomentum}
\end{equation}
In these three equations, the components of the usual Lorentz acceleration
${\bf (\nabla\times B)\times B}/4\pi$
have been split into two terms, the gradient of the magnetic pressure $P_{\rm mag}\equiv {\bf B}^2/8\pi$
and the components of the Lorentz tension $({\bf B\cdot\nabla}){\bf B}/4\pi$.
This is done to facilitate the numerical implementation of the momentum equation (see \S~3.1 below).

The entropy equation describes the transfer of entropy within the fluid:
\begin{equation}
\frac{\partial\epsilon}{\partial t} \ \ + \ \ {\bf\nabla\cdot}(\epsilon {\bf v}) \ \ = \ \
\frac{\eta}{4\pi\gamma \epsilon^{\gamma - 1}} \ \left| {\bf\nabla\times B}\right|^2 \ .
\label{entropy}
\end{equation}

The ideal--gas law relates the fluid pressure to the internal energy density and the entropy density:
\begin{equation}
P_{\rm fl} \ \ = \ \ (\gamma - 1)u \ \ = \ \ (\gamma - 1)\epsilon^\gamma \ .
\label{ideal}
\end{equation}
Since we adopt an ideal--gas equation of state, this equation can be used to rewrite
the conventional internal--energy equation of magnetohydrodynamics in the more efficient
form of eq.~({\ref{entropy}})
shown above. This calculation is new in the context of astrophysical MHD and it is
discussed in more detail in \S~2.1 below.

The divergence--free constraint for the magnetic field reads:
\begin{equation}
{\bf\nabla\cdot B} \ \ = \ \ 0 \ .
\label{delb}
\end{equation}

Finally, the induction equation for the evolution of the magnetic field reads:
\begin{equation}
\frac{\partial {\bf B}}{\partial t} \ \ - \ \ {\bf \nabla\times}({\bf v\times B}) \ \ = \ \
\eta {\bf\nabla^2 B} \ \ + \ \ {\bf S} \ ,
\label{induction}
\end{equation}
where ${\bf S}$ is the rate at which new field is generated by PR drag
(see \S~2.2 for details).

We further adopt the following simplifying assumptions:

(a) The fluid is axisymmetric about the $Z$--axis and remains axisymmetric in time.
This assumption allows us to eliminate the $\partial /\partial\phi$ derivatives from
the right--hand sides of equations~(\ref{rmomentum})--(\ref{zmomentum}) and~(\ref{induction})
and to simplify the divergence terms on the left--hand sides of
equations~(\ref{continuity})--(\ref{entropy}) and~(\ref{delb}), and in the curl term on the
left--hand side of equation~(\ref{induction}).

(b) The gravitational potential due to the external point--mass $M$ is spherically symmetric
and has the form
\begin{equation}
\Phi (r) \ \ = \ \ -\frac{G M}{r} \ ,
\label{potential}
\end{equation}
where $G$ is the gravitational constant
and $r = (R^2 + Z^2)^{1/2}$ is the spherical radius around the central point--mass.
We also choose units such that
$G = 1$ and $M = 1$ (this is done unconditionally
because the self--gravity of the fluid is ignored).

(c) The adiabatic exponent is taken to be $\gamma = 5/3$ in all model simulations.

(d) The source of poloidal magnetic field ${\bf S}$ is derived from a vector potential
${\bf E}$, as described in \S~2.2 below.

\subsection{The Entropy Equation}

The entropy equation (eq.~[\ref{entropy}] above) is derived by combining the
internal--energy equation of dissipative MHD
\begin{equation}
\frac{\partial u}{\partial t} \ \ + \ \ {\bf\nabla\cdot}(u {\bf v}) \ \ = \ \
-P_{\rm fl} {\bf \nabla\cdot v} \ \ + \ \ \frac{\eta}{4\pi} \left| {\bf\nabla\times B}\right|^2 \ .
\label{energy}
\end{equation}
with the ideal--gas equation of state (eq.~[\ref{ideal}] above).
For a constant adiabatic exponent $\gamma$,
let the entropy density variable $\epsilon$ of the ideal fluid be defined by
\begin{equation}
\epsilon \ \ \equiv \ \ u^{1/\gamma} \ ,
\label{epsilon}
\end{equation}
and substitute $u = \epsilon^\gamma$ and $P_{\rm fl} = (\gamma - 1)\epsilon^\gamma$ into eq.~(\ref{energy}).
Then the undesirable source term $-P_{\rm fl}{\bf \nabla\cdot v}$ is eliminated and the new equation for $\epsilon$
takes the Eulerian form shown also in eq.~(\ref{entropy}) above:
\begin{equation}
\frac{\partial\epsilon}{\partial t} \ \ + \ \ {\bf\nabla\cdot}(\epsilon {\bf v}) \ \ = \ \
\frac{\eta}{4\pi\gamma \epsilon^{\gamma - 1}} \ \left| {\bf\nabla\times B}\right|^2\ .
\label{entropy2}
\end{equation}
In this form, the only nonconservative (source) term remaining on the right--hand side
is due to Joule heating during
magnetic reconnection. On the other hand, eq.~(\ref{entropy2}) becomes strictly conservative, viz.,
\begin{equation}
\frac{\partial\epsilon}{\partial t} \ \ + \ \ {\bf\nabla\cdot}(\epsilon {\bf v}) \ \ \equiv \ \ 0 \ ,
\label{entropy3}
\end{equation}
when ${\bf B} = 0$ or when ${\eta} = 0$. Eq.~(\ref{entropy3}) shows that,
in the absence of magnetic fields (purely hydrodynamic case with ${\bf B} = 0$)
or in the absence of resistivity (ideal MHD case with $\eta = 0$),
the entropy density $\epsilon$ is a locally conserved quantity within the fluid.
This precise local conservation law was first derived by Tohline (1988)
for numerical simulations of purely hydrodynamical fluids with no magnetic fields.

Using the above equations (eq.~[\ref{entropy2}] or eq.~[\ref{entropy3}]) instead of eq.~(\ref{energy})
in numerical work offers a significant advantage because of the absence of
the ~$-P_{\rm fl}{\bf \nabla\cdot v}$ ~term that causes serious difficulties to the conservation of energy
when it is implemented in Eulerian discretization schemes 
(Stone \& Norman 1992a; Christodoulou, Cazes, \& Tohline 1997).

\subsection{The Source of Poloidal Field}

To ensure divergence--free conditions in the numerical implementation of the induction equation,
the source of the magnetic field ${\bf S}$ must be derived from a vector (electric) potential ${\bf E}$:
\begin{equation}
{\bf S} \ \ = \ \ {\bf \nabla\times E} \ .
\label{bdela}
\end{equation}
When the vector potential ${\bf E} = (0, E_\phi , 0)$ is purely toroidal,
then the field ${\bf S} = (S_R, 0, S_Z)$ is purely poloidal, in which case
\begin{equation}
S_R \ \ = \ \ -\frac{\partial E_\phi}{\partial Z} \ ,
\label{bpr}
\end{equation}
and
\begin{equation}
S_Z \ \ = \ \ \frac{1}{R}\frac{\partial}{\partial R}(R E_\phi) \ .
\label{bpz}
\end{equation}
For the function $E_\phi (R, Z)$, we adopt the definition
\begin{equation}
E_\phi (R, Z) \ \ \equiv \ \ F_S \frac{\Omega}{R} \ ,
\label{aphi}
\end{equation}
where $F_S$ is a constant with dimensions of magnetic flux.
Substituting then eq.~(\ref{aphi}) into eqs.~(\ref{bpr}) and~(\ref{bpz}), we find that
\begin{equation}
S_R \ \ = \ \ -F_S\frac{1}{R}\frac{\partial\Omega}{\partial Z} \ ,
\label{bpr2}
\end{equation}
and
\begin{equation}
S_Z \ \ = \ \ +F_S\frac{1}{R}\frac{\partial\Omega}{\partial R} \ .
\label{bpz2}
\end{equation}
Therefore, the source term in the induction equation (the last term in eq.~[\ref{induction}])
can be written in the following vector form:
\begin{equation}
{\bf S} \ \ = \ \ \left( -\frac{F_S}{R}\frac{\partial\Omega}{\partial Z}, \ \ \ 0, \ \ \
+\frac{F_S}{R}\frac{\partial\Omega}{\partial R} \right) \ .
\label{bp}
\end{equation}

The flux constant $F_S$ that appears in the above equations is related to the physical
parameters of the central accreting object and the PR drag, viz.,
\begin{equation}
F_S \ = \ \frac{L\sigma_T}{4\pi c e} \ ,
\label{fs1}
\end{equation}
in Gaussian cgs units, where $L$ is the luminosity of the central object, $\sigma_T$ is
the Thompson scattering cross--section, $c$ is the speed of light, and $e$ is the charge of electron.
Then, normalizing to the solar luminosity constant $L_\odot = 3.827\times 10^{33} \ {\rm erg \ s}^{-1}$,
this equation takes the form
\begin{equation}
F_S \ = \ 1.4\times 10^{7}\left(\frac{L}{L_\odot}\right) \ \mbox{Mx} \ ,
\label{fs2}
\end{equation}
where $1~{\rm Mx}\equiv 1~{\rm G~cm}^2$.
Finally, we scale $F_S$ to the magnetic--flux unit of the MHD code (which is $M\sqrt{G}$ since
both $M$ and $G$ have been normalized to 1), and we find that in the simulations we have to
choose $F_S$--values based on the dimensionless form
\begin{equation}
F_{S,NORM} \ \equiv \ \frac{F_S}{M\sqrt{G}} \
= \ 2.7\times 10^{-23}~\left(\frac{L/M}{L_\odot / M_\odot}\right) 
\simeq 10^{-18}\left(\frac{L}{L_{\rm Edd}}\right)\ ,
\label{fs3}
\end{equation}
where $M_\odot$ ($= 1.989\times 10^{33}~{\rm g}$) is one solar mass, the $L/M$ ratio
of the central object has been normalized by the corresponding solar constant, and
$L_{\rm Edd} = 1.3\times 10^{38} (M/M_{\odot})$ erg s$^{-1}$ is the Eddington luminosity.

Eq.~(\ref{fs3}) shows that the effect of PR drag is very small
indeed. Typical simulation values for $F_{S,NORM}$ ought to vary
between $\sim10^{-22}$ for white dwarfs with typical values of $L/M
= 10~L_\odot/M_\odot$ and $\sim10^{-19}$ for neutron stars and
active galactic nuclei with typical values of $L/M =
10^4~L_\odot/M_\odot$ (see Table 1 in CK for estimates of $L/M$ in
various accretion--powered sources). The magnetic field produced by
such small values of $F_{S,NORM}$ cannot reach dynamically
significant levels in our simulations given that they can be run
only for about 100 to 150 dynamical times. Since running model
evolutions for millions of orbits is not currently numerically
feasible, we need to ``enhance" artificially the source of the
magnetic field in our simulations. After some experimentation, we
decided to fix the flux constant to the value
\begin{equation}
F_{S,NORM} \ = \ 10^{-10}\ .
\label{fs4}
\end{equation}
This large value of the flux constant influences model evolutions in two different respects:
First, it results in a larger rate of magnetic field generation, in which case the influence of the magnetic
field to the global dynamics of the fluid will take place faster, namely, in just a few dynamical
times rather than millions of dynamical times. This is not necessarily a problem, as it simply
speeds up the global evolution of the simulated models. Second, it increases the rate of magnetic field
generation within each fraction of the dynamical time, in which case it essentially accelerates the local
response of the fluid elements and the magnetic field in every single timestep. This is a potential problem,
but we cannot avoid it, if we are to use the presently available computational resources.

\section{Numerical Techniques and Initial Model}

\subsection{Numerical Techniques}

We solve numerically the set of MHD equations~(\ref{continuity})--(\ref{potential}) and~(\ref{bp})
using an explicit, second--order accurate, multidimensional, finite--difference MHD code based
on van Leer (1977, 1979) monotonic interpolation and the method of characteristics (MoC) for wave propagation
(Stone \& Norman 1992a, b)
and on constrained transport (CT) for ensuring divergence--free conditions in the evolution of the magnetic field
(Evans \& Hawley 1988). The code has been tested over the past fifteen years by using publicly available
test suites (Stone \& Norman 1992a, b; Stone et al. 1992), as well as more specialized test cases relevant
to the problems under investigation (Christodoulou \& Sarazin 1996; Christodoulou, Cazes, \& Tohline 1997).

The numerical discretization scheme is similar to that described by Stone \& Norman (1992a, b), except
that we write finite differences for the new entropy equation (eq.~[\ref{entropy2}])
in place of the old internal--energy equation (eq.~[\ref{energy}]).
The MHD variables are staggered in the Eulerian grid, with scalar variables computed at cell centers
and vector components computed at cell faces.
The numerical integration is split into a source step
(in which all the terms on the right--hand sides of eqs.~[\ref{rmomentum}]--[\ref{entropy}] and~[\ref{induction}]
along with~[\ref{bp}] are computed in a certain order)
and a transport step (in which the fundamental MHD variables are advected consistently
from each particular grid cell to all adjacent cells).

While the transport step remains relatively straightforward (aided by reliable and well--tested numerical
techniques such as consistent advection, second--order directional splitting,
van Leer's interpolation, and MoC--CT
during computation of the divergences and the curls on the left--hand sides of
eqs~[\ref{continuity}]--[\ref{induction}]), the source step has become quite complicated due to the
introduction of the two new sources of magnetic field on the right--hand side of the induction equation
(eq.~[\ref{induction}]).
Figure 1 shows a flowchart of the various substeps and partial variable updates that are incorporated
in the MHD code in order to carry
out one complete timestep. The source step is finally completed within the transport step itself,
when the components
of the Lorentz tension $({\bf B\cdot\nabla}){\bf B}/4\pi$ are computed using the MoC and the velocity
components are subsequently updated.
A rotating reference frame is not used in the present work in order to reduce numerical complexity.
Artificial viscosity is not used either, because we do not want to smear out
sharp features that may develop during simulations. Even without artificial viscosity, however,
sharp discontinuities and wave fronts are being spread over 2--3 grid cells because of the various
approximations
built into the numerical scheme (mainly van Leer's algorithm and weighted averaging of variables
in adjacent cells in order to determine interpolated values).

For the present investigation, we assume, in addition, that all
variables are and remain axisymmetric in which case the
computational domain is restricted to just a vertical cross--section
in the $RZ$--plane. In this so-called ``2.5--dimensional" formalism,
the MHD variables retain all three spatial components although the
$\partial /\partial\phi$ derivatives are set to zero. In particular,
angular momentum and toroidal magnetic field, although constrained
to remain axisymmetric at all times, are still being updated
consistently by the $R$ and $Z$ gradients of the magnetic field.
``Free" boundary conditions are implemented at all the edges of the
computational domain, except along the inner vertical boundary
($R=0$), where ``symmetry" boundary conditions are imposed. Free
boundary conditions allow for fluid and magnetic field to exit the
computational domain with a minimum of artificial reflections at the
boundary. Symmetry boundary conditions guarantee that all radial
derivatives are exactly equal to zero on the $Z$--axis of the
coordinate system.

We choose to use equally--spaced mesh points in the $RZ$ computational grid so that each cell's extent is
the same in both the radial and the vertical direction, i.e., $\Delta R = \Delta Z$.
Time integration in the MHD code is explicit, and the largest allowed timestep $\Delta t_{\rm max}$ is
limited by the usual Courant condition for numerical stability in an explicit scheme:
\begin{equation}
\Delta t_{\rm max} \ \ = \ \ 0.5 \cdot  min\left(\frac{\Delta
R}{v_{\rm fms}} \ \ , \ \ \frac{(\Delta R)^2}{\eta}\right) \ ,
\label{timestep}
\end{equation}
where $min()$ denotes the smallest of the enclosed timescales defined within each grid cell
of size $\Delta R$: the propagation time of fast magnetosonic
waves with speed $v_{\rm fms} = [(\gamma P_{\rm fl} + 2 P_{\rm mag})/\rho]^{1/2}$ and the
diffusion time of the magnetic field.
Furthermore, $\Delta t_{\rm max}(n)$ at each step $n > 1$ is not allowed to increase by more than 30\%
relative to its value $\Delta t_{\rm max}(n-1)$ determined during the previous step.

\subsection{Initial model}

The initial disk model is an unmagnetized Papaloizou--Pringle (1984) torus rotating around a central
point--mass $M$ (located at the origin of the coordinate system) and embedded into a diffuse,
spherically--symmetric, isothermal, hydrostatic atmosphere. The fluid in the torus is a homoentropic
polytrope with index 1.5. The rotation profile $\Omega (R)$ is determined by specifying constant specific
angular momentum in the fluid:
\begin{equation}
\Omega (R) = \Omega_o\left(\frac{R_o}{R}\right)^2 \ ,
\label{omega}
\end{equation}
where $R_o$ is the location of the pressure maximum of the torus on its equatorial plane and
$\Omega_o\equiv\Omega (R_o)$. Since the self--gravity of the torus is ignored, the central point--mass
imposes Keplerian rotation at the radius $R_o$ of the pressure maximum, i.e., $\Omega_o^2 = GM/R_o^3$.
We choose units such that $G=M=R_o=1$, in which case $\Omega_o = 1$ as well, and we also set the maximum
density of the fluid arbitrarily to $\rho_{\rm max} \equiv \rho (R_o) = 10^{-10}$. The location of the
zero--pressure surface of the torus is then determined by a single parameter, the adiabatic sound speed
$c_o \equiv c(R_o)$ at the pressure maximum of the fluid. We set $c_o = 0.2$ which results in a
geometrically thick torus of moderate size: the inner and outer radii in the equatorial plane are located
at $R_{\rm in}=0.743$ and $R_{\rm out}=1.53$, respectively, and $(R_{\rm out} - R_{\rm in})/R_o \lsim 1$.

The spherically--symmetric isothermal atmosphere surrounding the torus is initially in hydrostatic
equilibrium with the central point--mass $M$. Its density profile $\rho_a (r)$ is determined
by solving the equation of hydrostatic equilibrium in spherical coordinates
\begin{equation}
\frac{1}{\rho_a}\frac{dP_a}{dr} \ \ + \ \ \frac{d\Phi}{dr} \ \ = \ \ 0 \ ,
\label{hydrostatic}
\end{equation}
where $P_a = c_a^2\rho_a$ is the pressure and $c_a$ is the constant isothermal sound speed.
We find that
\begin{equation}
\rho_a (r) \ \ = \ \ \rho_{\rm min}\cdot \ {\rm exp}\left\{\frac{GM}
{c_a^2}\left(\frac{1}{r} - \frac{1}{r_{\rm max}}\right)\right\} \ .
\label{atmosphere}
\end{equation}
In this equation, we use again the normalization $G = M = 1$ as well as a value of $c_a = 10 c_o = 2$
for the isothermal sound speed (i.e., we assume that the atmosphere is 100 times hotter than the
pressure maximum of the torus). We also choose the integration constant $\rho_{\rm min}(r_{\rm max})$ as follows:
We wish to ensure that the MHD code will be capable of evolving self--consistently all the low--density
values of the atmosphere. For this reason, the lowest initial density in the atmosphere, $\rho_{\rm min}$,
must be larger than the lowest cutoff value that the MHD code can resolve (which is set to
$10^{-7}\rho_{\rm max} = 10^{-17}$, i.e., 7 orders of magnitude lower than the density maximum in the torus).
Therefore, we choose $r_{\rm max}$ to be the outermost radial point in the equatorial plane of the grid
($r_{\rm max} = 1.3 R_{\rm out} = 2.0$), and we also adopt $\rho_{\rm min}(r_{\rm max})=3\times 10^{-16}$ as a reasonable
compromise---a value significantly lower than all the initially resolved density values of the fluid in the
torus ($\rho > 10^{-14}$) but 30 times higher than the absolutely lowest density cutoff implemented in the MHD
code ($\rho = 10^{-17}$).

\section{Numerical Simulations}

In the simulations of the above initial model, we choose to measure time in initial orbital periods
at the pressure maximum $R_o$ of the torus. Since $\Omega_o = 1$, then one initial
orbital period at $R_o$ corresponds to a time of ~$t_o = 2\pi /\Omega_o = 2\pi$ ~and our normalized
time variable $\tau$ is defined in terms of the orbital period $t_o$ by the equation
\begin{equation}
\tau \equiv \frac{t}{t_o} = \frac{t}{2\pi} \ .
\label{time}
\end{equation}

With the gravitational potential of the central point--mass set to the form $\Phi (r) = -1/r$
(eq.~[\ref{potential}] with $G = M = 1$), the adiabatic exponent set to the value of $\gamma = 5/3$,
and the flux constant set by necessity to $F_{S,NORM} = 10^{-10}$ (eq.~[\ref{fs4}]),
model simulations are parameterized by a single remaining free parameter,
the resistivity $\eta$ that appears in eqs.~(\ref{entropy}) and~(\ref{induction}).
Furthermore, we rewrite $\eta$ in terms of the spatial resolution of the computational grid as
\begin{equation}
\eta \ \ \equiv \ \ \kappa \ (\Delta R)^2  \ , \label{kappa}
\end{equation}
and we choose freely the value of the resistive frequency $\kappa$ instead of the value of $\eta$.

In the simulations described below, we use 64 equally spaced grid zones in each direction because
this modest numerical resolution ($\Delta R = \Delta Z = 3.10784\times 10^{-2}$)
allows for model evolutions to be followed for more than 150 orbits
when necessary. At this resolution and including input/output operations and graphics, the 2.5--D
MHD code takes an average of $42\pm 8$ milliseconds per timestep on a 1.6 GHz Intel T2050
microprocessor, while typical model evolutions last for 1 to 1.5 million timesteps and 12 to 18 hours
of real time.
Finally, the resistive
frequency $\kappa$ is varied in the interval $0\leq\kappa\leq 10$, where $\kappa =0$ corresponds to
the ideal MHD case and $\kappa\geq 1$ corresponds to models in which the diffusion of the magnetic field
proceeds on local dynamical timescales or faster. Our standard, moderately diffusive model has $\kappa = 0.1$
and divides the parameter space to models with low resistivities ($\kappa\leq 0.01$) and models with high
resistivities ($\kappa\geq 1$).

Most of the figures that follow show contour plots and vector fields in the $RZ$--plane of each model
at various times. These figures have three panels. Panel {\it a} shows mass density contours as solid lines,
angular momentum density contours as dashed lines, and poloidal momentum densities as vectors.
Panel {\it b} shows contours of the toroidal magnetic field (solid lines in the $+\phi$ direction and
dashed lines in the $-\phi$ direction) and vectors of the poloidal magnetic field. In each of these two
panels, contours are drawn down to the 5\% level of the corresponding maximum value and arrows
are drawn down to 10\% of the largest magnitude. In addition, very small vectors with magnitudes
between 1\% and 10\% of the maximum value are replaced by dots in order to indicate in which regions
of the grid the vector fields tend to spread. Finally, panel {\it c} shows poloidal--field lines with
no regard to field magnitude and captures the detailed structure of the very weak field as well.
The various snapshots were chosen specifically to exhibit some of the most interesting large--scale
features in each model evolution.

\subsection{Standard Model With $\kappa =0.1$}

In the beginning of the simulation, poloidal loops of magnetic field are generated in the torus by the
secular PR source term. The field is stronger near the surface of the torus because the surrounding
isothermal atmosphere does not rotate and the gradients of the angular velocity are largest on the
surface (see eq.~[\ref{bp}]). The differential rotation of the fluid twists the poloidal loops quickly
and generates a prominent toroidal--field component around the surface of the fluid (Figs.~2 and 3).
The field that
is created by this mechanism is responsible for destabilizing the orbiting fluid through an MRI but
the instability does not take hold immediately because field diffusion works against it, even at this
moderate level of diffusivity. The MRI causes rarefaction waves to propagate radially out
within the fluid (Figs.~2a and 3a). These waves slowly carry outward the angular momentum while the azimuthal
magnetic flux accumulates near the inner edge of the torus, as was predicted by Christodoulou,
Contopoulos, \& Kazanas (1996, 2003). After $\sim$~2 orbits, the MRI has managed to destabilize only
the inner edge and a little matter enhanced with
azimuthal flux is inflowing toward the central point--mass (Fig.~4). This is also when
the first jets are seen to emanate from the nuclear region.
Inside the torus, the instability is suppressed efficiently by field diffusion 
(compare Figs.~3c, 4c)
and this allows the torus to survive for many subsequent orbits.

The inflowing matter builds another torus in the vicinity of the
central point--mass (Fig.~5). This nuclear torus continues to
receive inflowing matter and azimuthal flux from the inner edge of
the original torus. The main characteristics near the equatorial
pressure maximum of the nuclear torus are summarized in Table~1,
where large values of the ratio $B_Z/B_\phi$ signify that an axial
outflow has already occurred and the toroidal field has been
released from the nuclear torus. For $\kappa = 0.1$, the diffusivity
is not sufficiently strong to drive the magnetic flux out of the
nuclear region and no significant wind--like outflow is observed in
the vicinity of the central object (the diffusion ``bubbles" seen in
Fig.~5c between the two tori are quite weak and devoid of azimuthal
flux, as Fig.~5b indicates). When the nuclear azimuthal flux becomes
just a few percent of the equipartition value, it is released
vertically in a collimated jet outflow (Figs.~5 and 6) reminiscent
of the ``plasma--gun" mechanism described by Contopoulos~(1995).
Such outflows are usually bipolar, but occasionally they are
asymmetric when the field is expelled preferentially in one or the
other direction (Figs.~8 and 10). The asymmetry develops on the
equator of the nuclear torus because the interface between toroidal
fields with opposite polarities is unstable (it is fluttering up and
down). The instability is seen in Figs.~5b and~9b, where this
interface is curved throughout the region between the central
point--mass and the inner edge of the original torus. On occasion,
the nuclear toroidal field becomes weaker when it is expelled from the
center and then we can see the complex structure of the much weaker
field in and around the original torus (Figs.~7b and 9b).

\begin{table}[t]
\begin{center}
\begin{minipage}{7in}
\caption{}
\begin{tabular}{cccccccc}
\multicolumn{8}{c}{\sc Nuclear Disk in the Standard Model With $\kappa = 0.1$} \\
\hline\hline
$\tau$      & $\rho /\rho_{\rm max}$ & $P_{\rm fl}$             & $\beta$ & $B_\phi$            &  $B_Z/B_\phi$  & $v_\phi$ &  $v_Z/v_\phi$  \\
\hline\hline
  1.93      &   0.0023           & $1.0\times 10^{-13}$ &    0.1  & $5.3\times 10^{-6}$ &   0.5          & 5.4      &    3.3   \\
  4.59      &   0.48             & $6.5\times 10^{-10}$ &   50    & $4.6\times 10^{-6}$ &   3.7          & 0.3      &    0.3   \\
 20.85      &   6.6              & $1.0\times 10^{-8}$  &  847    & $4.6\times 10^{-6}$ &   3.7          & 1.2      &    1.0   \\
 29.61      &   7.9              & $1.2\times 10^{-8}$  & 1773    & $2.5\times 10^{-7}$ &  52.0          & 0.007    &  235     \\
 59.72      &  46.2              & $6.7\times 10^{-8}$  &   39    & $1.8\times 10^{-4}$ &   0.6          & 0.6      &    0.9   \\
 92.46      &  26.3              & $4.3\times 10^{-8}$  & 1200    & $1.9\times 10^{-6}$ &  15.4          & 0.007    &    5.1   \\
110.93      &  15.2              & $2.6\times 10^{-8}$  &   56    & $3.3\times 10^{-5}$ &   3.0          & 0.03     &   39     \\
\hline\hline
\multicolumn{8}{l}{{\sc Notes}.---Plasma $\beta\equiv P_{\rm fl}/P_{\rm mag}$, ~~$\rho_{\rm max}\equiv 10^{-10}$.}\\
\end{tabular}
\end{minipage}
\end{center}
\end{table}

The poloidal magnetic flux
\begin{equation}
\Psi (R, 0) \equiv 2\pi \int_{0}^{R} B_Z (R^\prime)  R^\prime dR^\prime \ \ ,
\label{phirz}
\end{equation}
is monitored over the entire equatorial plane of the computational grid (Fig.~11).
Because the PR source generates complete poloidal loops, the total poloidal flux over the equator
of the grid is zero (loops crossing the equator in one direction have to turn back eventually
and cross in the opposite direction).

The conservation of $\Psi (R_{\rm max}, 0)$, where $R_{\rm max}$ is the radial edge of the grid,
is broken after 18 orbits when magnetic field from the surface
of the torus reaches the outer edge of the grid and outflows. Then poloidal loops open up at
$R = R_{\rm max}$ (Figs.~6c, 7c) and $\Psi (R_{\rm max}, 0)$ becomes permanently positive (Fig.~11). 
In the nuclear torus,
the enclosed flux switches polarity several times and also undergoes high--frequency oscillations
due to the episodic evolution of the nuclear magnetic field (see panels $c$ in Figs.~5--10).
At all the other equatorial radii, and especially in the regions between the two tori and within
the original fluid, the poloidal flux increases linearly with time and this increase continues
for more than 80 orbits (Fig.~11), when large amounts of mass and angular momentum reach the radial
edge of the computational grid.
Such a steady, gradual increase of the poloidal field was predicted by CK and CKC
for moderate and high levels of diffusivity and owes its linear character to the time independence
of the PR source term ${\bf S}$ that was included in the induction equation (eq.~[\ref{induction}]).

\subsection{Low Resistivity Models With $\kappa\leq 0.01$}

In models with $\kappa\leq 0.01$, the diffusion of the field is
strong enough to dampen the MRI and to delay the organized inflow of matter
for at least 10 orbits (e.g., Figs.~12a and~13a). Slow diffusion and the fluttering instability
cause the toroidal field to spread away from the surface of the
fluid where it is created. Some of this field is ejected
into the surrounding atmosphere but another part of it diffuses
into the fluid of the torus (Fig.~12b). Eventually, the inner
edge of the torus is destabilized by the MRI, loses part of its
angular momentum, and an unbroken stream (a "sheet") of inflowing matter is
created that is threaded by strong toroidal magnetic field (Figs.~12b, 14b).
In the low resistivity models, a nuclear disk is not formed.
Instead, lumps of matter with embedded field that reach the center ahead of the inflowing stream 
are ejected vertically in asymmetric outflows (e.g., Fig.~13a, b).

The entire evolution of these models is reminiscent of the corresponding ideal MHD
model ($\kappa = 0$;~Fig.~15) except that the ideal MHD fluid is dominated by strong 
oblique shocks that distort the torus (Fig.~15a) and that are not observed in models 
with $\kappa\sim 0.01$ (Fig.~14a). We note that the lump of fluid seen at the center
in Fig.~15a will not be the seed for the formation of a nuclear disk; it will soon be ejected
vertically and it will clear the center for more lumps to come in ahead of the organized inflowing 
stream that needs another 6 or 7 orbits to get into the same area.

The poloidal magnetic flux at different equatorial radii of the model with $\kappa = 0.01$ is shown
in Fig.~16. Although the flux is initially growing linearly with time, this growth is quickly
terminated after about 23 orbits. This is because the toroidal field remains nearly frozen into
the fluid and does not unwind (e.g., Fig.~14b, c). Eventually magnetic
reconnection and the fluttering instability along the equator of the grid limit field growth,
while the new field generated by the PR source is not significant in magnitude to make a difference.
This behavior is intimately linked to the inability of the low--resistivity models to form a nuclear disk.
We have determined by additional simulations that all models with $\kappa\leq 0.06$ present the same
characteristics, while two models with $\kappa = 0.065$ and $\kappa = 0.07$ exhibit nuclear--disk formation 
within $3-4$ orbits and uninterrupted central flux amplification for 20 orbits when these runs were terminated.

\subsection{High Resistivity Models With $\kappa\geq 1$}

In models with $\kappa\geq 1$, field diffusion occurs over dynamical timescales
and the field quickly spreads out to the surrounding atmosphere and inward to the fluid
of the original torus. The MRI is damped very efficiently in these simulations and the original torus
survives for more than 140 orbits (Figs.~17--20) when large amounts of outflowing matter and angular momentum
have crossed the outer radial edge of the computational grid. For the first few orbits, the dominant field
is toroidal and it is being built up continually by the differential rotation of the fluid in the original torus.
Most of this field is carried into the nuclear region where it becomes anchored in the newly formed
nuclear torus and diffuses away from it in all directions (Figs.~17b,~18b and~20b).
Inside the original torus, the field is limited efficiently by magnetic reconnection; it remains very weak,
and this is why its does not appear in the contour plots of the same figures.

\begin{table}[t]
\begin{center}
\begin{minipage}{7in}
\caption{}
\begin{tabular}{cccccccc}
\multicolumn{8}{c}{\sc Nuclear Disk in the High Resistivity Model With $\kappa = 1$} \\
\hline\hline
$\tau$      & $\rho /\rho_{\rm max}$ & $P_{\rm fl}$             & $\beta$ & $B_\phi$            &  $B_Z/B_\phi$  & $v_\phi$ &  $v_Z/v_\phi$  \\
\hline\hline
 11.01      &   0.3              & $5.2\times 10^{-10}$ &   114   & $1.1\times 10^{-6}$ &   9.8          & 0.06     &    3.7     \\
 37.66      &  48.7              & $3.2\times 10^{-9}$  &    88   & $1.8\times 10^{-5}$ &   1.4          & 0.7      &    0.9     \\
100.11      &   6.7              & $1.1\times 10^{-8}$  &    49   & $6.2\times 10^{-5}$ &   0.7          & 0.9      &    1.0     \\
141.91      &   3.8              & $6.4\times 10^{-9}$  &   180   & $4.1\times 10^{-6}$ &   7.2          & 0.1      &    2.3     \\
\hline\hline
\multicolumn{8}{l}{{\sc Notes}.---Plasma $\beta\equiv P_{\rm fl}/P_{\rm mag}$, ~~$\rho_{\rm max}\equiv 10^{-10}$.}\\
\end{tabular}
\end{minipage}
\end{center}
\end{table}

The nuclear torus continues to receive the inflowing matter and azimuthal flux.
The main characteristics near its equatorial pressure maximum are summarized in Table~2.
After the first 30 orbits, this torus ends up rotating faster than the nuclear torus of the standard model.
The fluid is cooler and the axial outflows are weaker and appear much later (at $\tau\sim 100$).
Magnetic pressure support remains always at a level of at least $\sim 1\%$ of the fluid pressure,
and as a result, the nuclear fluid is always less dense than that of the standard model.

Interestingly, in models with $\kappa\sim 10$, there is no sign of inflowing matter or the
development of the MRI for more than $10-15$ orbits, which implies that the instability is
suppressed efficiently by field diffusion. But in models with $\kappa\sim 1$, matter and
azimuthal flux inflow does occur over short timescales and the evolution proceeds initially as in
the standard model with a nuclear torus forming in $2-4$ orbits. But the magnetic field diffuses easily
out of this nuclear structure and this process weakens the development and the input power of the jets
in all the models with $\kappa \geq 1$. For $\kappa = 1$, some mild jets ($v_Z\sim 1-2$ and
$B_Z\sim B_\phi$) are finally observed after about 100 orbits (Fig.~19b, c).

The equatorial poloidal flux grows again linearly in time at all radii outside of the nuclear
torus, just as in the standard model. This is shown in Fig.~21 for the model with $\kappa = 1$.
The conservation of $\Psi (R_{\rm max}, 0) = 0$ is now broken sooner because of the higher rates
of diffusivity. Fig.~21 shows that after just 8 orbits the field loops reach the outer edge of the grid 
and open up (see also Fig.~17c); then $\Psi (R_{\rm max}, 0)$ becomes permanently positive. 
Furthermore, the poloidal flux within the nuclear torus remains positive at all times
(which indicates the presence of a large--scale organized poloidal field; panels $c$ in Figs.~18--20) 
and undergoes again high--frequency oscillations due to radial fluctuations of the material 
to which the poloidal--field lines are attached.

\section{Summary and Discussion}

In this work, we have performed detailed numerical, multidimensional, MHD simulations of
the Poynting--Robertson (PR) battery, a mechanism that is capable of generating cosmic magnetic fields
in the vicinity of luminous, accreting, compact and stellar objects. The PR effect was included
in the simulations of an accretion--disk model orbiting a central point--mass by introducing a
continuous source of poloidal magnetic field into the induction equation (eq.~[\ref{induction}]).
At the same time, the differential rotation of the disk model provided an elemental source of toroidal
magnetic field by twisting dynamically the poloidal field lines. The fluid in the accretion disk and in
a surrounding tenuous nonrotating atmosphere was resistive and allowed for the magnetic field to diffuse
away from the areas where it was originally produced. In all models, a large--scale accretion flow was
established from the initial disk model toward the central point--mass by the action of a
magnetorotational instability (MRI) in the orbiting fluid. Two of the above features,
the PR current that acts as a continuous source of weak magnetic field and the global accretion flow that
may cause its amplification by drawing field of a single polarity to the center, are what sets the PR
battery apart from previously proposed and critically reviewed mechanisms of field generation and
amplification such as the Biermann (1950) battery and the turbulent dynamo process (see also \S~1).

The present simulations constitute a first attempt toward studying the global magnetohydrodynamics of
resistive large--scale accretion flows and the possible amplification or saturation of the generated
magnetic flux in the presence of various degrees of magnetic dissipation. The latter is controlled by
a free parameter, the resistive frequency $\kappa$ (eq.~[\ref{kappa}]) which is a direct measure of
the resistivity $\eta$ of the fluid and inversely proportional to the electrical conductivity $\sigma$.
We have found that a value of $\kappa\approx 0.06$ (corresponding to a diffusion timescale
$\tau_{\rm diff}\approx 16$ local dynamical times) is the critical value that separates two types of 
physically different model evolutions:
\begin{enumerate}
\item Models with moderate and high resistivities ($\kappa > 0.06$)
exhibit strong field amplification that continues uninterrupted for over 100 orbits (Figs.~11 and 21).
In about $2-4$ orbits, the inflowing matter
creates a nuclear torus near the central point--mass and the magnetic field that is transported into the
nucleus by accretion and by diffusion becomes anchored onto this torus. When the nuclear toroidal field
becomes strong, it unwinds and produces episodic bipolar jet--like outflows, in addition
to the diffusing field bubbles that are observed to emerge from the center when $\kappa\geq 1$.
The equatorial field is unstable to fluttering and this instability is responsible for the occasional appearance
of markedly asymmetric vertical jets and for the ejection of magnetic field into the surrounding atmosphere. 
All of these details are illustrated in Figs.~\ref{fig2}--\ref{fig10}
for our standard model with $\kappa = 0.1$ and in Figs.~\ref{fig17}--\ref{fig20} for the $\kappa = 1$ model.
\item Models with low--resistivities ($\kappa\leq 0.06$)
exhibit some moderate field amplification for about 20 orbits, but then the magnetic field quickly saturates
to dynamically insignificant levels because of the weak diffusion and the absence of unwinding of the toroidal
component, as the magnetic field remains nearly frozen into the matter. The accretion flow carries its
magnetic field toward the central point--mass but it does not create a nuclear torus. Eventually magnetic
reconnection, the fluttering instability, and some asymmetric ejections of magnetized lumps
limit the growth of the field to less than 3 orders of magnitude above the values seen early in the
model evolutions (Fig.~16).
All of these details are illustrated in Figs.~\ref{fig12}--\ref{fig14} for the $\kappa = 0.01$ model
and in Fig.~\ref{fig15} for the ideal MHD model with $\kappa = 0$.
\end{enumerate}
The above results are in agreement with those discussed by CK and CKC on the basis of qualitative
arguments and more idealized model calculations. The present work provides further evidence
in support of our original conclusions and we are confident that the proposed battery mechanism
will prove important to the theory of generation of cosmic magnetic fields.

The critical value of the inverse magnetic Prandtl number determined by CK, namely $({\cal P}_m)^{-1}\simeq 2$,
also appears to be in agreement with the critical value of $\kappa\approx 0.06$ determined from the 
present simulations, if an allowance is made for a rough, order-of-magnitude estimate of the effective 
viscous timescale $\tau_{\rm vis}$ associated with turbulent, MRI--driven inflow from the initial torus under 
ideal--MHD conditions (when the dynamics is not altered by resistive slipping of the magnetic field through
the matter): In our $\kappa = 0$ simulation with frozen--in magnetic field, the inflowing stream 
of matter has not reached the center after 19 orbits (Fig.~15) and the continuing evolution shows that this 
is still the case after 26 orbits when the stream is getting close to the center. Based on this observation, 
we estimate that $\tau_{\rm vis}\approx 26\tau_{\rm dyn}$, where $\tau_{\rm dyn}$ is the local dynamical time 
in the initial torus. Since the critical $\kappa\approx 0.06$ implies that 
$\tau_{\rm diff}\approx 16\tau_{\rm dyn}$, then the critical inverse magnetic 
Prandtl number in the resistive simulations is
\begin{equation}
({\cal P}_m)^{-1} = \frac{\tau_{\rm vis}}{\tau_{\rm diff}}\simeq 1.6\ ,
\label{invp}
\end{equation}
and field amplification occurs for $\kappa > 0.06$~ or, equivalently, for $({\cal P}_m)^{-1} > 1.6$.
We note that the above value of $\tau_{\rm vis}$ implies also an effective value of 
$\alpha_{\rm mag}\gsim 0.04$ for 
the analogue of the Shakura--Sunyaev (1973) parameter of the accretion flow initiated by the MRI in the 
ideal--MHD model. 
This is just a rough estimate and as such it is not out of line compared to values determined previously from 
simulations of the MRI in the ideal--MHD limit ($\alpha_{\rm mag}\sim 0.1$; Hawley \& Krolik [2002] and references 
therein). But notice that the effective $\alpha_{\rm mag}$--parameter increases dramatically to a value of 
$\alpha_{\rm mag}\approx 0.3-0.5$ in the $\kappa > 0.06$ models in which a robust nuclear disk forms in just a few orbits.
We have to conclude then that a sufficient amount of magnetic diffusivity appears to be the cause of
dynamical nuclear--disk formation in the above resistive 
models.\footnote{This conclusion should not be confused with the conclusions of Stone \& Pringle (2001)
and Hawley \& Balbus (2002) who see the fluttering instability and the formation of the nuclear disk but
find no purely axial, collimated outflow and no substantial differences in their models when 
a small amount of artificial resistivity is included to smear out current sheets. These simulations
were essentially carried out using ideal MHD; as such, they can see the intrinsic instability of the
equatorial toroidal field and the large--scale structure of the accreted fluid, but they cannot capture 
the influence of moderate or large amounts of anomalous resistivity. Furthermore, the inflowing stream 
in these ideal--MHD simulations reaches the center in just 2 orbits. 
Such inflow is too fast, essentially dynamical, and it does not appear in line with the values of the 
effective $\alpha_{\rm mag}$--parameter ($\sim 0.05-0.1$) reported for the fluid in the original torus when 
it is destabilized by the MRI. Stone \& Pringle (2001) claim that global stresses of the 
radial magnetic field are responsible for this early outward transport of angular momentum and the 
associated inflow that occurs before the MRI actually becomes nonlinear. But no such global
stresses are observed in our $\kappa = 0$ simulation in which the magnetosonic rarefaction waves
take time to transverse the fluid, to redistribute the conserved quantities locally, and finally 
to drive the MRI into the nonlinear regime.}

All the simulations of models with $\kappa > 0$ show that field diffusion works against the MRI and this
instability is damped with increasing success as the value of $\kappa$ is increased. This result is known
and well--understood (e.g., Fleming, Stone, \& Hawley 2000; Fleming \& Stone 2003). When the magnetic
field is allowed to slip through the matter, then the field lines cannot hold on to specific fluid elements
and facilitate their exchange of angular momentum and azimuthal magnetic flux.
However, the MRI is not eliminated from any model with
a reasonable value of $\kappa$ and the weakened modes continue to transport some of the angular momentum
to larger radii and matter with enhanced azimuthal magnetic flux toward the central point--mass
(see also Christodoulou, Contopoulos, \& Kazanas 1996, 2003). In the moderate and high--resistivity models
($\kappa\geq 0.1$), the transfer of these conserved quantities is gradual and this allows the original accretion
tori to survive for hundreds of orbits (in our simulations, it takes $80-140$ orbits for substantial
amounts of matter and angular momentum to cross the outer radial edge of the grid, a distance only twice
as large as the characteristic size of the initial torus).

In our model evolutions, we strengthened artificially the PR source
because the current state of computing does not allow us to run MHD
models with a weak PR source and wait for millions of dynamical
times to see whether the magnetic field will be amplified or not
(\S~2.2). Even with an artificially enhanced PR source, however, the
initial magnetic field is $1-2$ orders of magnitude smaller than
that utilized to induce an MRI in previous MHD simulations (e.g.
Hawley 2000; Hawley, Balbus, \& Stone 2001): At early times
($\tau\gsim 0.1$), the poloidal field grows at the inner edge of
the initial torus to $\sim 10^{-9}$, a value that results in a
plasma $\beta\sim 5\times 10^4$. Within the first orbit, the
toroidal field also catches up in magnitude and at later times, all
field components are amplified as the magnetic field is drawn into
the nuclear region. The amplification of the poloidal flux is
eventually interrupted (at $\tau\sim 20$) in the
low--resistivity models, as discussed above and in \S~4.2. The
amplification of the toroidal flux in the $\kappa > 0.06$ models is
also interrupted episodically by the repeated unwinding of the
toroidal field that accumulates in the nuclear torus. This effect is
a version of the ``plasma--gun" expulsion suggested by Contopoulos
(1995): when the toroidal field grows close to equipartition in the
nuclear torus, it can no longer be confined; it is released
dynamically in the vertical direction and its stresses act to
confine the poloidal--field component close to the symmetry axis of
the torus. This situation is evident in many of the diagrams (panels
{\it c}) shown in \S~4.1 and \S~4.3, where the poloidal field lines
are nearly vertical at small radii and the ``funnels" along the
$Z$--axis are extremely narrow. At the same time, the {\it
b}--panels of the same figures depict the substantial degree of
collimation imposed by the toroidal field to the low--density matter flowing
within these funnels. These results are of interest because they
demonstrate that the anomalous resistivity in the accretion flows
and the plasma--gun mechanism may be responsible for producing the
highly collimated jets observed in a variety of accretion--powered
galactic and extragalactic objects (see, e.g., ~Bridle \& Perley
1984; Mirabel \& Rodr\'{i}guez 1999; and Wilson, Young, \& Shopbell
2001).

Furthermore, our results indicate that the magnetic diffusivity
plays a more important role in accretion disks than previously
thought. For moderate or large values of this parameter (or,
equivalently, for values of the resistive frequency $\kappa >
0.06$), there is a clear tendency in the models to generate and
maintain strong, well--ordered, large--scale poloidal magnetic
fields which couple to the rotation of the nuclear flow and result
in matter expulsion along the rotation axis. In contrast, no such
features are seen at large scales in models with $\kappa\leq 0.06$.
Therefore, our models suggest that the well--known and well--defined
dichotomy of accretion--powered objects (e.g., ~Xu, Livio, \& Baum
1999; ~Ivezi\'c {\em et al.} 2004) to those exhibiting powerful jets
(``radio--loud") and those lacking such structures (``radio--quiet")
may be related to and should be sought in the physics that
determines the value of this particular macroscopic parameter of the
accretion flows that power the emission of these objects. The
calculations presented here are only a preliminary step toward
exploring this notion.


\acknowledgments

This work was supported in part by a Chandra grant.

\newpage

\section*{FIGURE CAPTIONS}

\figcaption{Flowchart for the completion of one timestep in the MHD code.
\label{fig1}
}

\figcaption{Standard model with $\kappa = 0.1$ at time $\tau = 0.76$.
The first rarefaction wave propagates outward in the fluid and a substantial toroidal field
with amplitude $B_\phi = 6.1\times 10^{-7}$ has been built
by differential rotation on the surface of the torus.
{\it Panel a}: Mass density contours as solid lines,
angular momentum density contours as dashed lines, and poloidal momentum densities as vectors.
{\it Panel b}: Contours of the toroidal magnetic field (solid lines in the $+\phi$ direction and
dashed lines in the $-\phi$ direction) and vectors of the poloidal magnetic field. In each
of these two panels,
contours are drawn down to the 5\% level of the corresponding maximum value and arrows
are drawn down to 10\% of the largest magnitude. Also, very small vectors with magnitudes
between 1\% and 10\% of the maximum value are replaced by dots in order to indicate in which regions
of the grid the vector fields tend to spread.
{\it Panel c}: Poloidal--field lines irrespective of field magnitude; the detailed structure of
the very weak field can be seen here as well. The data are padded with zeroes at $R=0$ in order to
delineate the behavior of the field lines near the $Z$--axis.
\label{fig2}
}

\figcaption{As in Fig.~2 but for $\tau = 1.19$.
The inner edge of the torus is destabilized by the MRI.
\label{fig3}
}

\figcaption{As in Fig.~2 but for $\tau = 1.93$.
Matter and field have flowed into the nuclear region and an episodic vertical jet--like outflow has
developed ($|v_Z|\approx 18$ at the base of the jet, as opposed to $v_\phi\approx 5.4$). The unwinding
of the nuclear toroidal field results in a substantial axial field near the $Z$--axis (see Table~1).
\label{fig4}
}

\figcaption{As in Fig.~2 but for $\tau = 4.59$.
High--angular momentum fluid has flowed to larger radii away from the outer edge of the torus,
another (nuclear) torus has formed near the central point mass by inflowing matter, while a prominent
jet has developed in the vertical direction away from the nucleus ($|v_Z|\sim 1-2$ at the base of the jet).
\label{fig5}
}

\figcaption{As in Fig.~2 but for $\tau = 20.85$.
The nuclear torus has become denser than the original torus ($\rho = 6.6\times 10^{-10}$),
while the jet--like outflow appears to be very well collimated and bipolar.
\label{fig6}
}

\figcaption{As in Fig.~2 but for $\tau = 29.61$.
The original torus has flattened substantially due to inflowing and outflowing matter,
while the fluttering instability has expelled the toroidal field from the nuclear torus
that appears to be weakly magnetized ($B_\phi = 2.5\times 10^{-7}$; see also Table~1).
Panel {\it b} ~then shows the structure of the relatively weak magnetic field
($B_\phi\sim 10^{-6}$) that has spread into the original torus and in the surrounding atmosphere.
\label{fig7}
}

\figcaption{As in Fig.~2 but for $\tau = 59.72$.
The original torus has separated into two regions and the outer region is moving outward.
The nuclear torus has become very dense, hot, and strongly magnetized (Table~1). This torus
appears to also support an asymmetric jet--like outflow with a strong magnetic field ($P_{\rm mag}\sim P_{\rm fl}$)
embedded into the diffuse ($\rho\sim 10^{-12}$) outflowing matter.
\label{fig8}
}

\figcaption{As in Fig.~2 but for $\tau = 92.46$.
The original torus continues to feed matter to the nuclear region and to move radially outward,
while another vertical outflow ($|v_Z|\sim 2$ at its base) is taking place in the nuclear torus.
\label{fig9}
}

\figcaption{As in Fig.~2 but for $\tau = 110.93$.
The nuclear disk has expelled much of its own angular momentum in a wind and it has also developed another
asymmetric vertical outflow. Only 9.7\% of the initial mass and 2.0\% of the initial angular momentum
remain within the computational grid at this time.
\label{fig10}
}

\figcaption{Poloidal magnetic flux $\Psi (R, 0)$
on the equatorial plane of the grid, integrated out to different radii,
for the standard model with $\kappa = 0.1$. The integrated flux beyond the nuclear torus
increases linearly with time for over 80 orbits, while the flux within
the nuclear torus oscillates at very high frequencies and switches polarity several times.
\label{fig11}
}

\figcaption{Low resistivity model with $\kappa = 0.01$ at time $\tau = 7.50$.
Two rarefaction waves carry angular momentum outward in the torus, a "sheet" of inflowing
matter has developed toward the nucleus, and the fluttering instability has pushed field
into the surrounding atmosphere. The toroidal field presents quite a complex distribution,
but it is weak in magnitude ($\sim 10^{-7}$ or smaller).
\label{fig12}
}

\figcaption{As in Fig.~12 but for $\tau = 10.38$.
Matter in the nuclear region does not get organized in a disk; instead it is ejected
asymmetrically in the vertical direction ($v_Z\approx 5$) along with its embedded toroidal
field ($B_\phi\sim 10^{-6}$), while a weak axial--field component ($B_Z\sim 10^{-7}$) also
develops at small radii.
\label{fig13}
}

\figcaption{As in Fig.~12 but for $\tau = 16.30$.
No nuclear disk has developed and the asymmetric vertical outflow continues in the inner region,
while the inflowing sheet of material appears to be threaded by strong magnetic field
(all components are $\sim 10^{-6}$).
Only 3.5\% of the initial mass and 3.7\% of the initial angular momentum has flowed out
of the computational grid at this time.
\label{fig14}
}

\figcaption{Ideal MHD model with $\kappa = 0$ at time $\tau = 18.86$.
No field diffusion occurs in this model and
the field remains permanently frozen into the fluid.
This model evolution is similar to the low resistivity model shown in Fig.~14 above,
except for the strong oblique shocks observed here within the fluid of the original torus
and at the tip of the inflowing sheet.
Only 3.1\% of the initial mass and 3.2\% of the initial angular momentum has flowed out
of the computational grid at this time.
\label{fig15}
}

\figcaption{Poloidal magnetic flux $\Psi (R, 0)$
on the equatorial plane of the grid, integrated out to different radii,
for the low resistivity model with $\kappa = 0.01$. The integrated flux no longer increases
linearly with time at times $\tau > 23$.
\label{fig16}
}

\figcaption{High resistivity model with $\kappa = 1$ at time $\tau = 11.01$.
A nuclear torus has formed from inflow and a strong magnetic field
($B_Z\sim 10^{-5}$, $B_\phi\sim 10^{-6}$) is anchored onto it (see also Table~2). Two field "bubbles"
have expanded out of the center and have diffused obliquely into the surrounding atmosphere.
\label{fig17}
}

\figcaption{As in Fig.~17 but for $\tau = 37.66$.
The nuclear torus has become very dense ($\rho\approx 5\times 10^{-9}$) and a vertical outflow
has developed ($|v_Z|\sim 1$ at its base) in addition to the obliquely expanding bubbles.
\label{fig18}
}

\figcaption{As in Fig.~17 but for $\tau = 100.11$.
The original torus has been flattened by the MRI, while the strongest field ($B_Z\approx 8\times 10^{-5}$)
participates in a collimated jet--like outflow ($|v_Z|\sim 2$ at its base)
anchored at the nuclear torus.
\label{fig19}
}

\figcaption{As in Fig.~17 but for $\tau = 141.91$.
The original torus has spread toward the outer edge of the grid as rarefaction waves continue to
redistribute angular momentum, the nuclear torus has transported outward much of its own angular momentum
in a wind, the fluttering interface instability has disrupted the sheet of inflowing matter,
and a vertical jet is seen along with magnetic--field bubbles that diffuse obliquely out of the center.
Only 8.6\% of the initial mass and 7.0\% of the initial angular momentum remain
within the computational grid at this time.
\label{fig20}
}

\figcaption{Poloidal magnetic flux $\Psi (R, 0)$
on the equatorial plane of the grid, integrated out to different radii,
for the high resistivity model with $\kappa = 1$. As in the standard model
($\kappa = 0.1$ in Fig.~11), the integrated flux beyond the nuclear torus
increases linearly with time for over 140 orbits, while the flux within
the nuclear torus oscillates at very high frequencies but maintains a positive polarity.
\label{fig21}
}



\begin{thebibliography}{}

\bibitem[Balbus \& Hawley(1991)]{bh1}
    Balbus, S. A., \& Hawley, J. F. 1991, \apj, 376, 214

\bibitem[Balbus \& Hawley(1998)]{bh2}
    Balbus, S. A., \& Hawley, J. F. 1998, Rev. Mod. Phys., 70, 1

\bibitem[Biermann(1950)]{bman}
        Biermann, L. 1950, Z. Naturforsch, 5a, 65

\bibitem[Bisnovatyi-Kogan \& Blinnikov(1977)]{B-KB}
    Bisnovatyi-Kogan, G. S., \& Blinnikov, S. I. 1977, \aa, 59, 111

\bibitem[Bisnovatyi-Kogan {\em et al.}(2002)]{B-Ketal}
    Bisnovatyi-Kogan, G. S., Lovelace, R. V. E., \& Belinski, V. A. 2002, \apj, 580, 380

\bibitem[Bridle \& Perley(1984)]{bp}
        Bridle, A. H., \& Perley, R. A. 1984, \araa, 22, 319

\bibitem[Christodoulou, Cazes \& Tohline(1997)]{cct}
    Christodoulou, D. M., Cazes, J. E., \& Tohline, J. E. 1997, New Astronomy, 2, 1

\bibitem[Christodoulou, Contopoulos \& Kazanas(1996)]{cck1}
    Christodoulou, D. M., Contopoulos, J., \& Kazanas, D. 1996, \apj, 462, 865

\bibitem[Christodoulou, Contopoulos \& Kazanas(2003)]{cck2}
    Christodoulou, D. M., Contopoulos, J., \& Kazanas, D. 2003, \apj, 586, 372

\bibitem[Christodoulou \& Sarazin(1996)]{cs}
    Christodoulou, D. M., \& Sarazin, C. L. 1996, \apj, 463, 80

\bibitem[Contopoulos (1995)]{c05}
    Contopoulos, J. 1995, \apj, 450, 616

\bibitem[Contopoulos \& Kazanas (1998)]{ck}
    Contopoulos, I., \& Kazanas, D. 1998, \apj, 508, 859 (CK)

\bibitem[Contopoulos, Kazanas \& Christodoulou(2006)]{ckc}
    Contopoulos, I., Kazanas, D., \& Christodoulou, D. M. 2006, \apj, 652, 1451 (CKC)

\bibitem[Evans \& Hawley(1988)]{eh}
    Evans, C. R., \& Hawley, J. F. 1988, \apj, 332, 659

\bibitem[Fleming \& Stone(2003)]{fs}
    Fleming, T. P., \& Stone, J. M. 2003, \apj, 585, 908

\bibitem[Fleming, Stone \& Hawley(2000)]{fsh}
    Fleming, T. P., Stone, J. M., \& Hawley, J. F. 2000, \apj, 530, 464

\bibitem[Hawley(2000)]{h}
        Hawley, J. F. 2000, \apj, 528, 462

\bibitem[Hawley \& Balbus(2002)]{hb}
        Hawley, J. F., \& Balbus, S. A. 2002, \apj, 573, 738

\bibitem[Hawley, Balbus, \& Stone(2001)]{hbs}
        Hawley, J. F., Balbus, S. A., \& Stone, J. M. 2001, \apj, 554, L49

\bibitem[Hawley \& Krolik(2002)]{hk}
        Hawley, J. F., \& Krolik, J. H. 2002, \apj, 566, 164

\bibitem[Igumenshchev, Narayan, \& Abramowicz(2003)]{ina}
        Igumenshchev, I. V., Narayan, R., \& Abramowicz, M. A. 2003, \apj, 592, 1042

\bibitem[Ivezi\'c {\em et al.}(2004)]{i}
        Ivezi\'c, \v{Z}., Richards, G. T., Hall, P. B., Lupton, R. H., Jagoda, A. S., Knapp, G. R., Gunn, J. E.,
        Strauss, M. A., Schlegel, D., Steinhardt, W., \& Siverd, R. J. 2004, ASP Conf. Ser., 311, 347

\bibitem[Kulsrud {\em et al.}(1997)]{KCOR97}
        Kulsrud, R. M., Cen, R., Ostriker, J. P., \& Ryu, D. 1997, \apj, 480, 481

\bibitem[Mirabel, \& Rodr\'{i}guez(1999)]{mr}
        Mirabel, I. F., \& Rodr\'{i}guez, L. F. 1999, \araa, 37, 409

\bibitem[Narayan \& Yi (1994)]{ny}
    Narayan, R., \& Yi, I. 1994, \apj, 428, L13

\bibitem[Papaloizou \& Pringle(1984)]{pp}
    Papaloizou, J. C. B., \& Pringle, J. E. 1984, \mnras, 208, 721

\bibitem[Shakura \& Sunyaev(1973)]{ss}
    Shakura, N. I., \& Sunyaev, R. A. 1973, A\&A, 24, 337

\bibitem[Stone et al. (1992)]{shen}
    Stone, J. M., Hawley, J. F., Evans, C. R., \& Norman, M. L. 1992, \apj, 388, 415

\bibitem[Stone \& Norman(1992a)]{sn1}
    Stone, J. M., \& Norman, M. L. 1992a, \apjs, 80, 753

\bibitem[Stone \& Norman(1992b)]{sn2}
    Stone, J. M., \& Norman, M. L. 1992b, \apjs, 80, 791

\bibitem[Stone \& Pringle(2001)]{springle}
    Stone, J. M., \& Pringle, J. E. 2001, \mnras, 322, 461

\bibitem[Tohline(1988)]{t}
    Tohline, J. E. 1988, unpublished

\bibitem[Vainshtein \& Rosner(1991)]{vr91}
Vainshtein, S. I., \& Rosner, R. 1991, \apj, 376, 199

\bibitem[van Leer(1977)]{v1}
    van Leer, B. 1977, J. Comp. Phys., 23, 276

\bibitem[van Leer(1979)]{v2}
    van Leer, B. 1979, J. Comp. Phys., 32, 101

\bibitem[Wilson, Young, \& Shopbell(2001)]{wys}
        Wilson, A. S., Young, A. J.,  \&  Shopbell, P. L. 2001, \apj, 547, 740

\bibitem[Xu, Livio, \& Baum(1999)]{xlb}
        Xu, C., Livio, M., \& Baum, S. 1999, \aj, 118, 1169

\end{thebibliography}
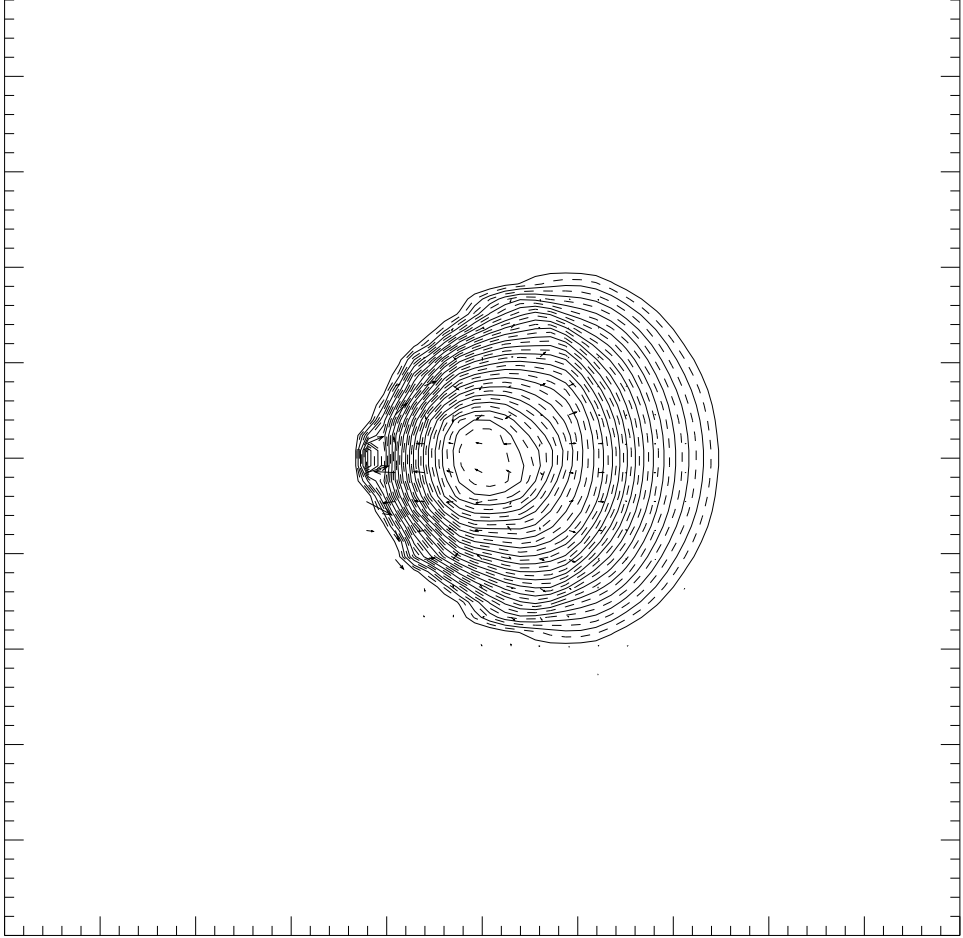
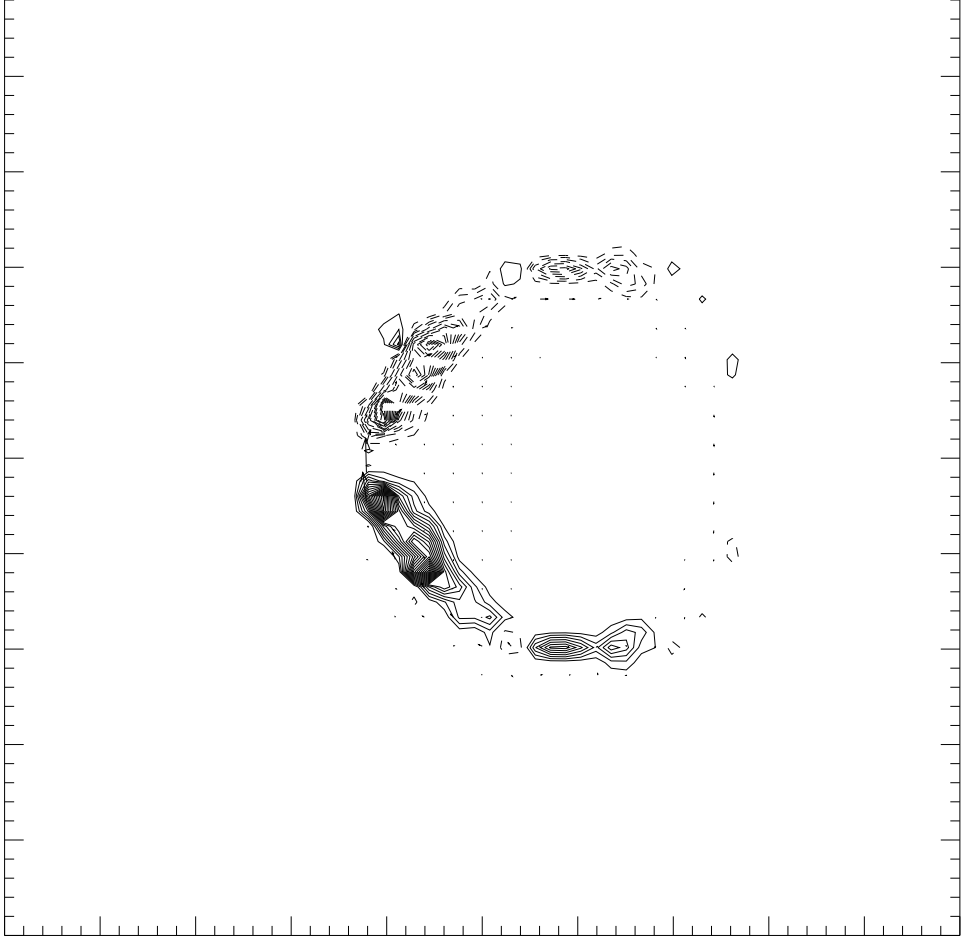
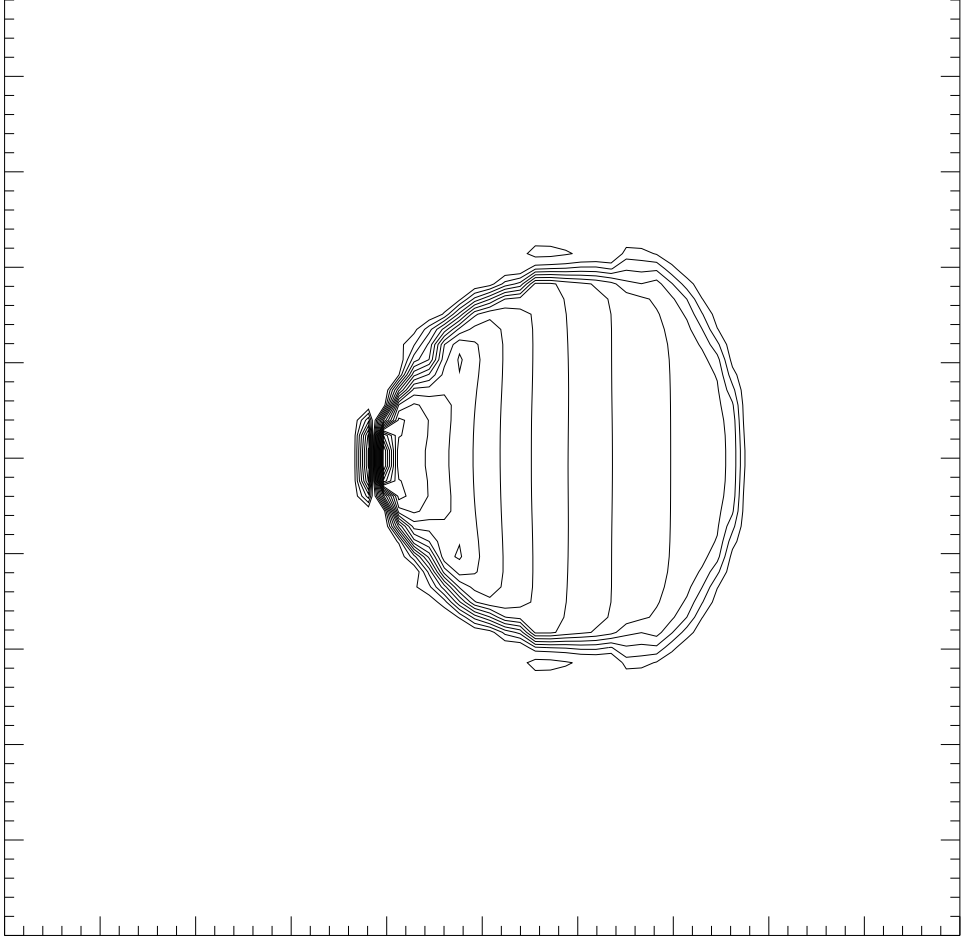
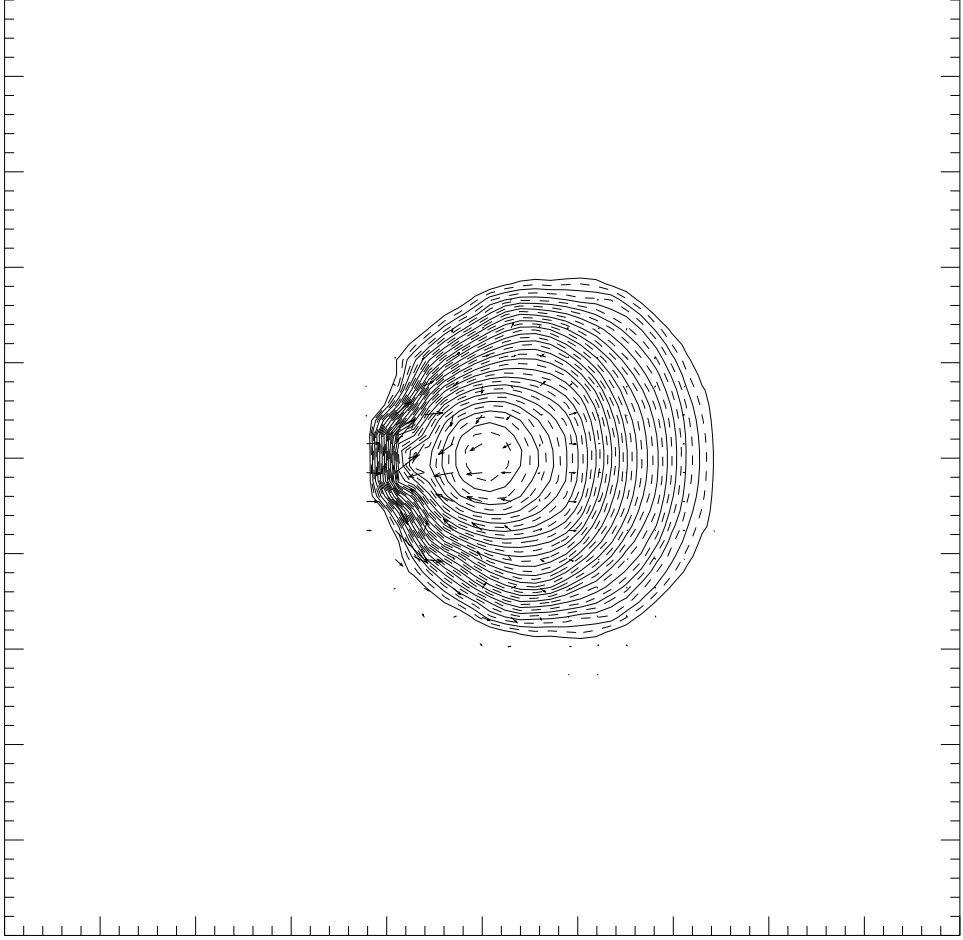
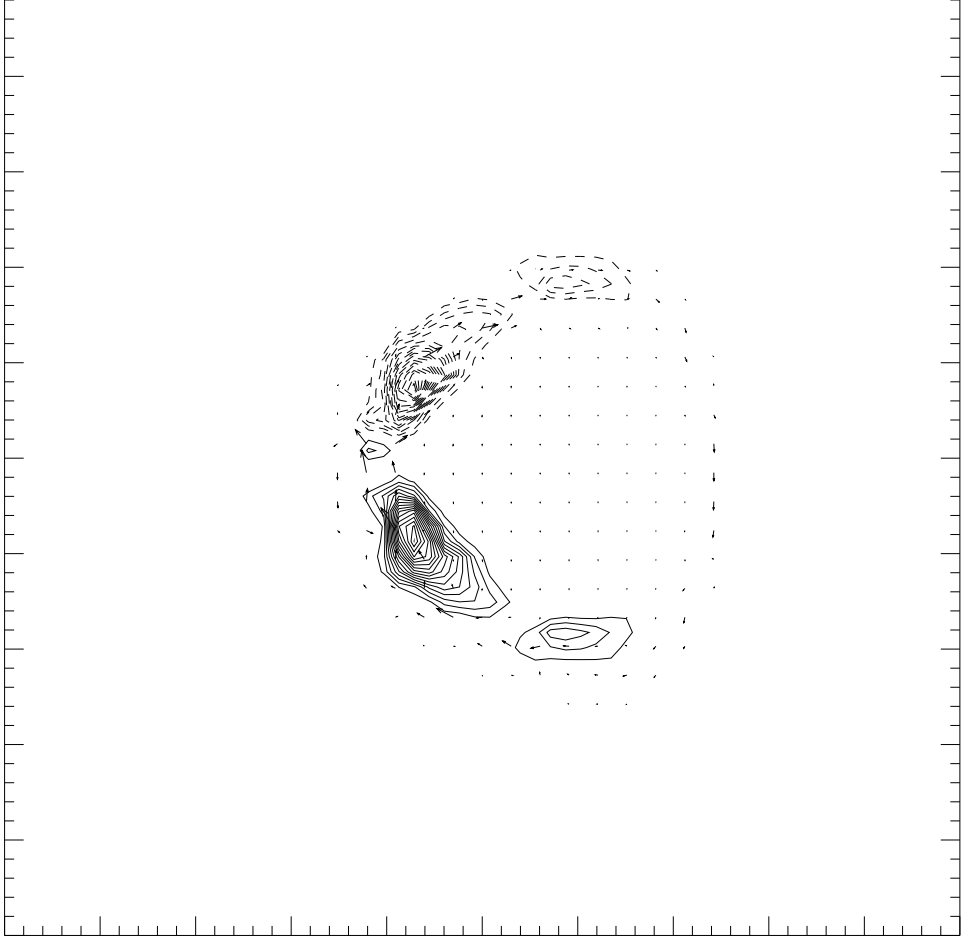
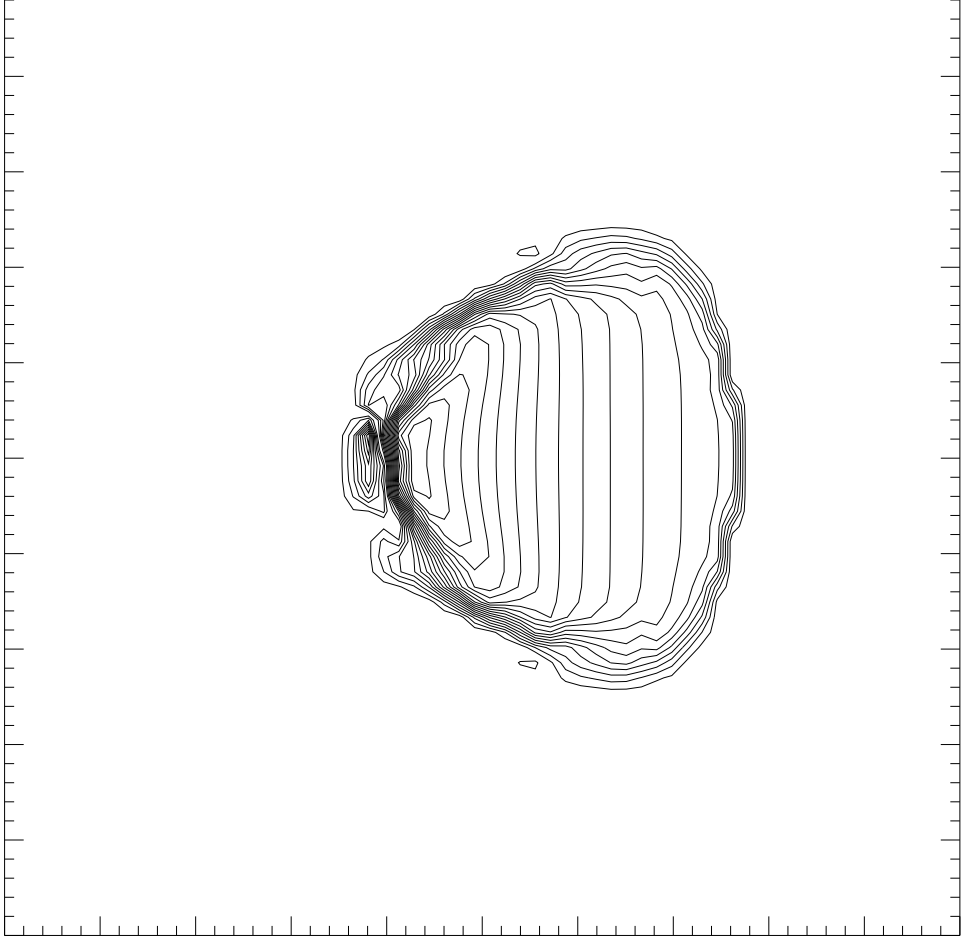
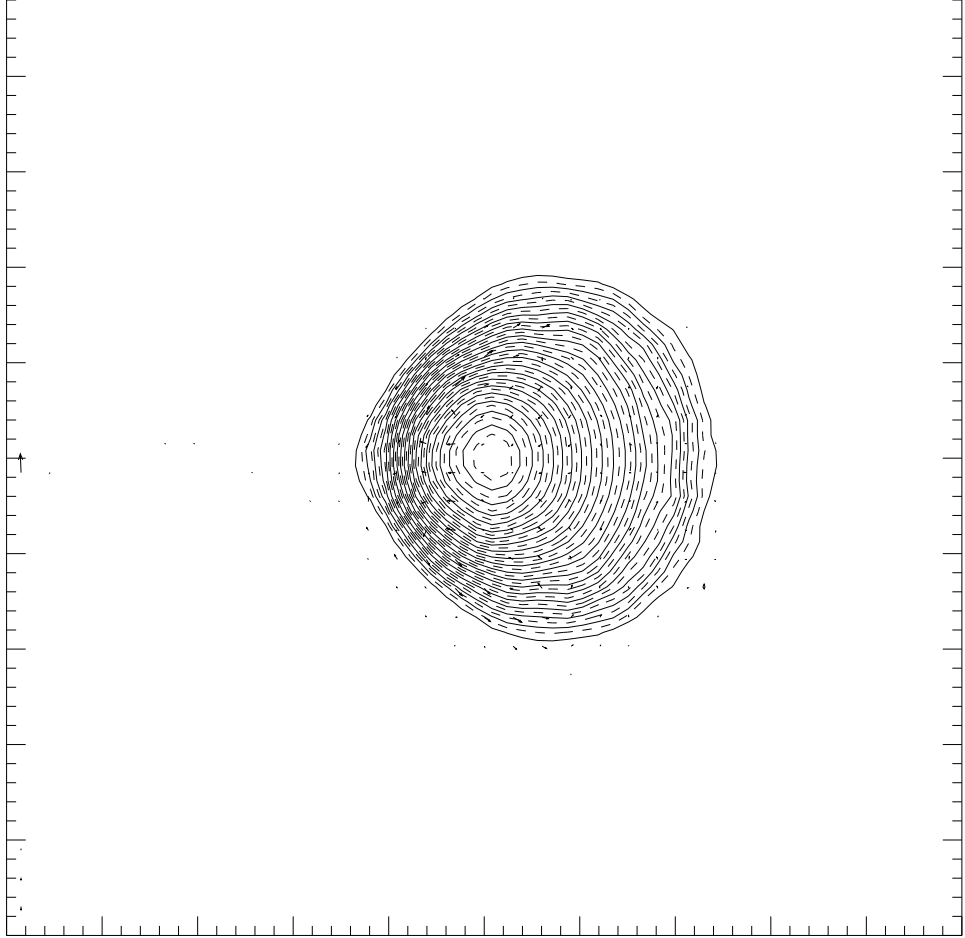
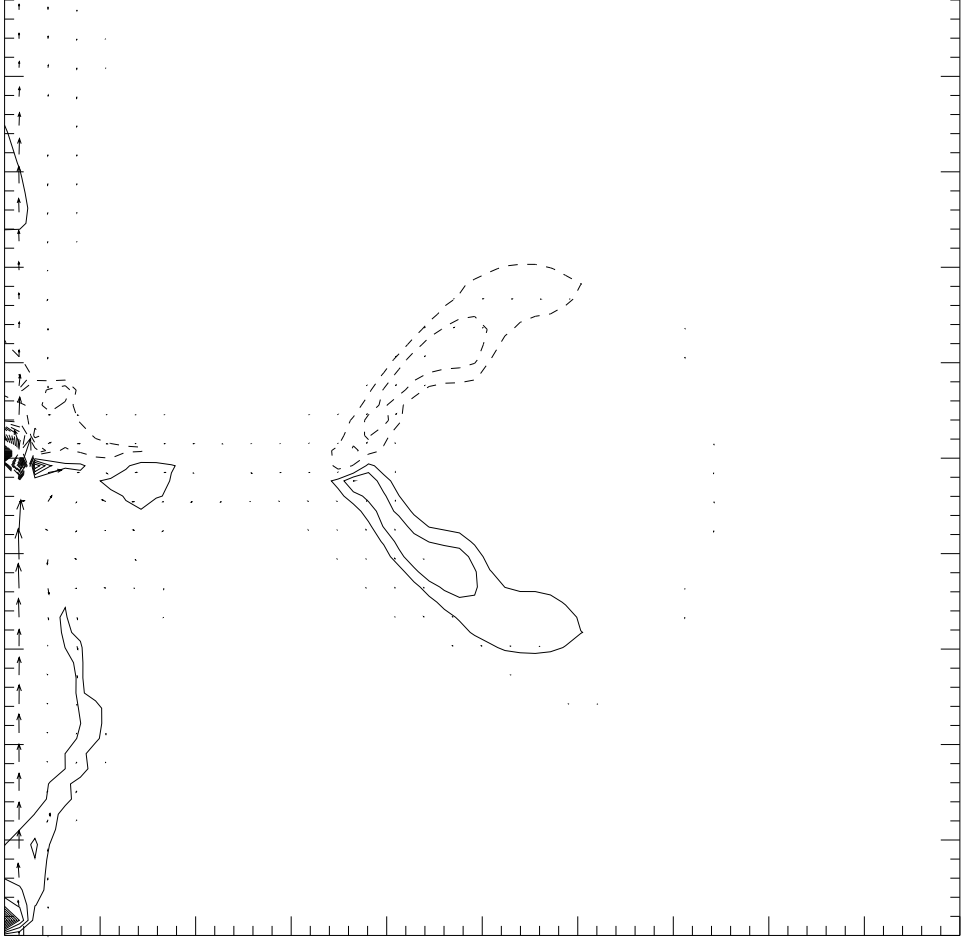
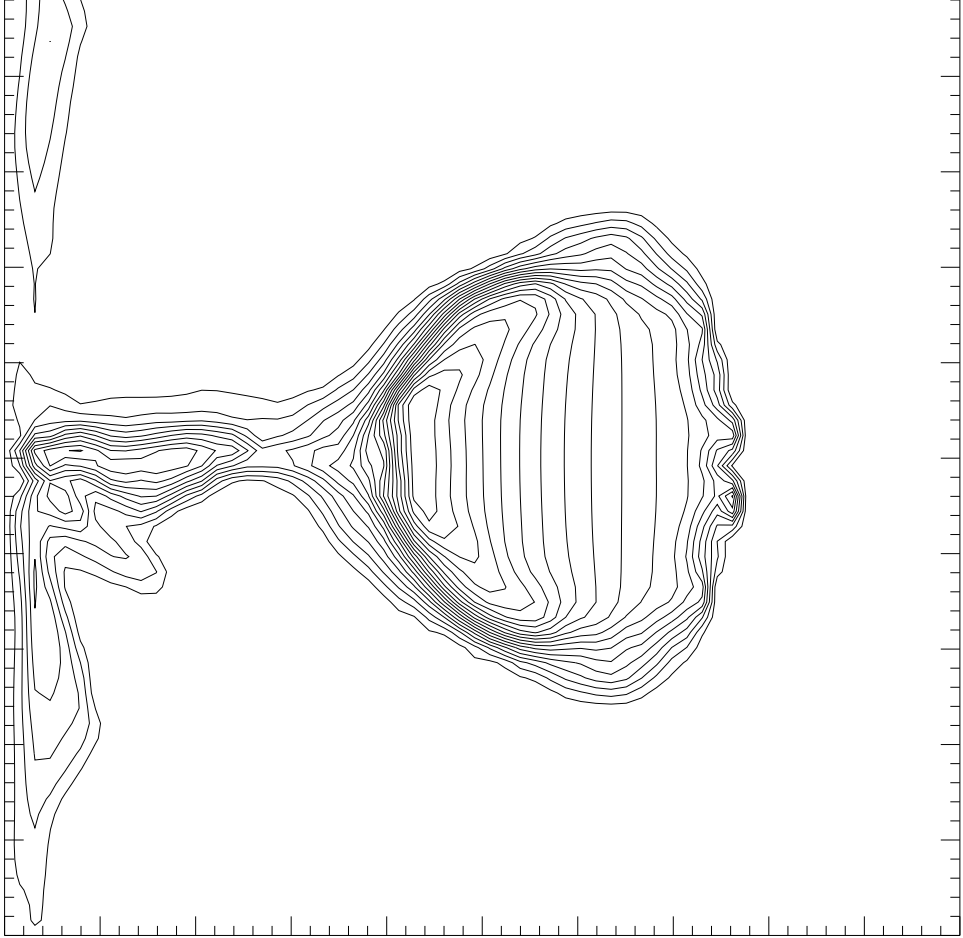
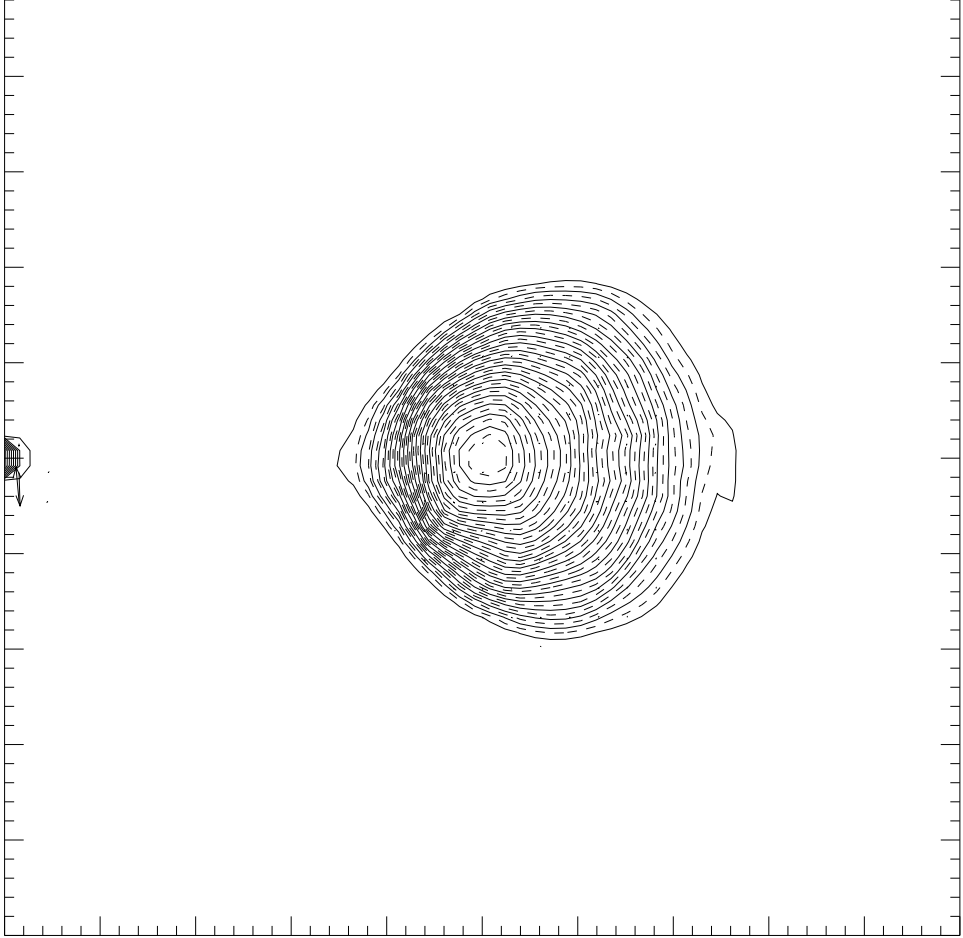
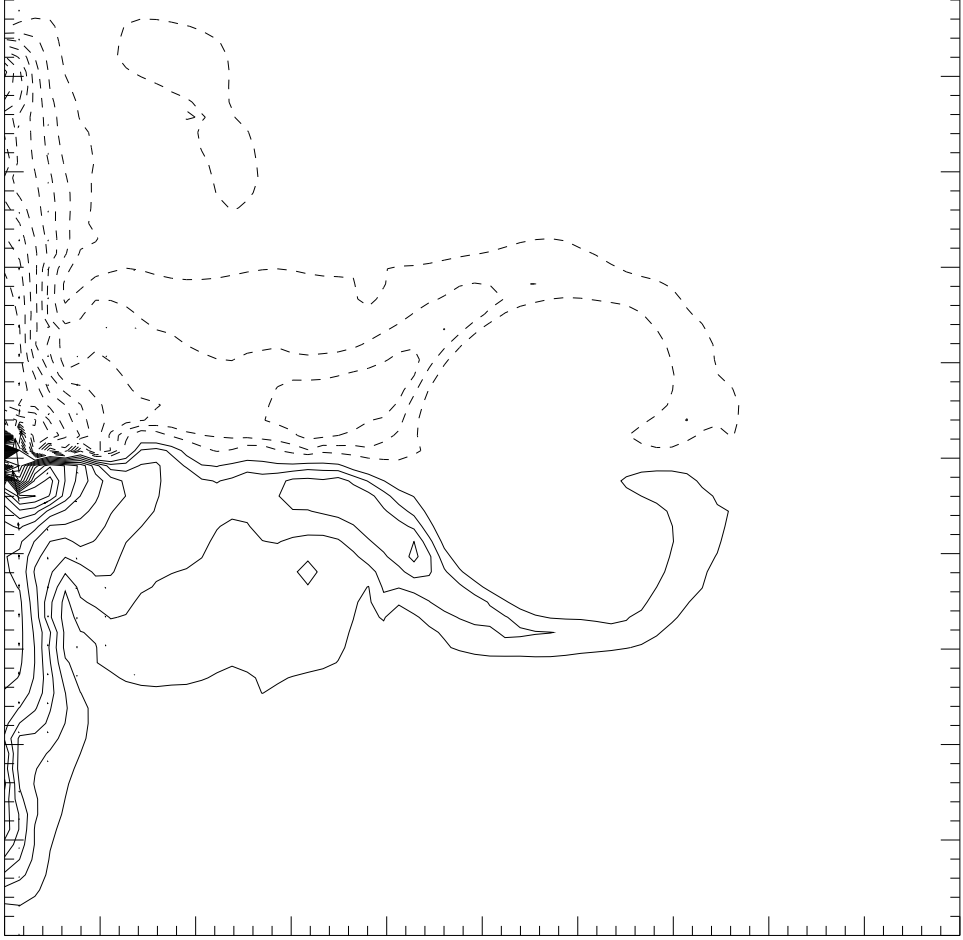
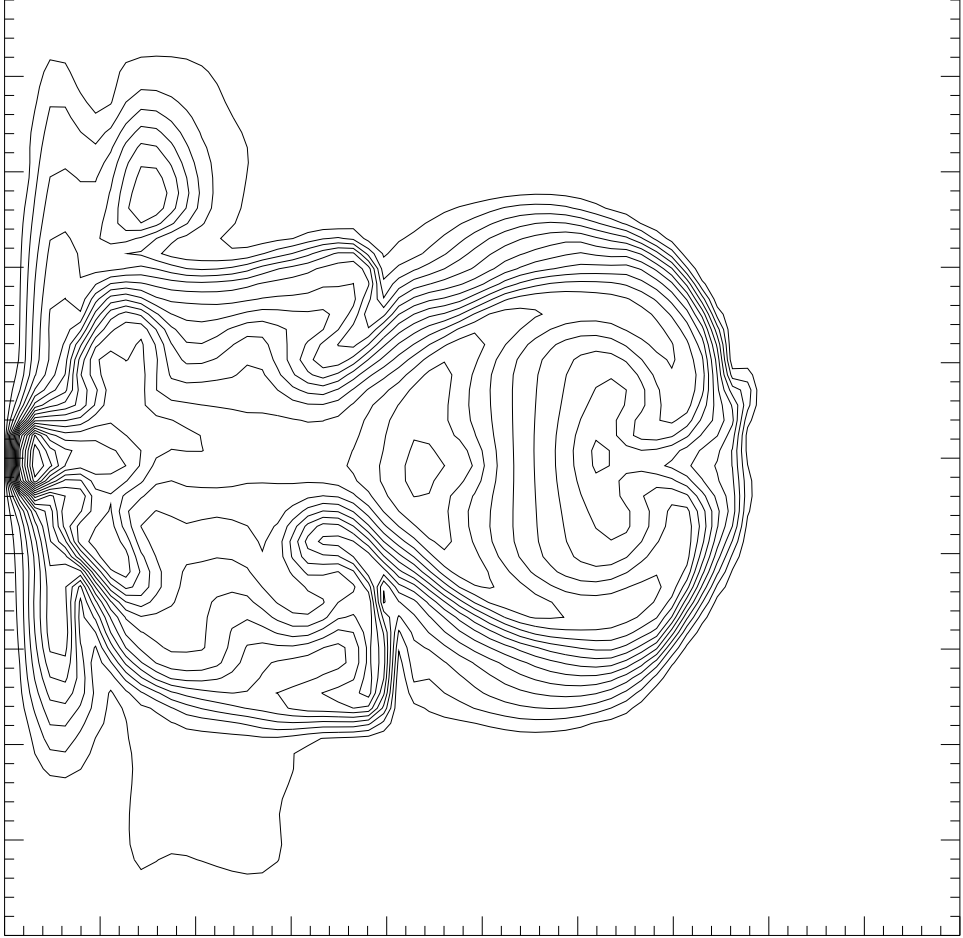
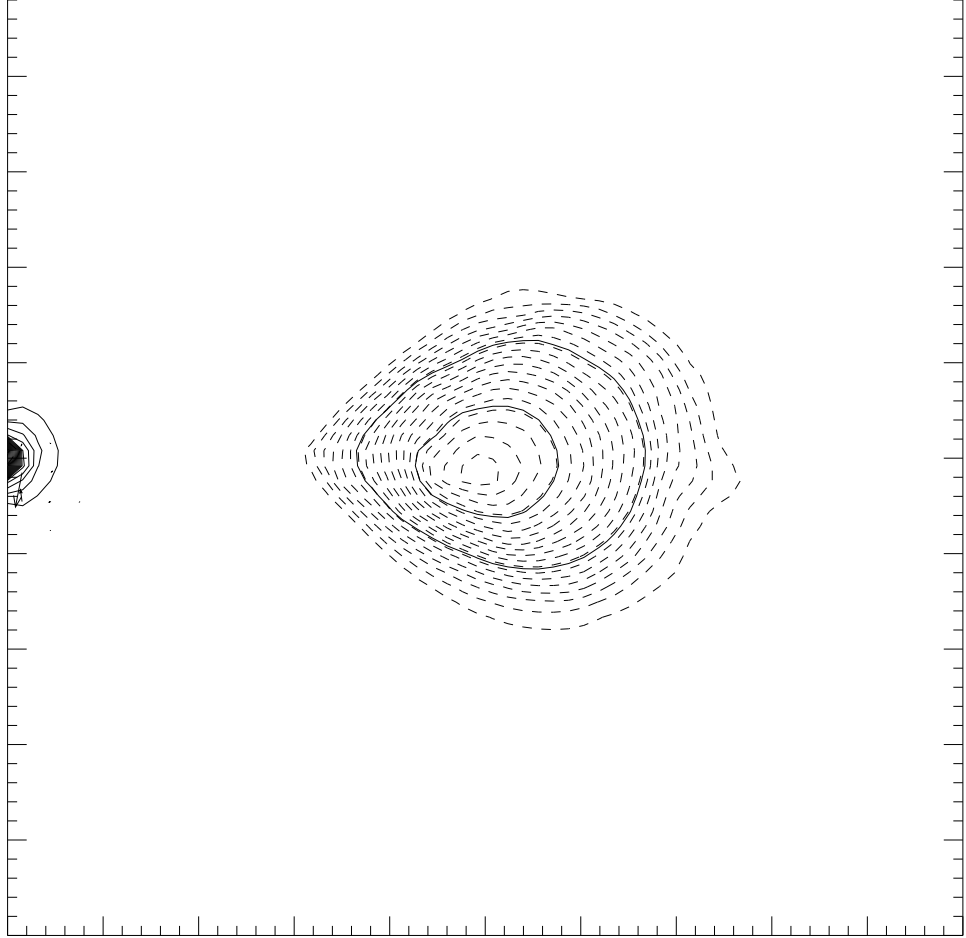
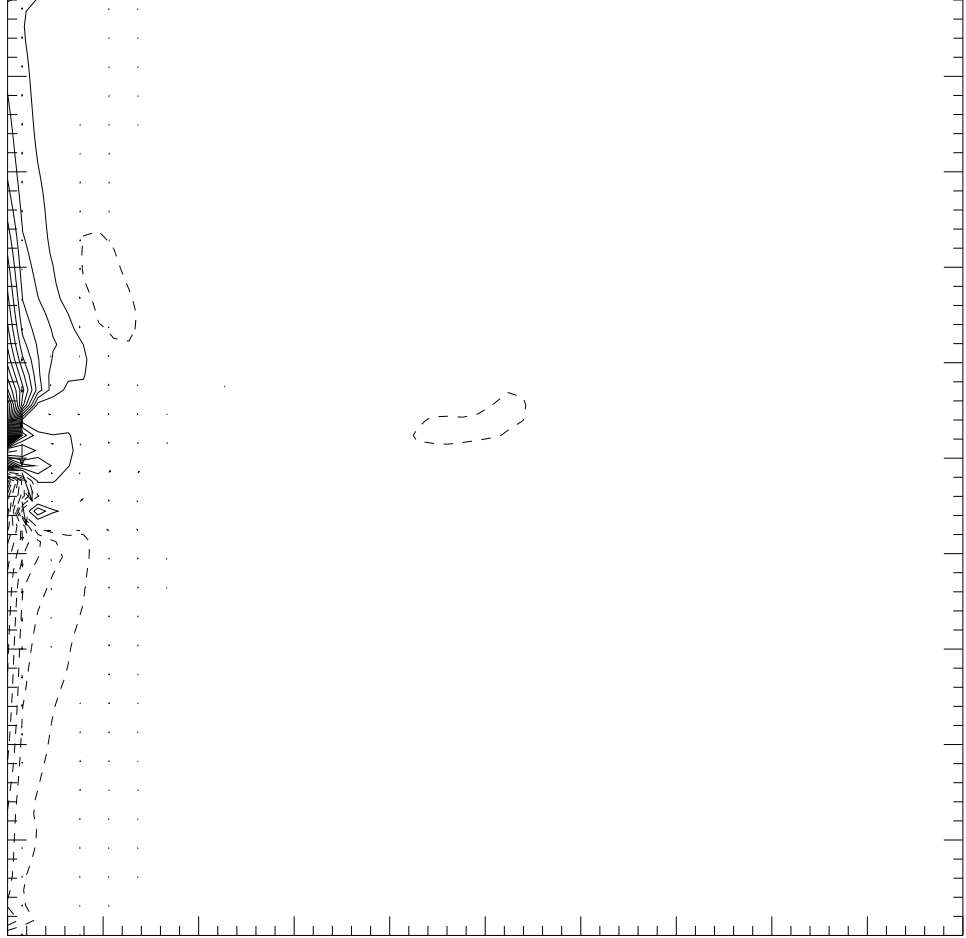
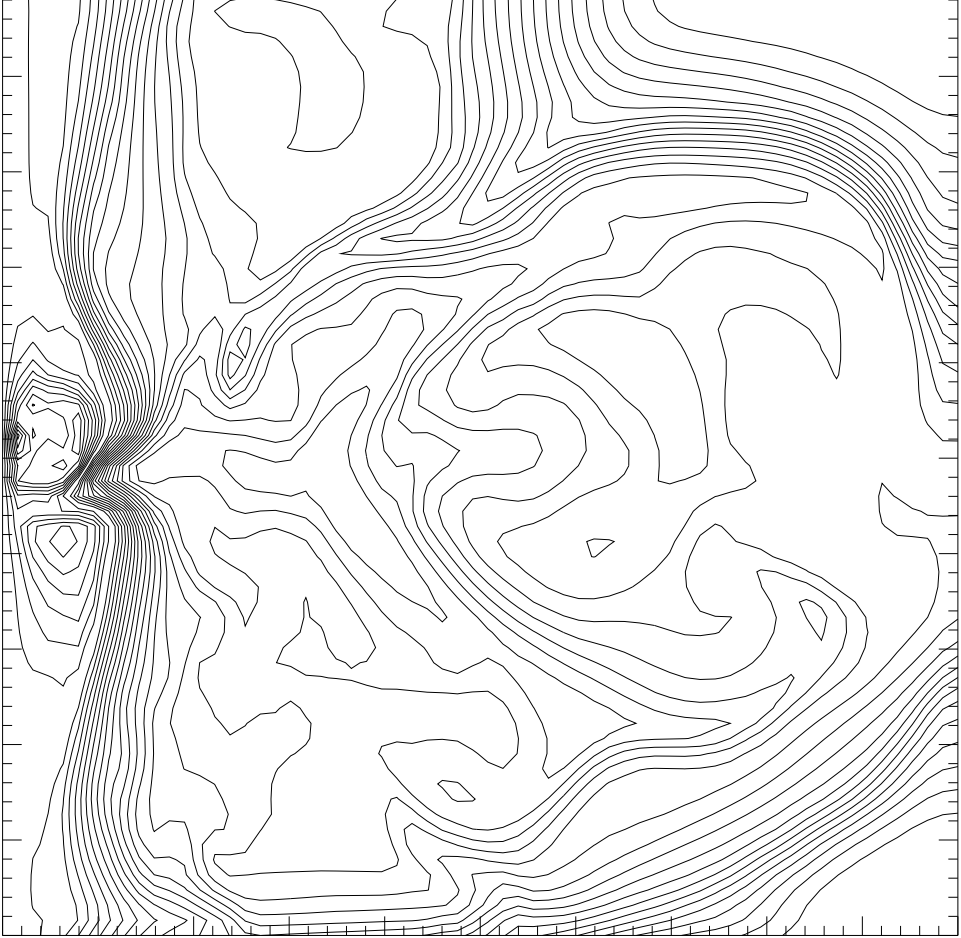
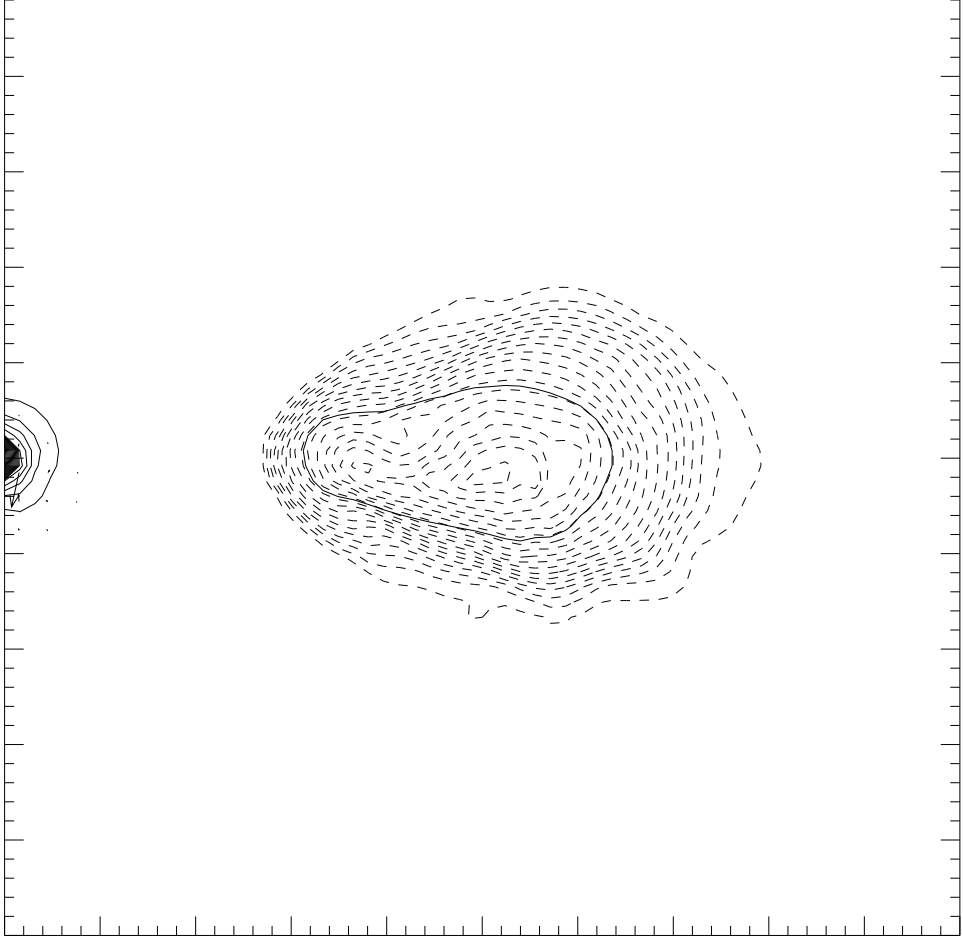
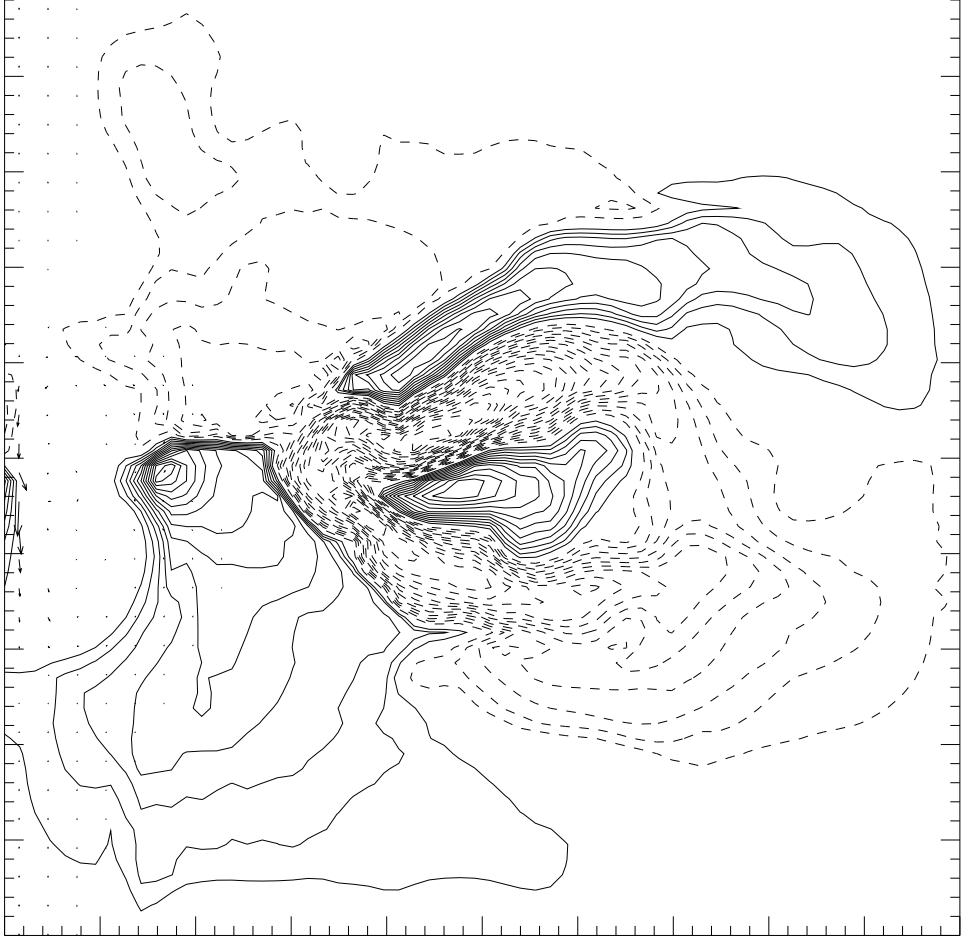
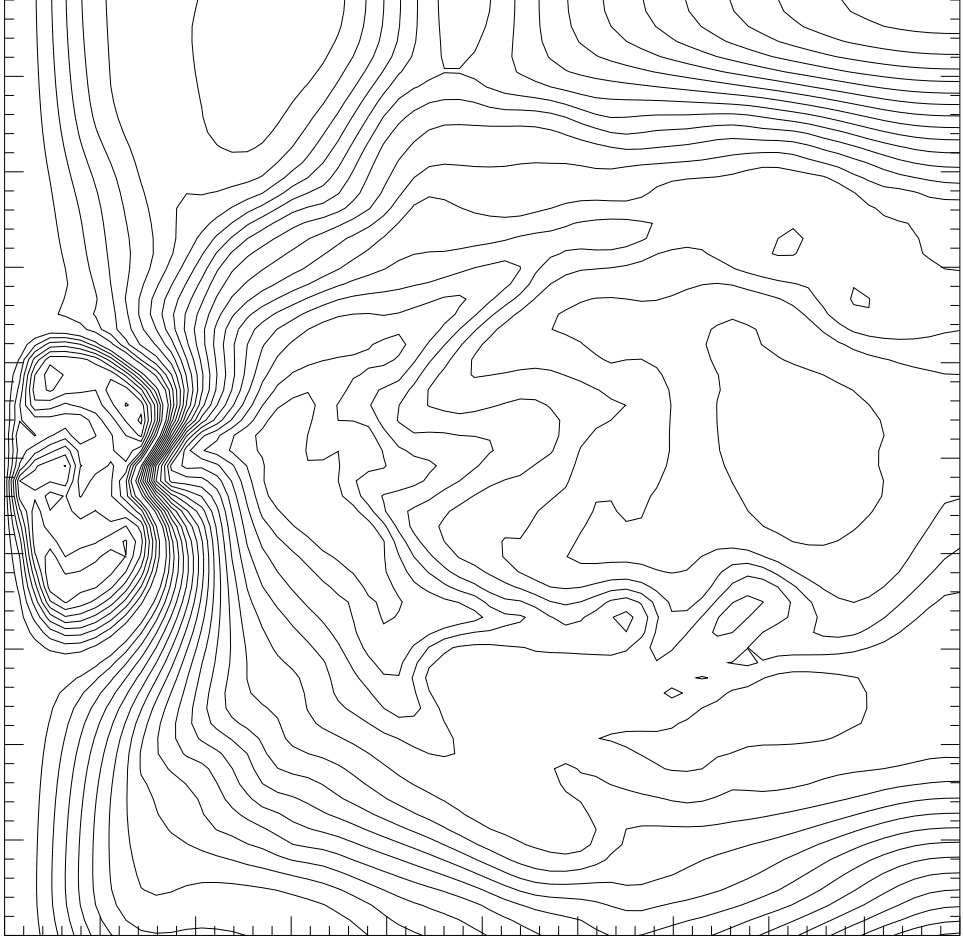
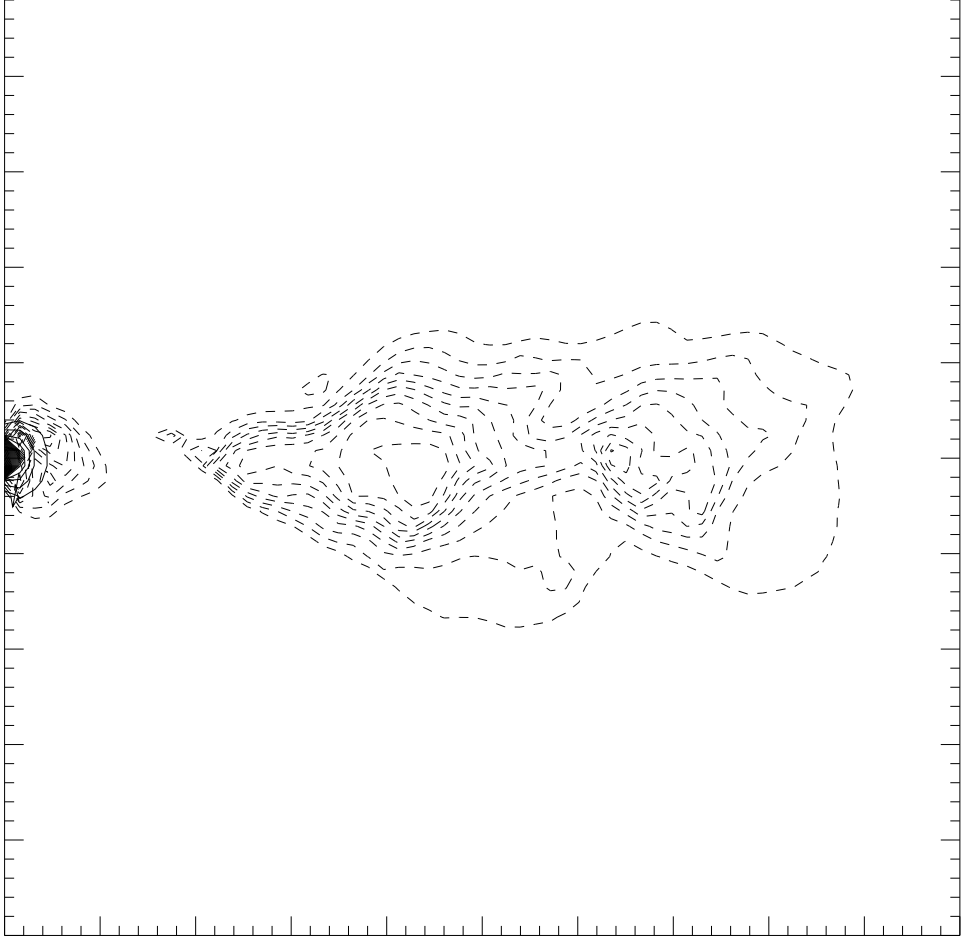
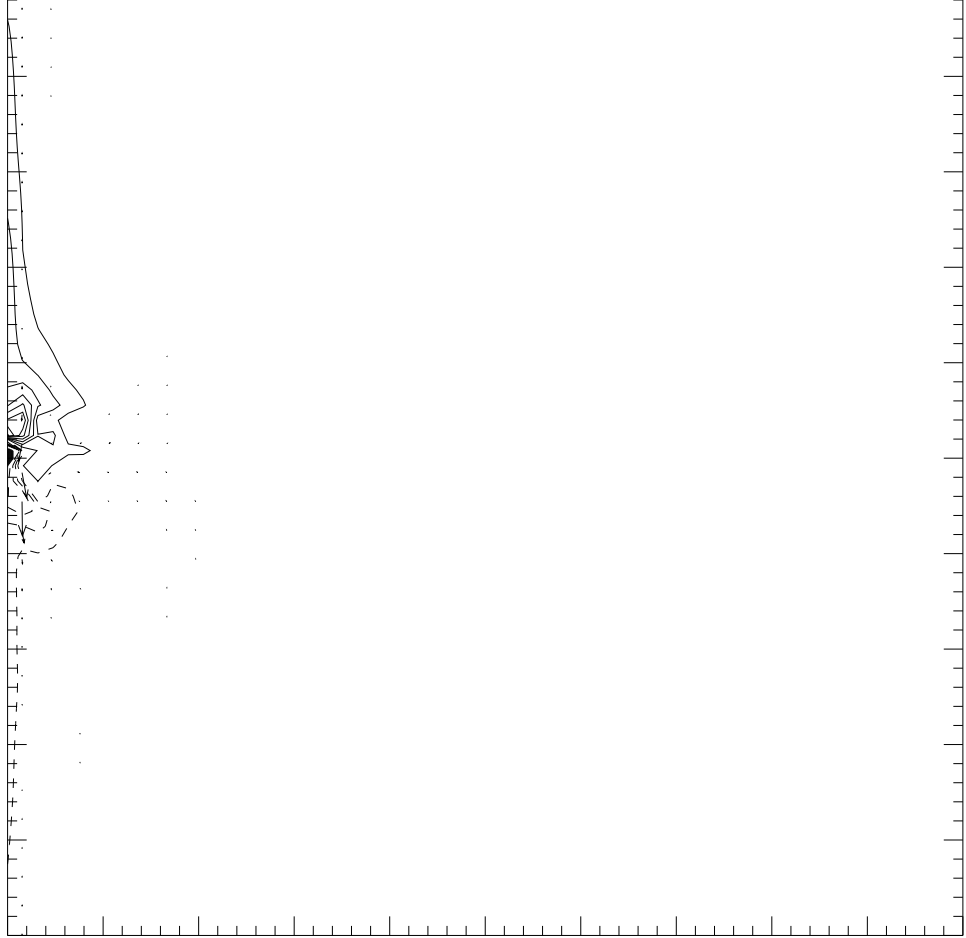
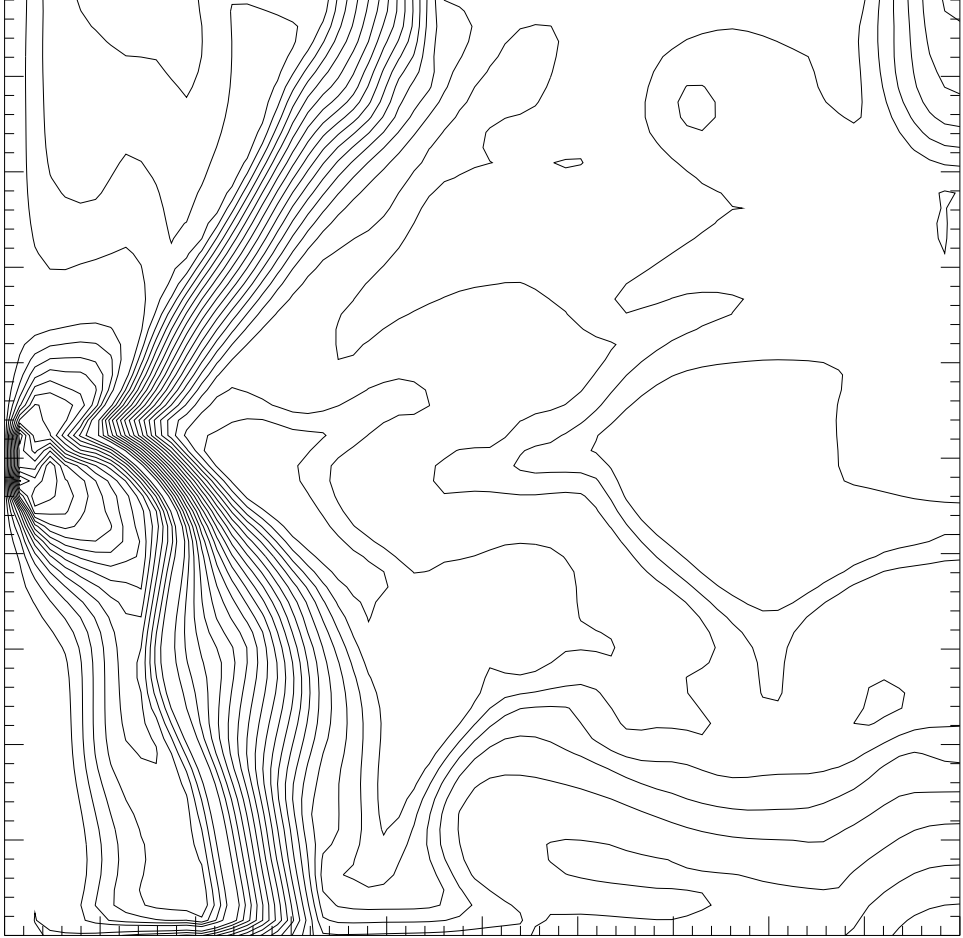
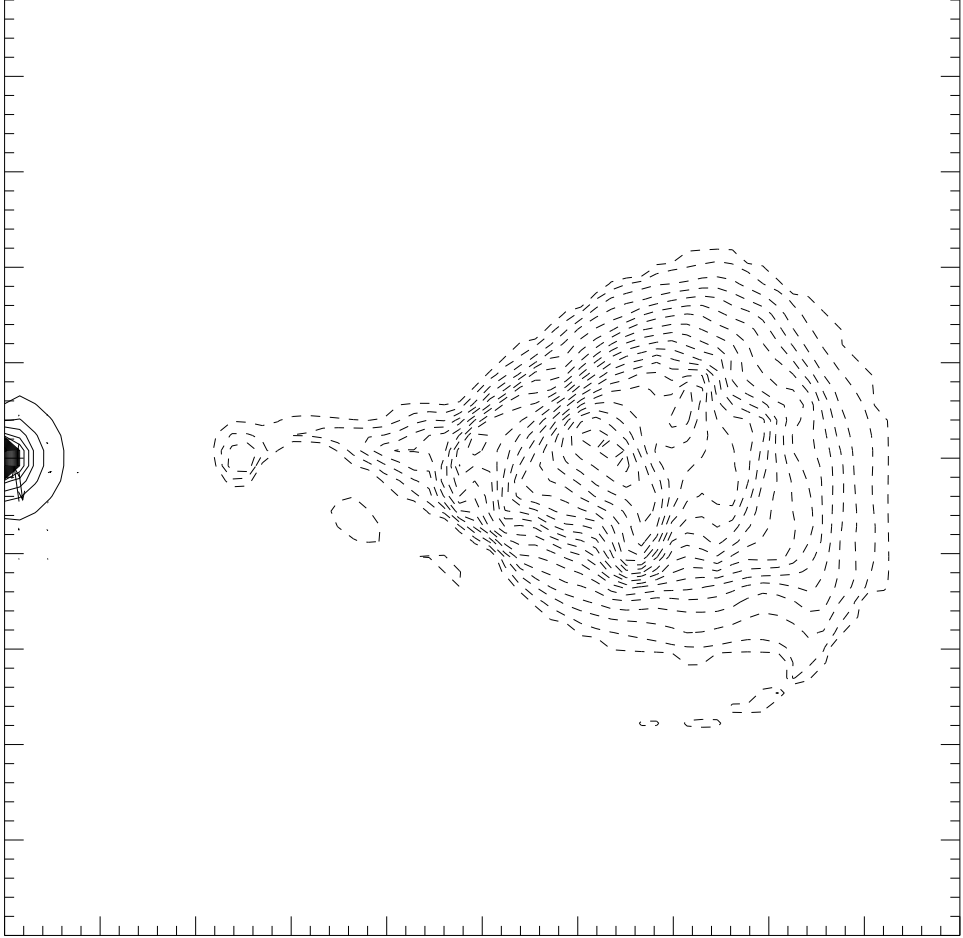
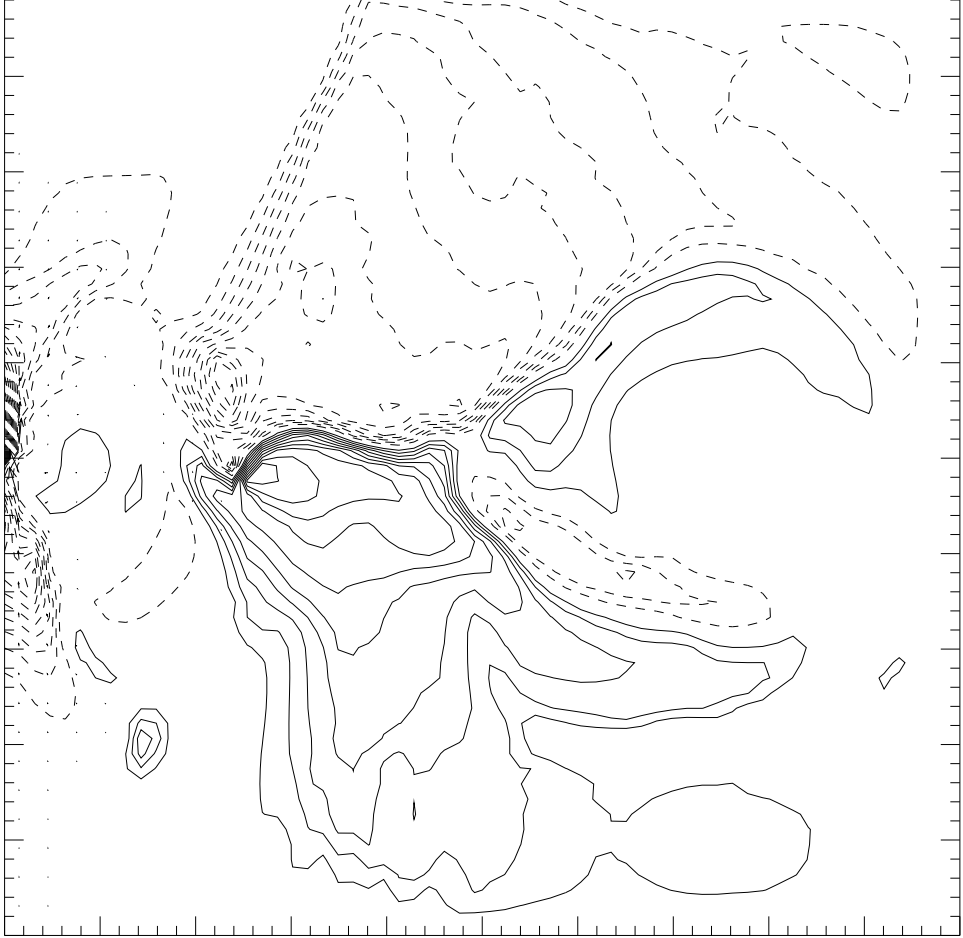
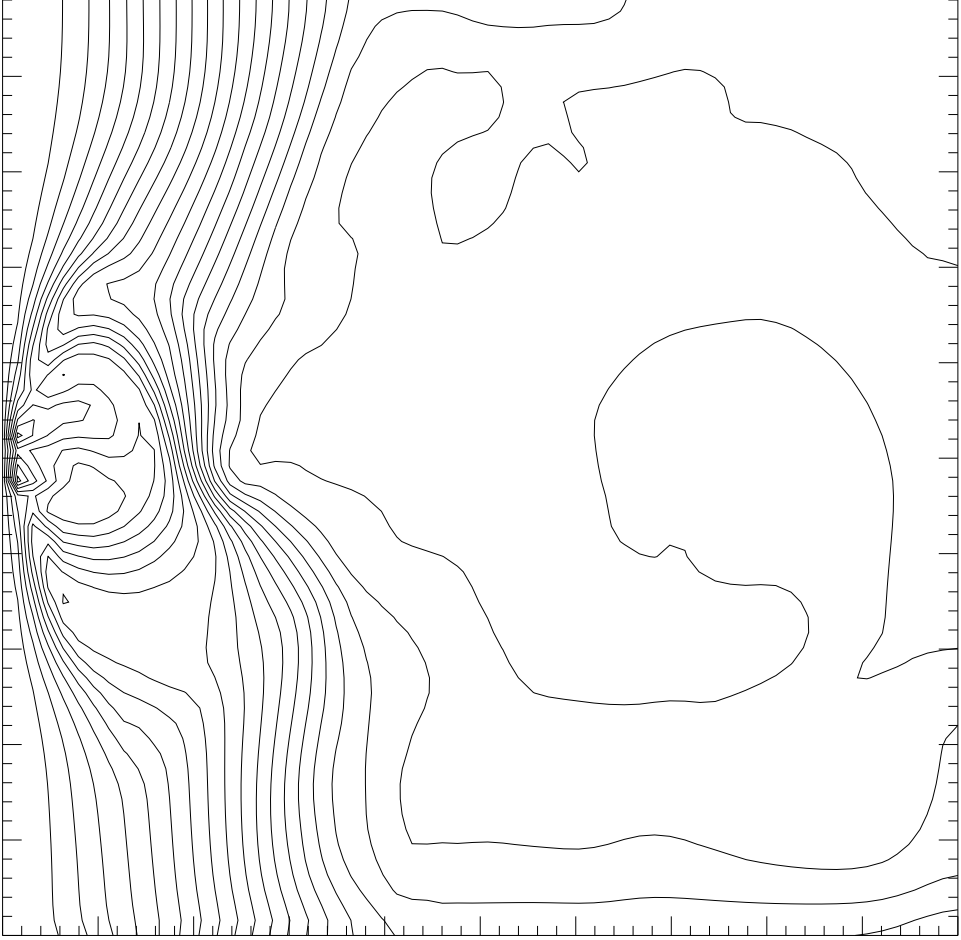
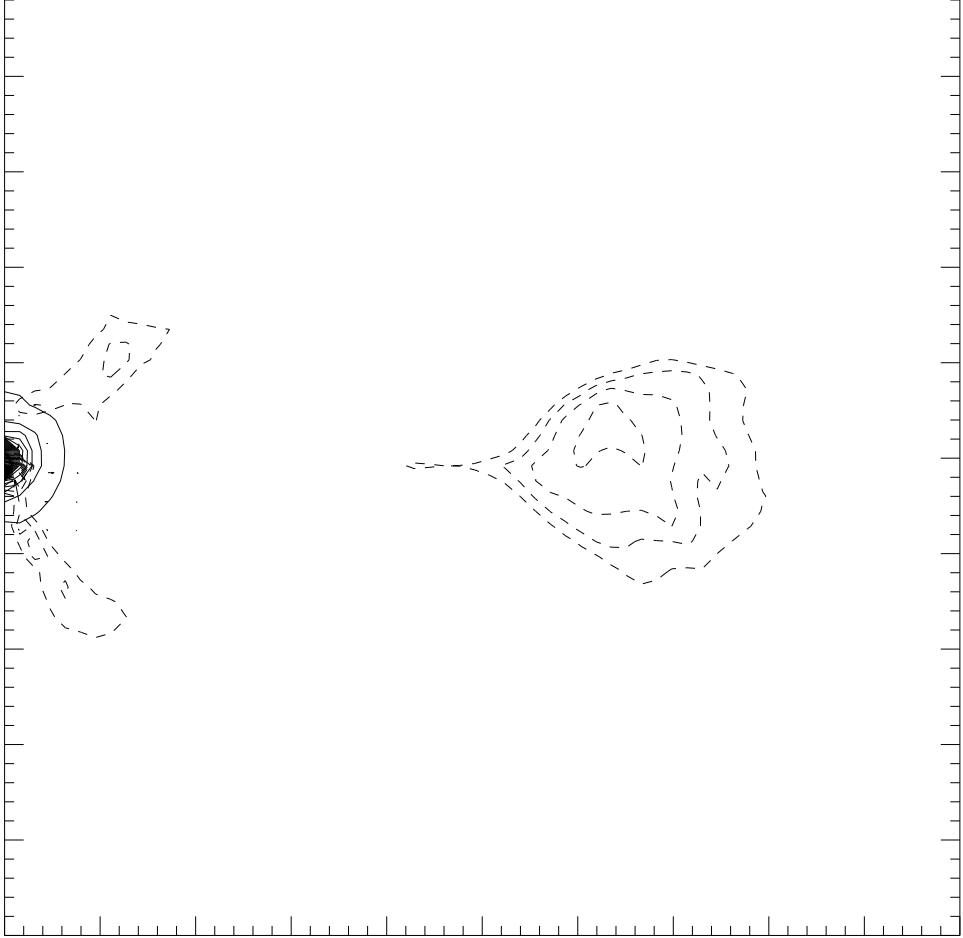
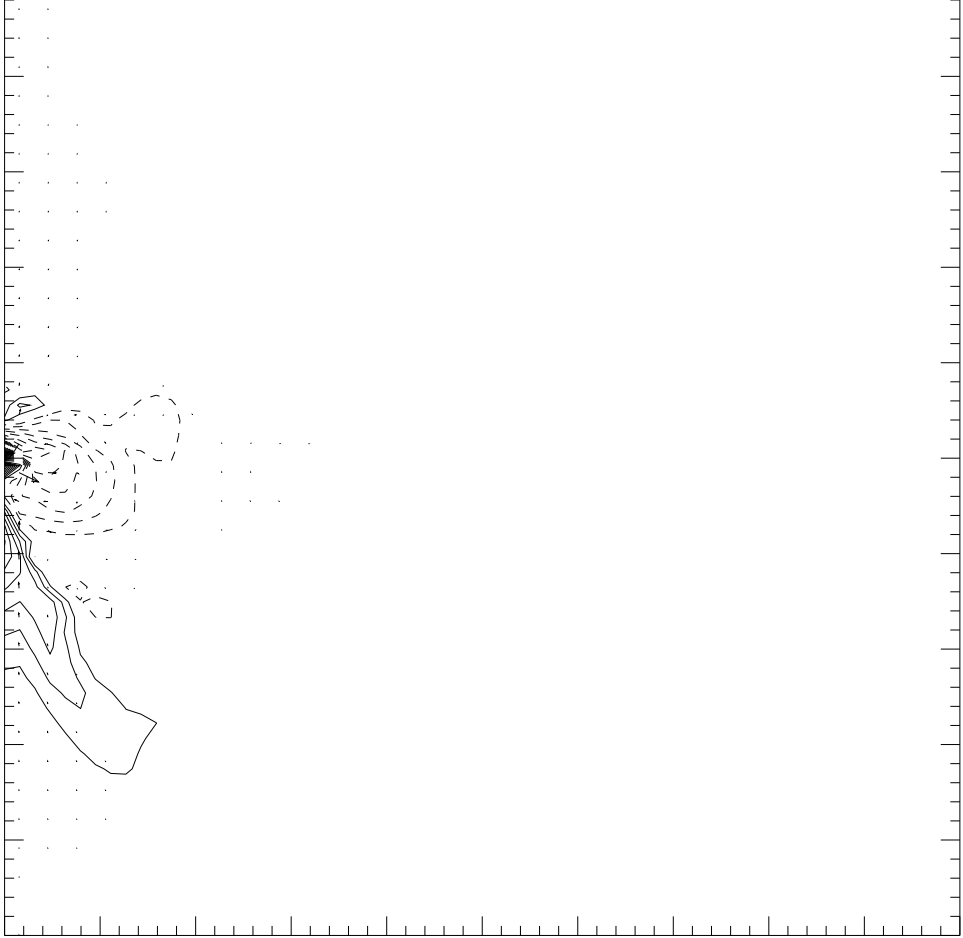
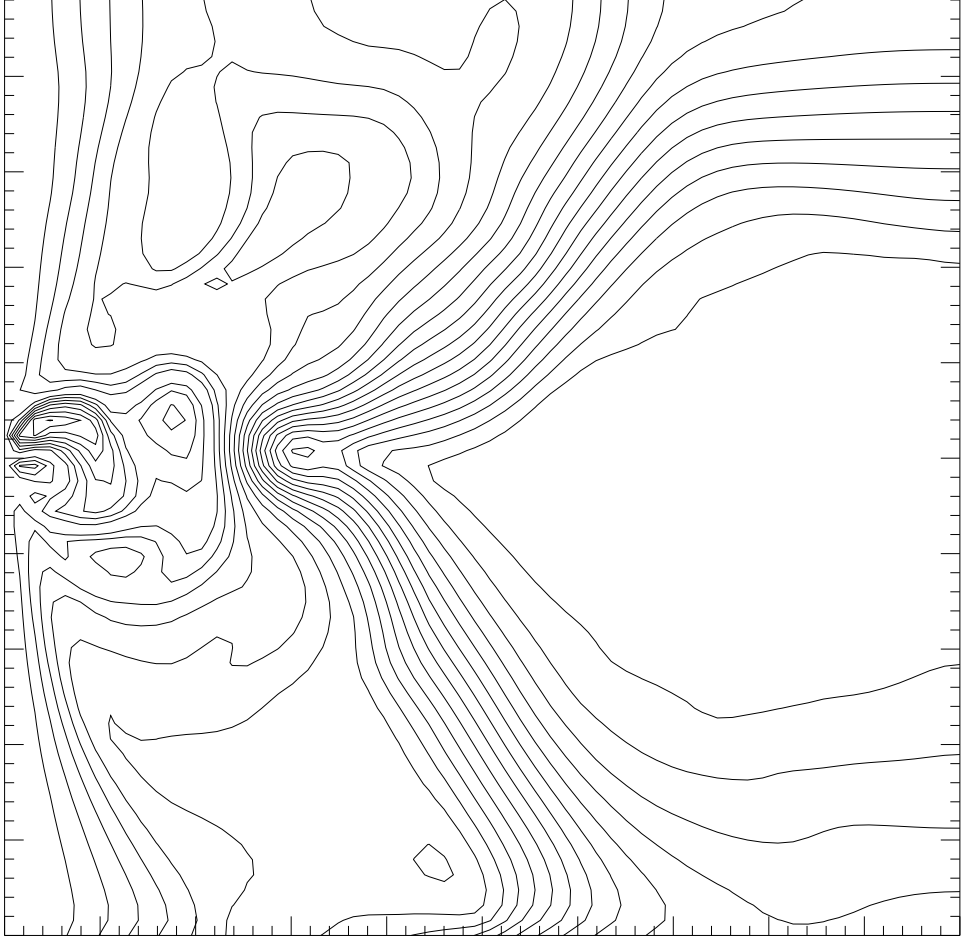
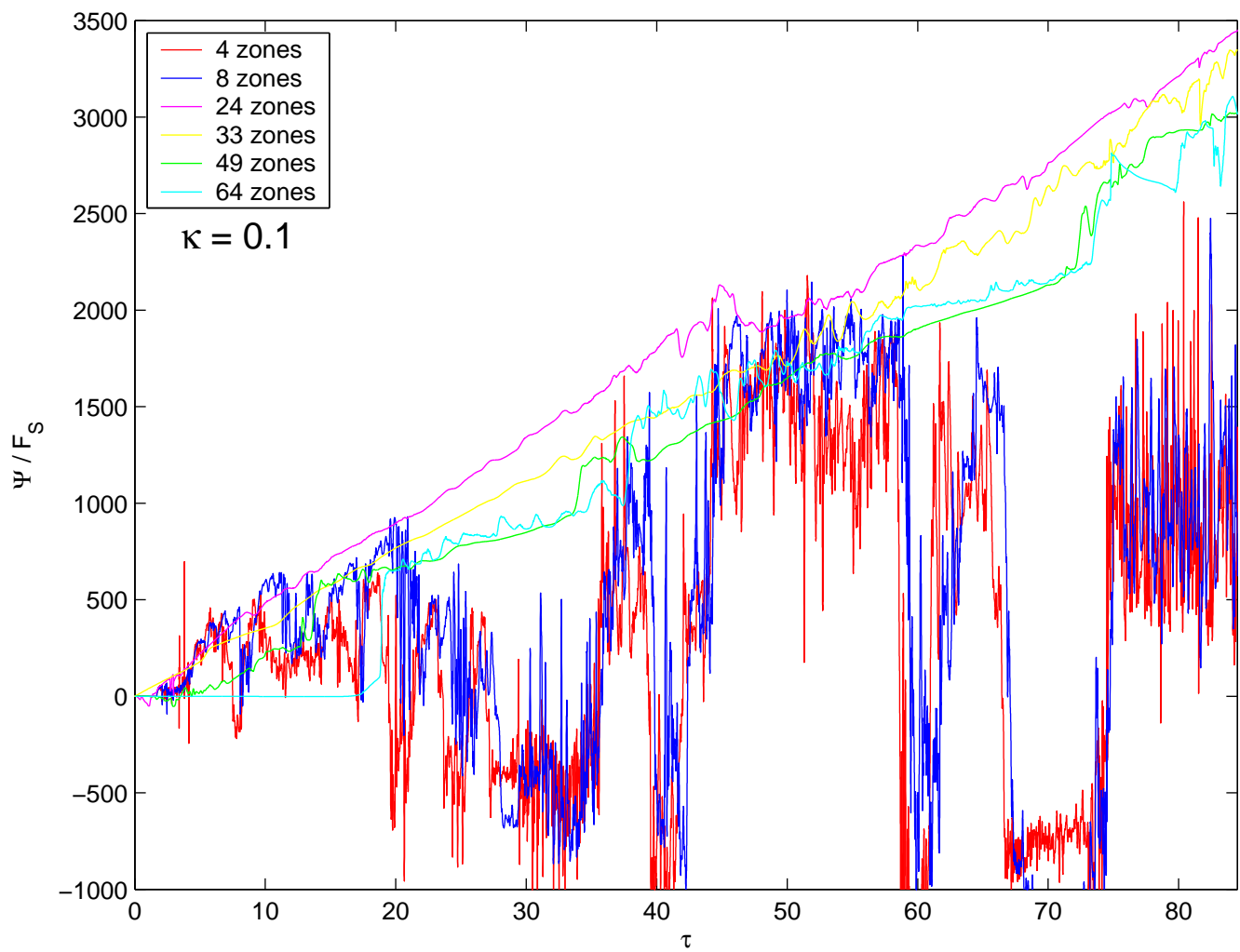
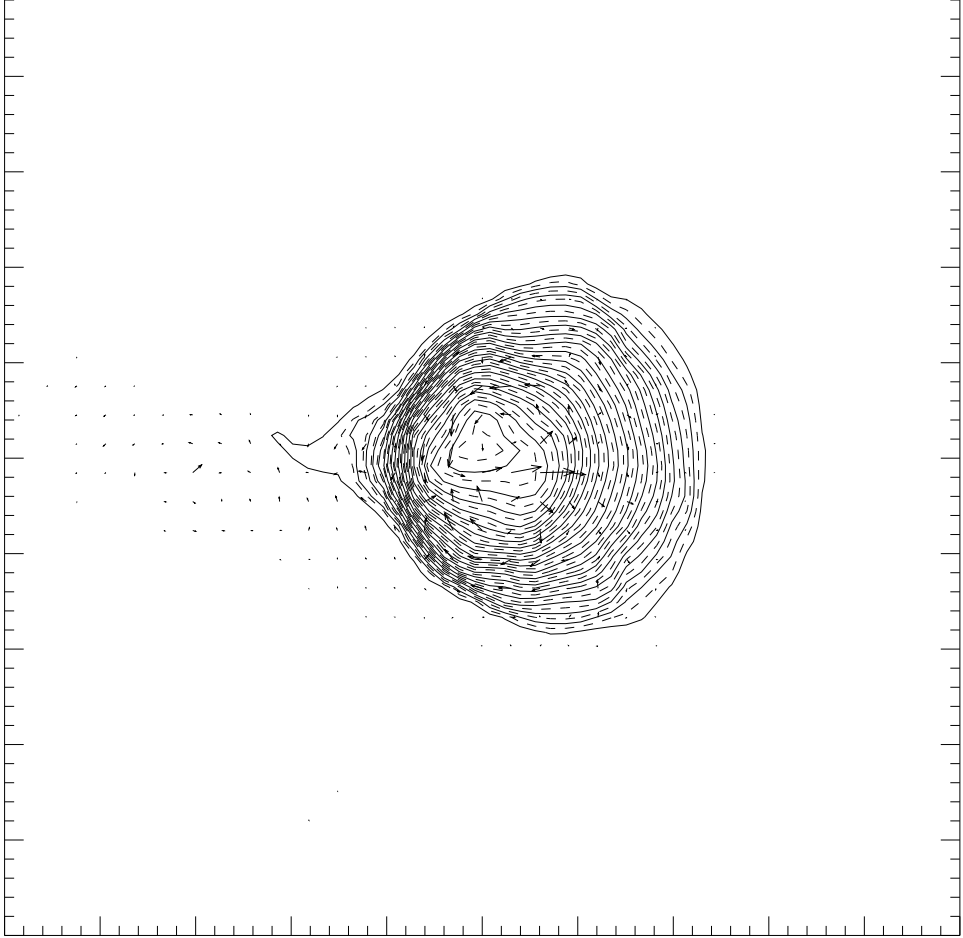
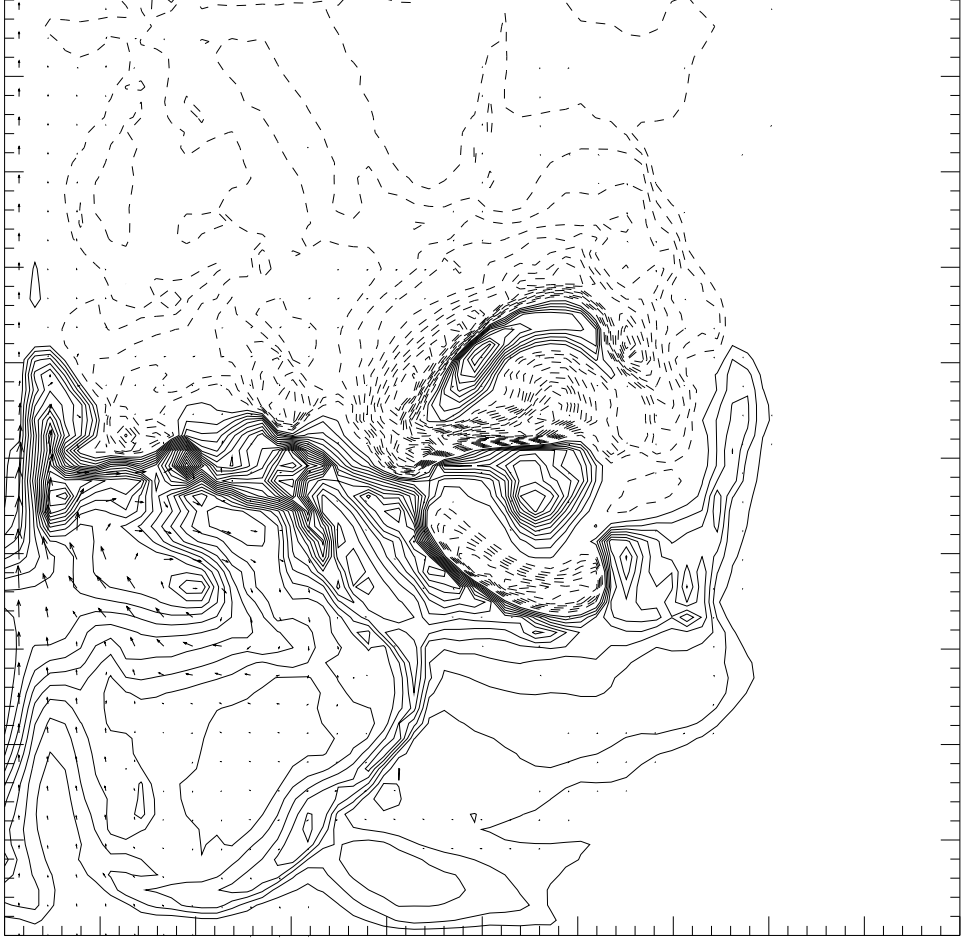
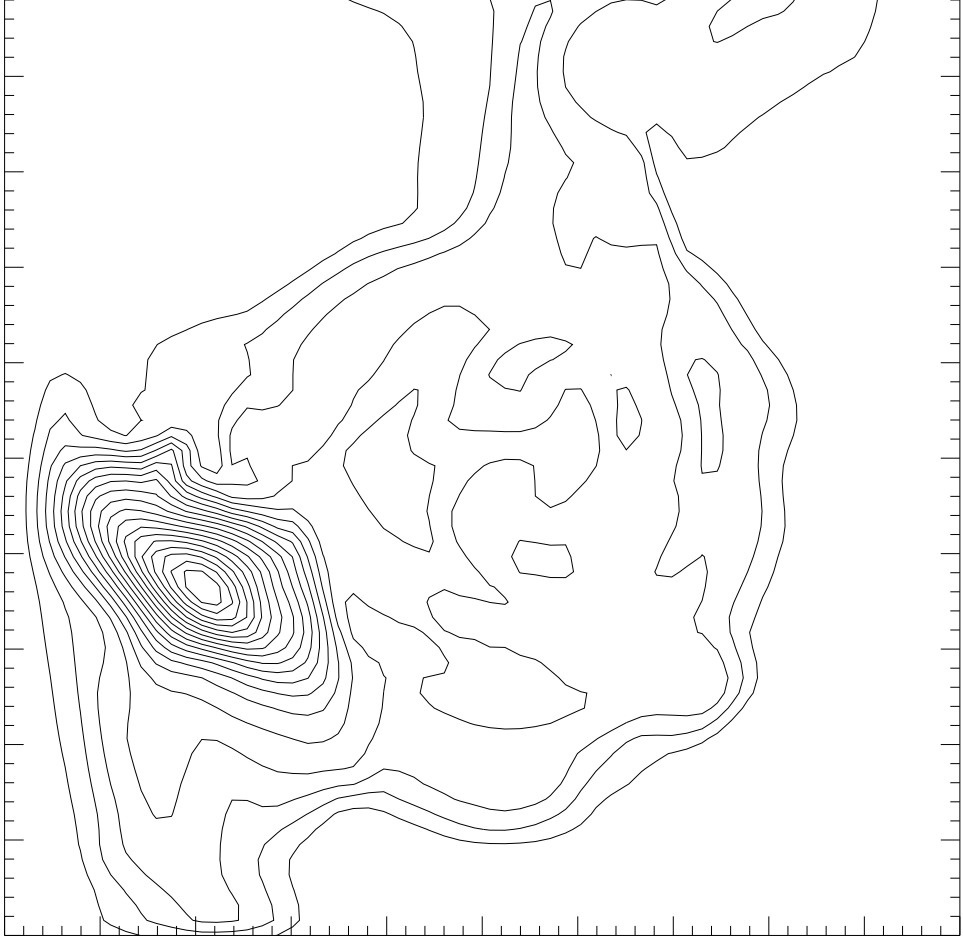
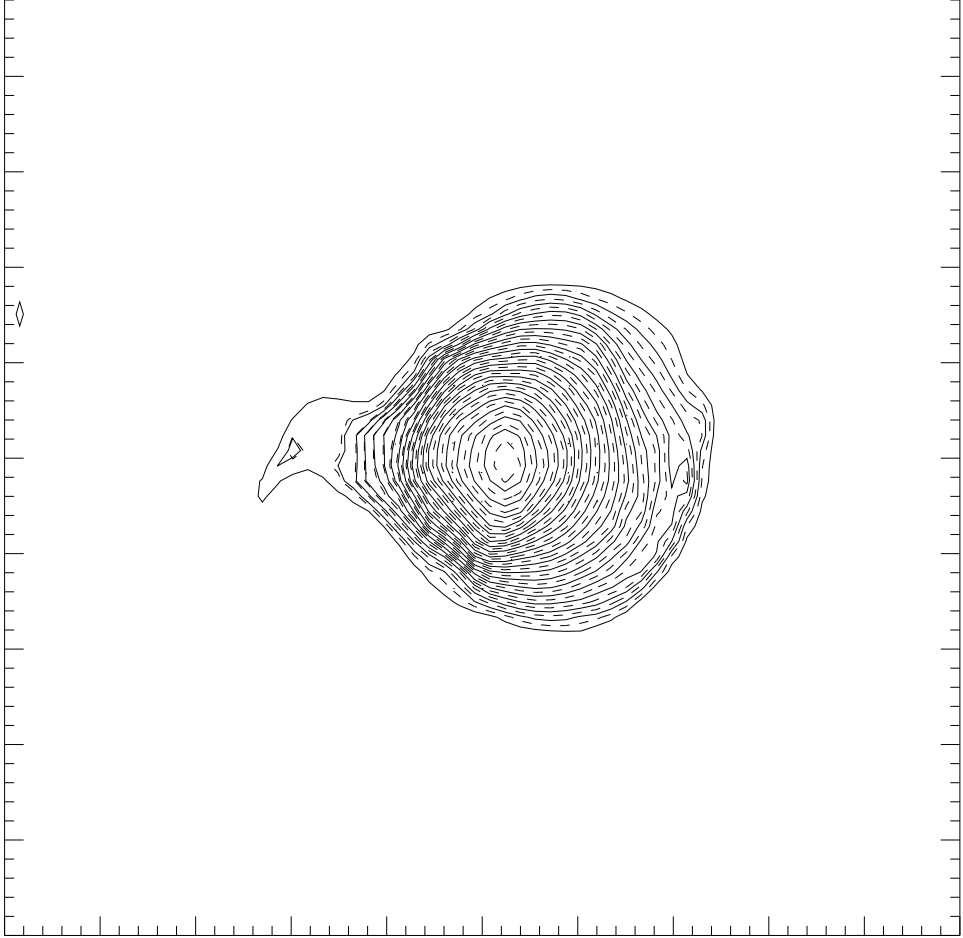
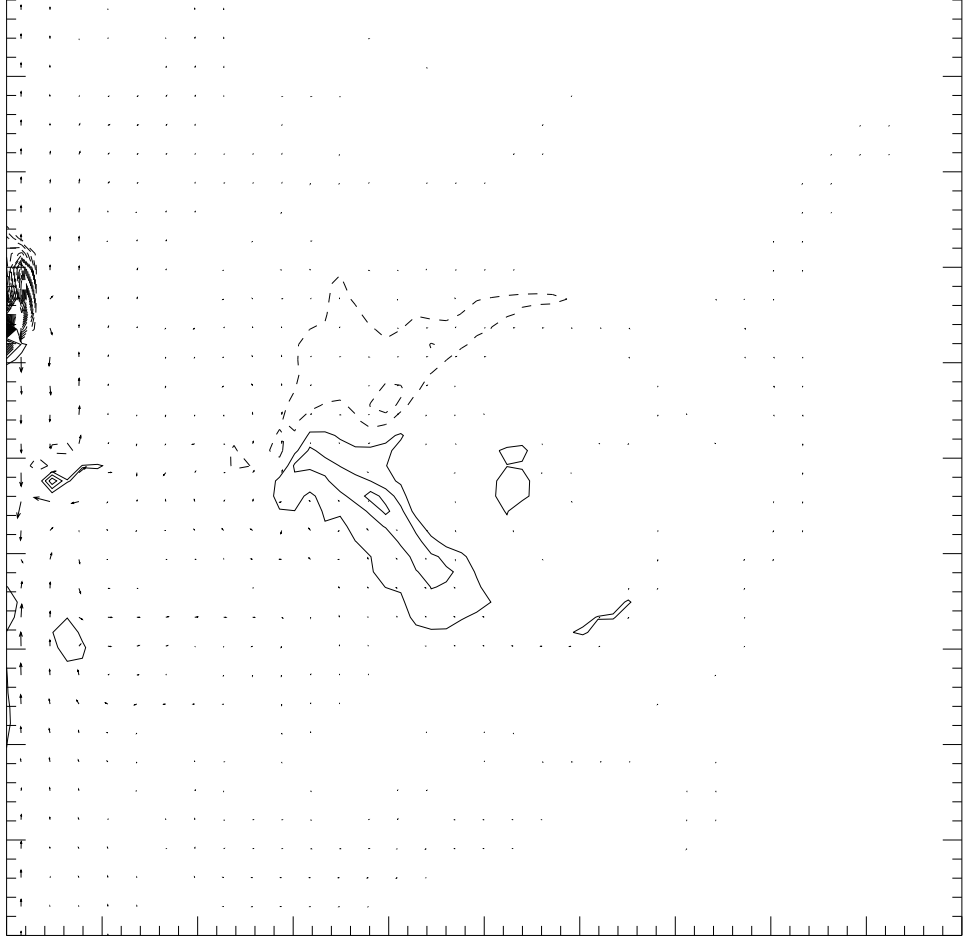
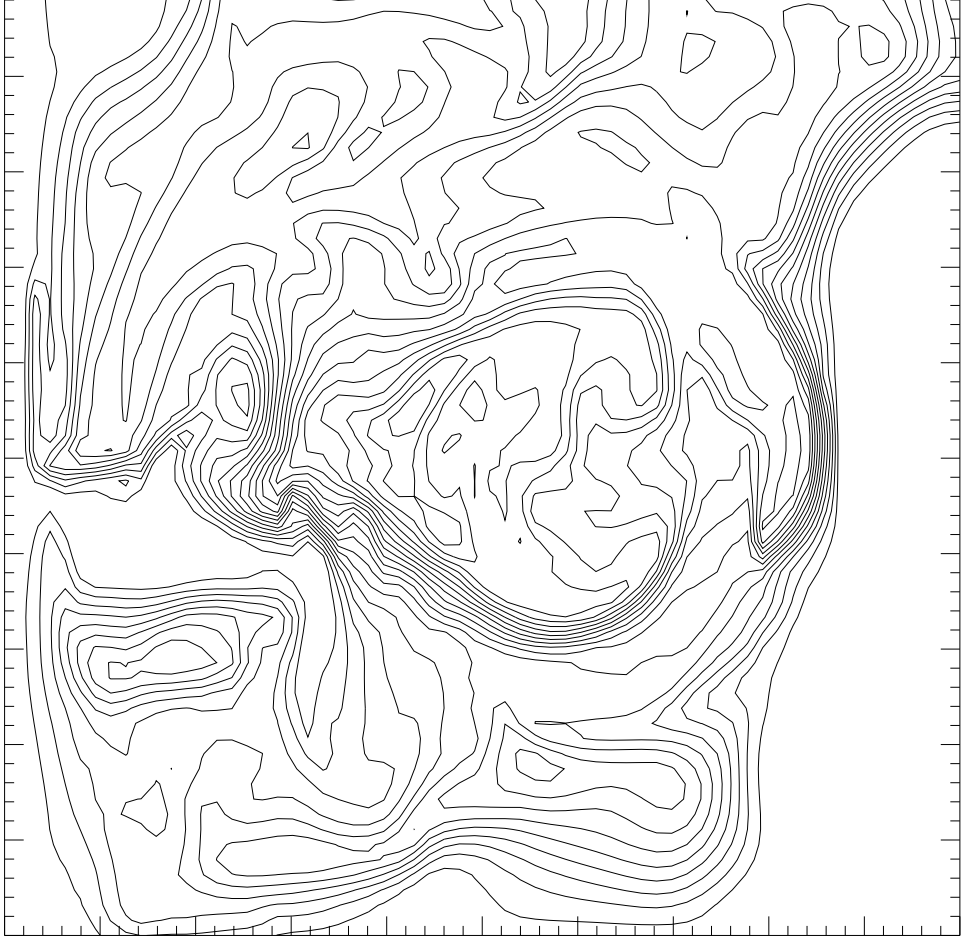
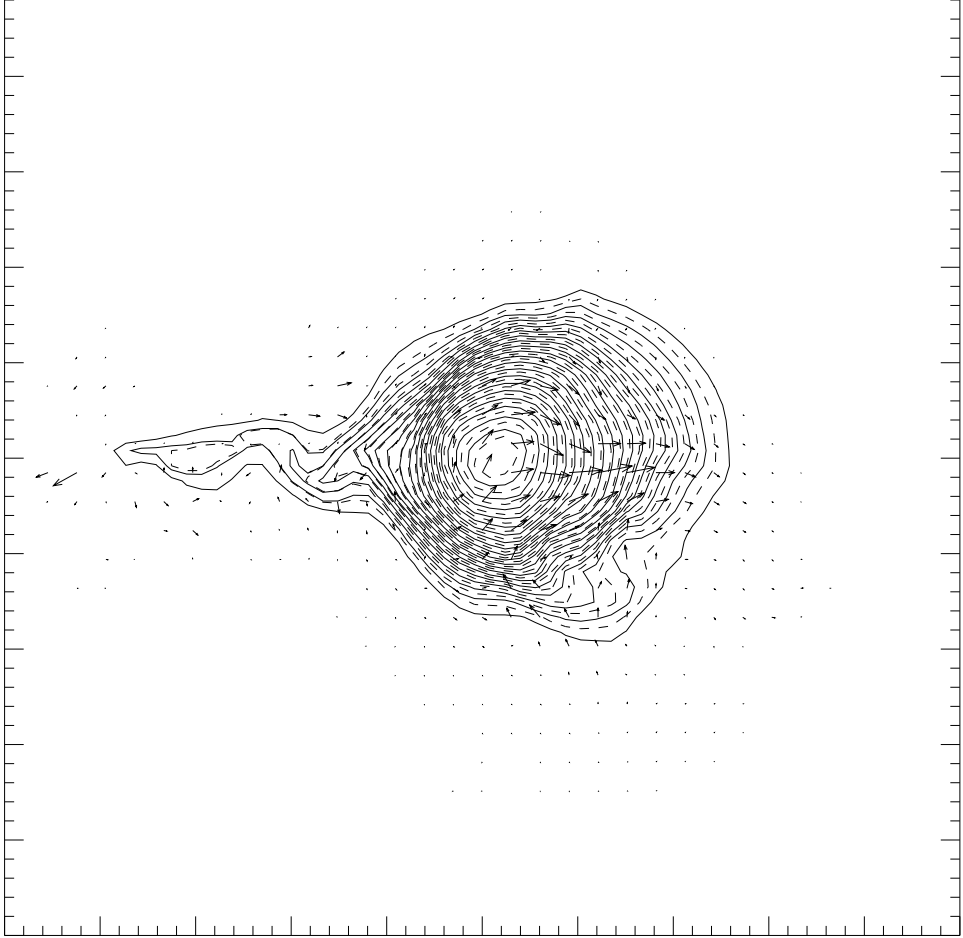
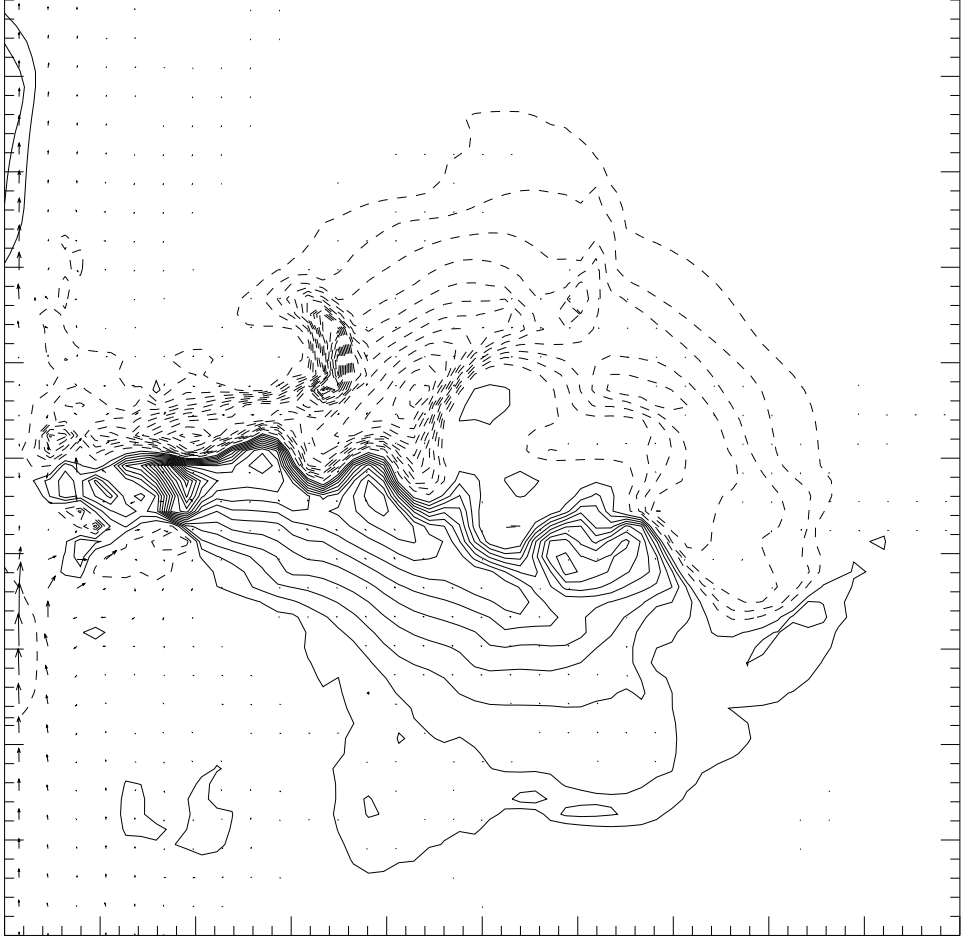
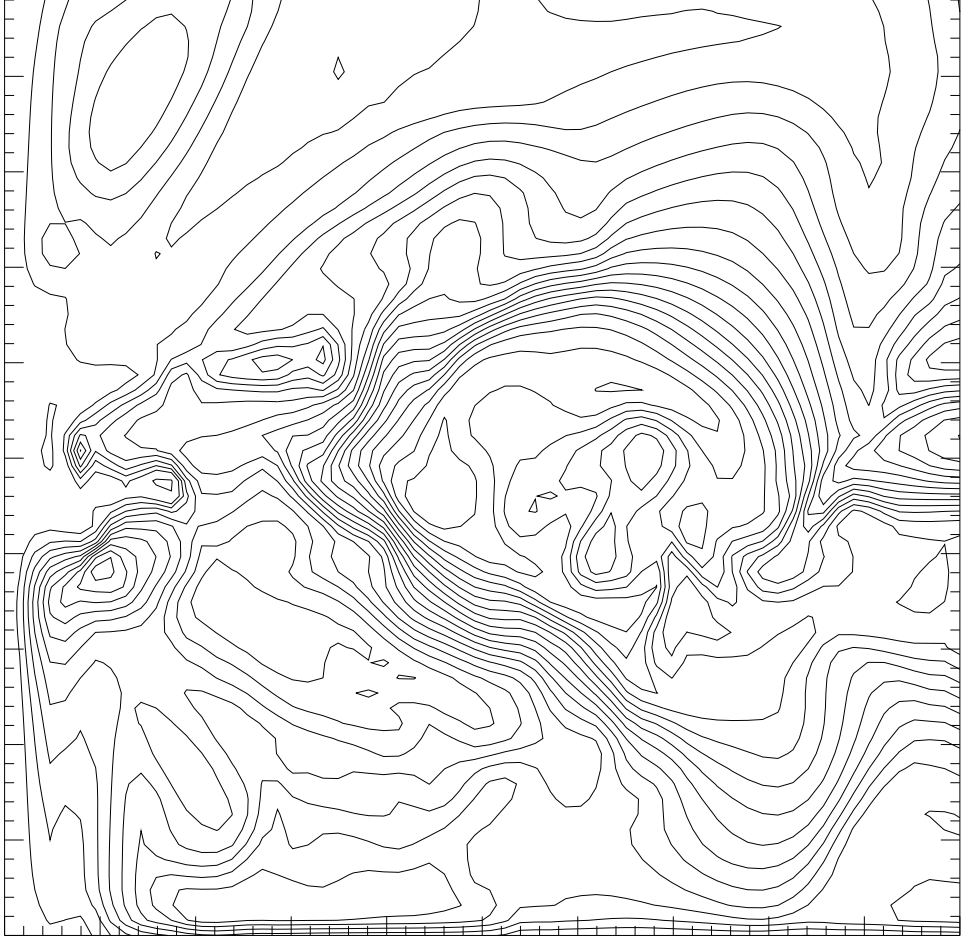
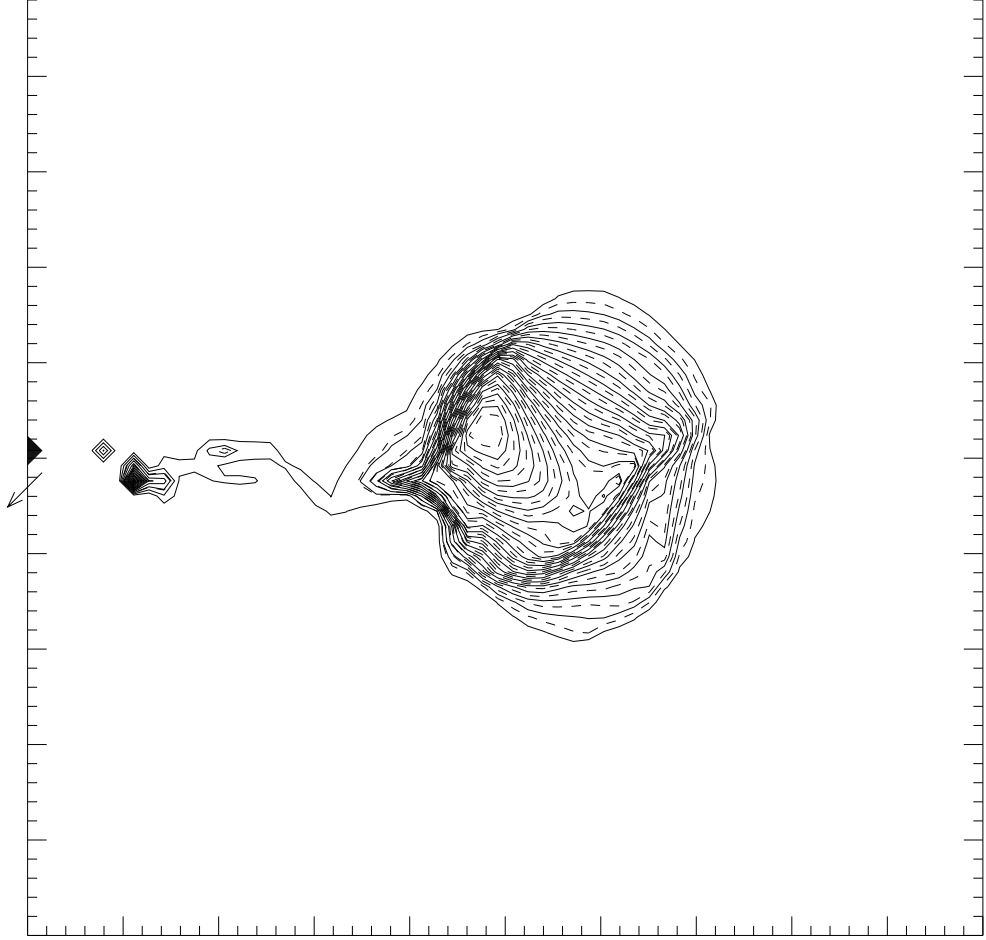
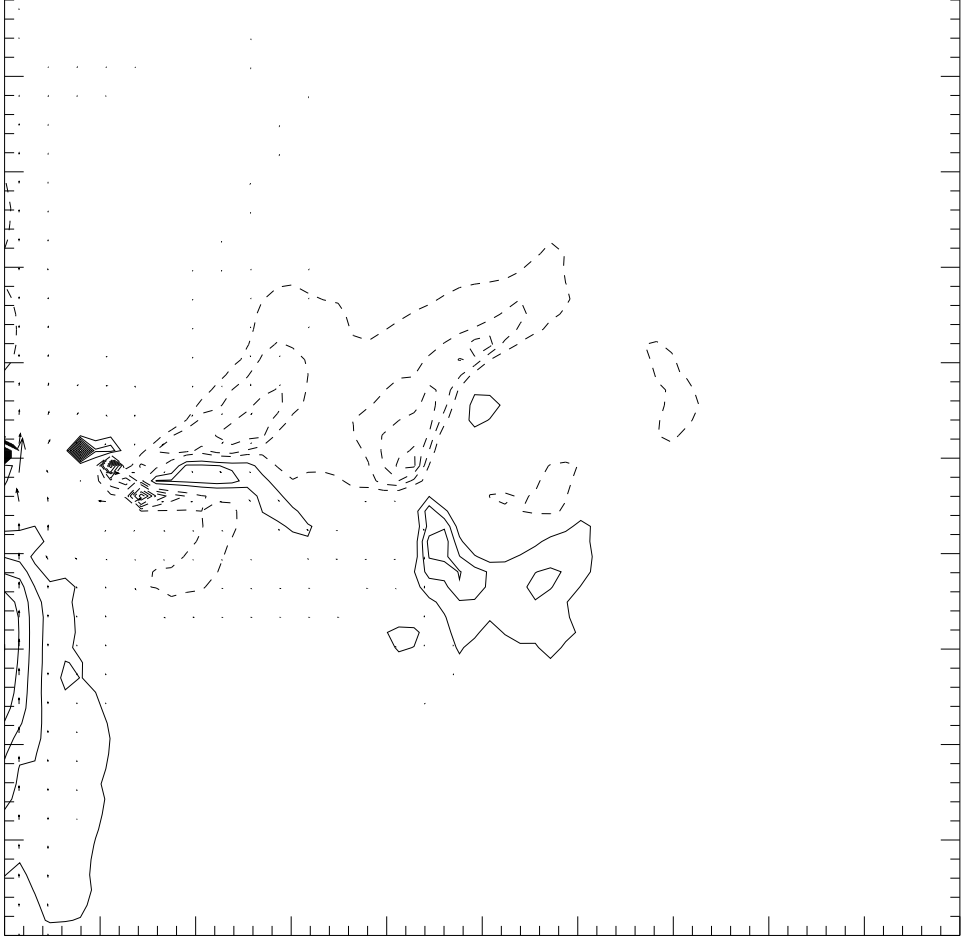
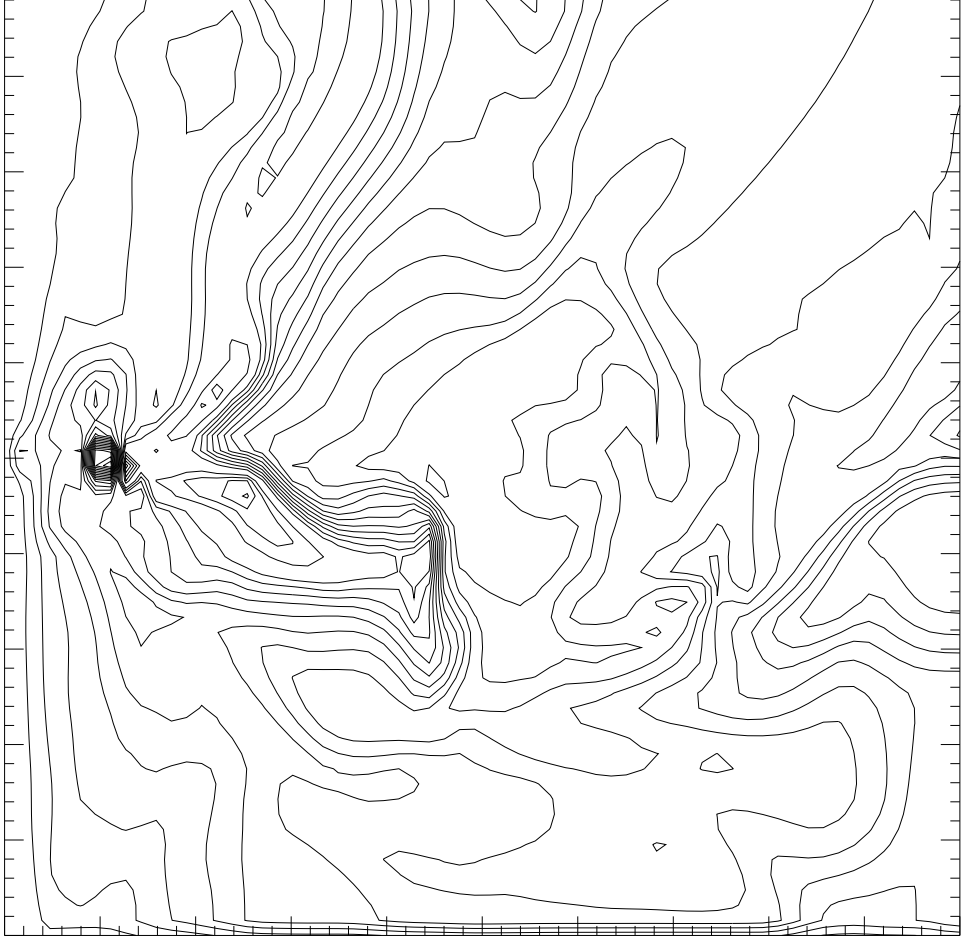
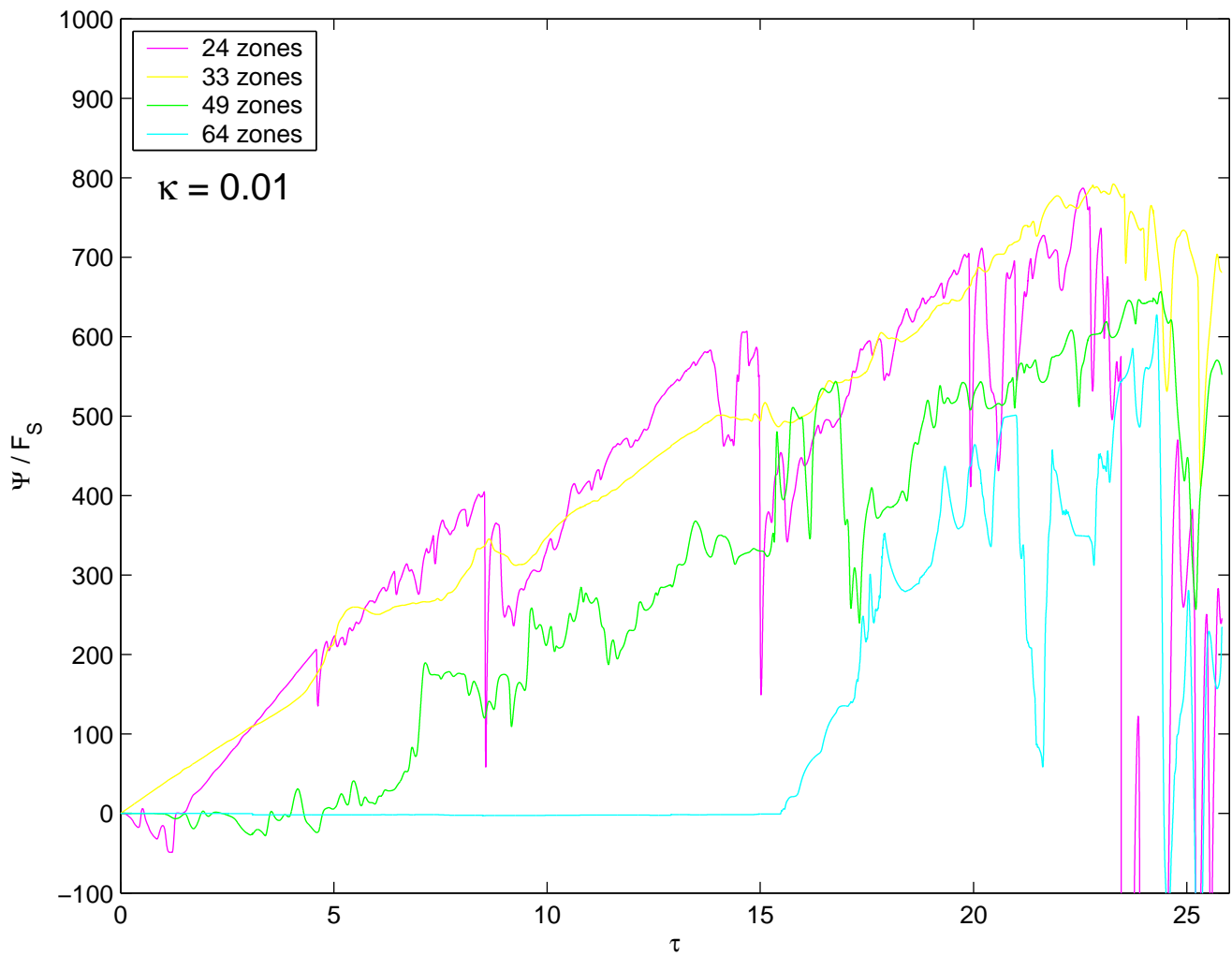
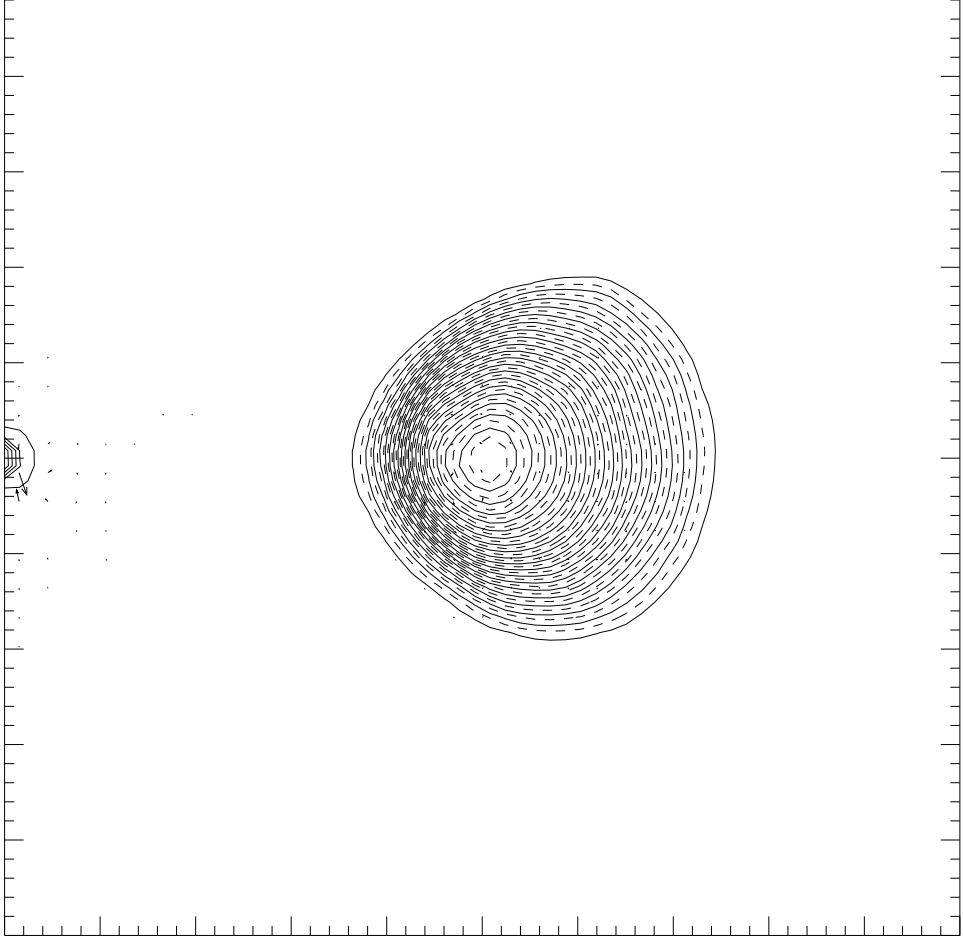
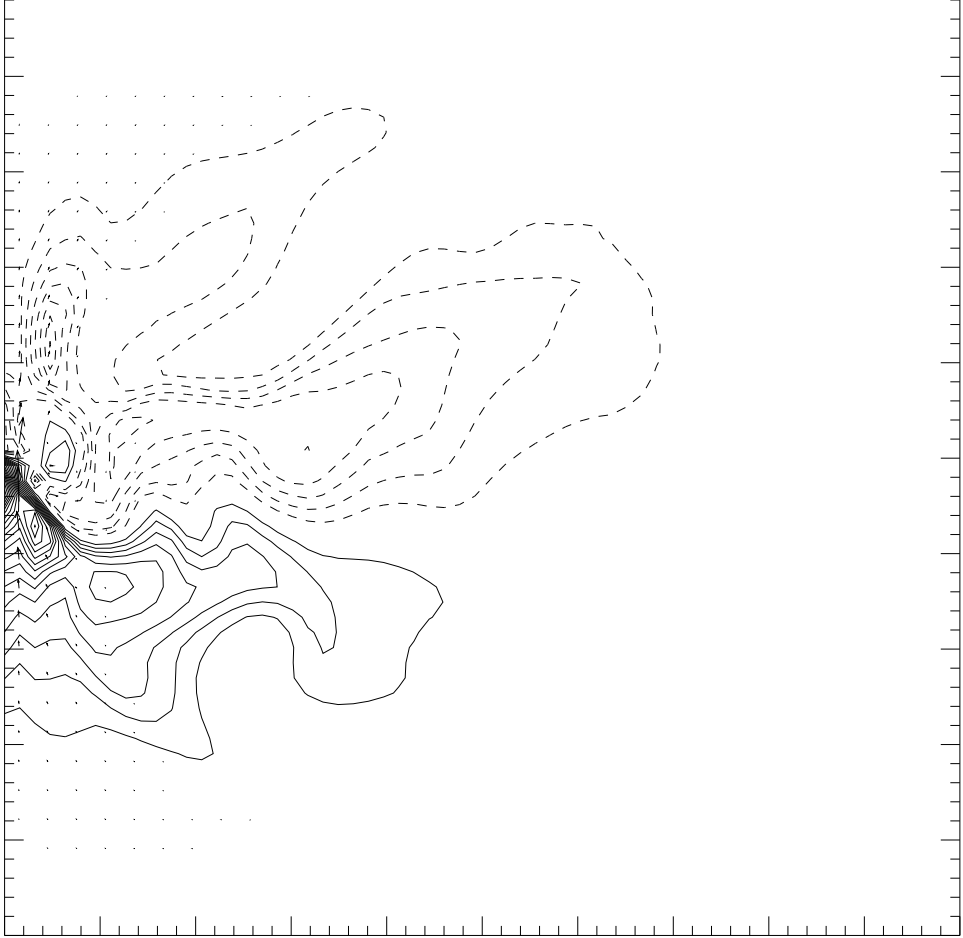
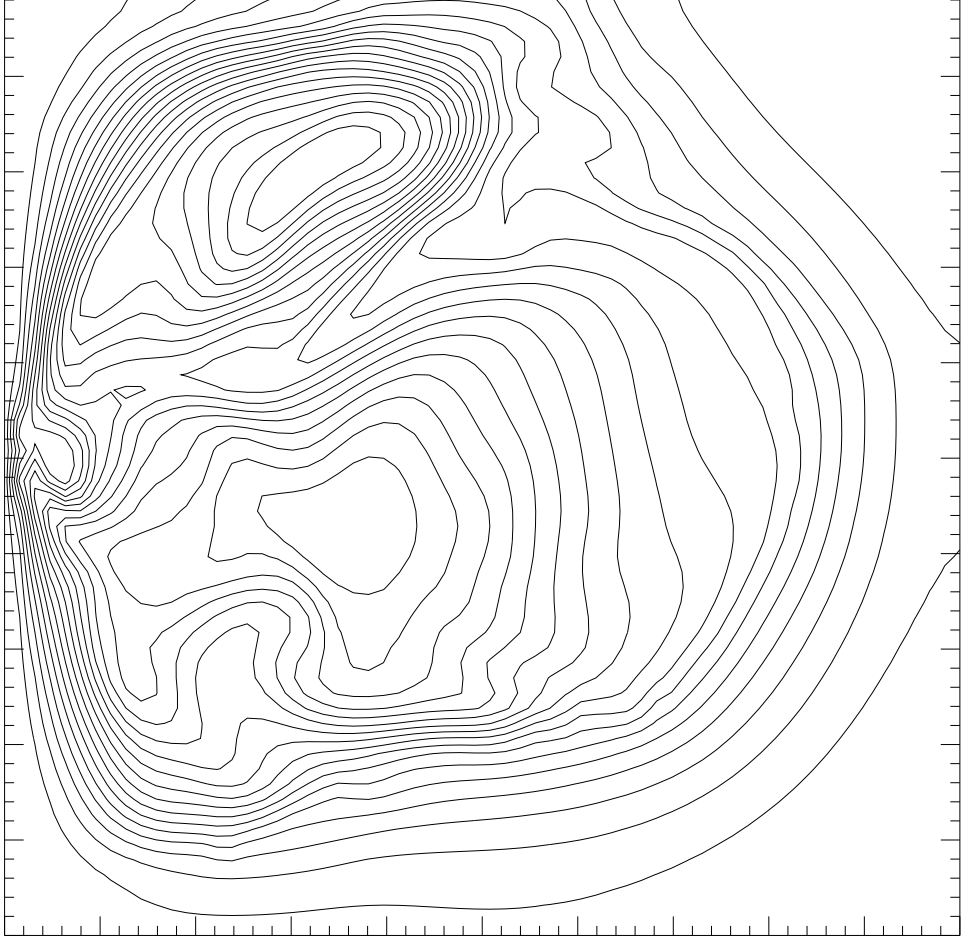
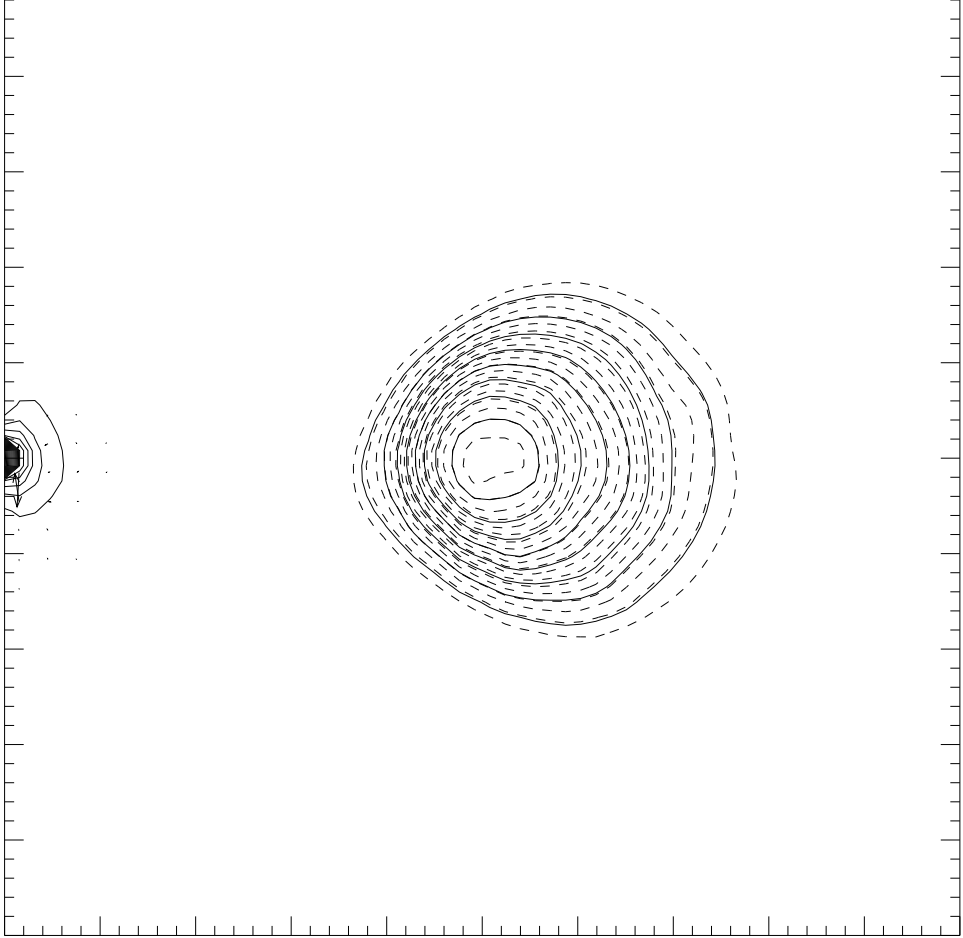
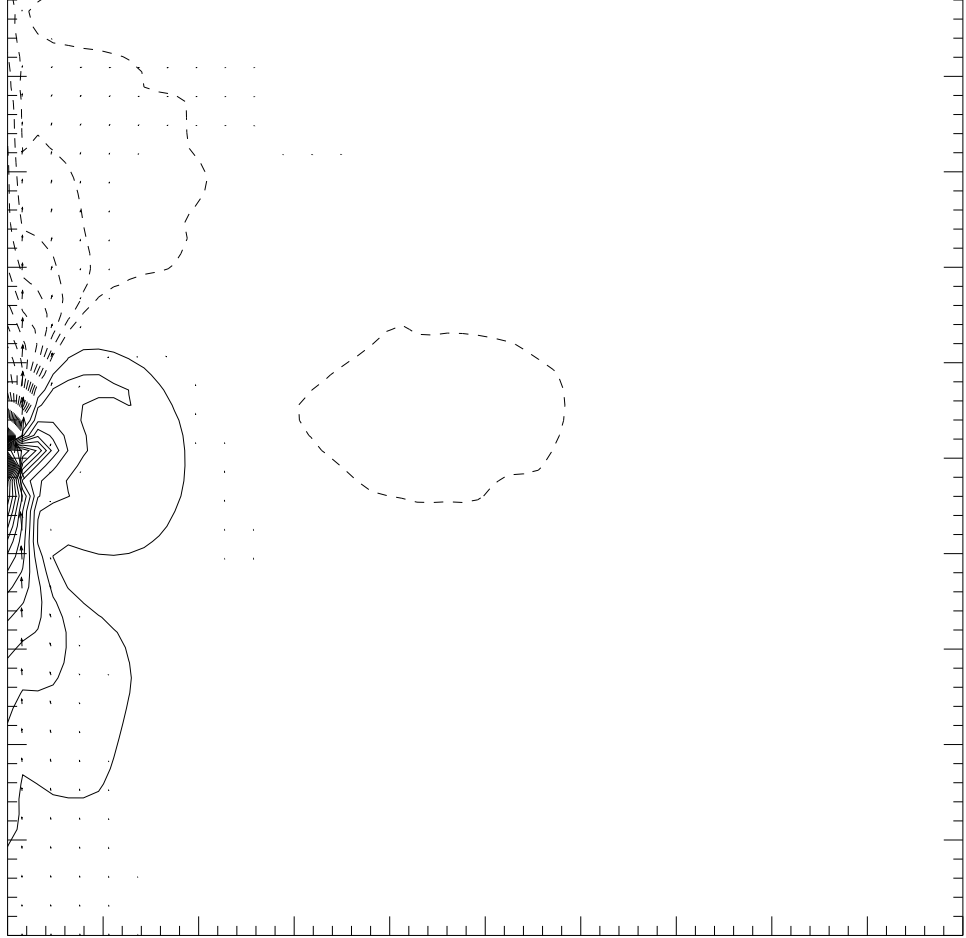
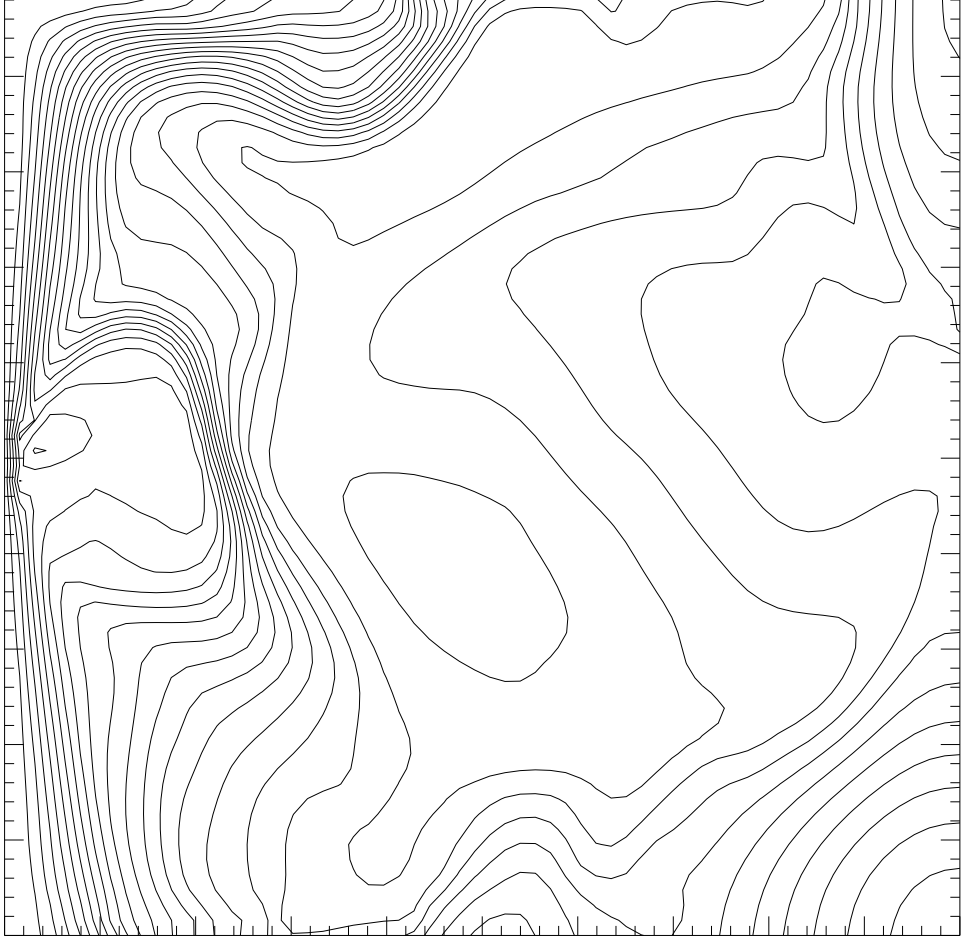
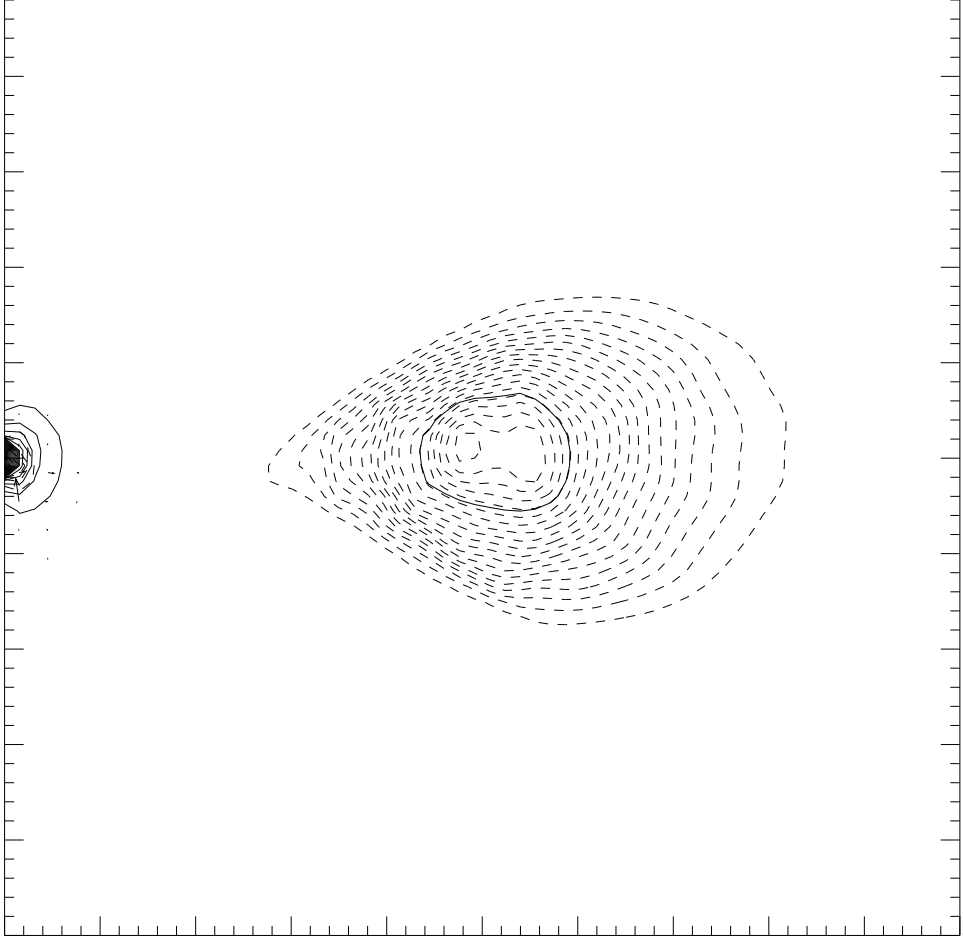
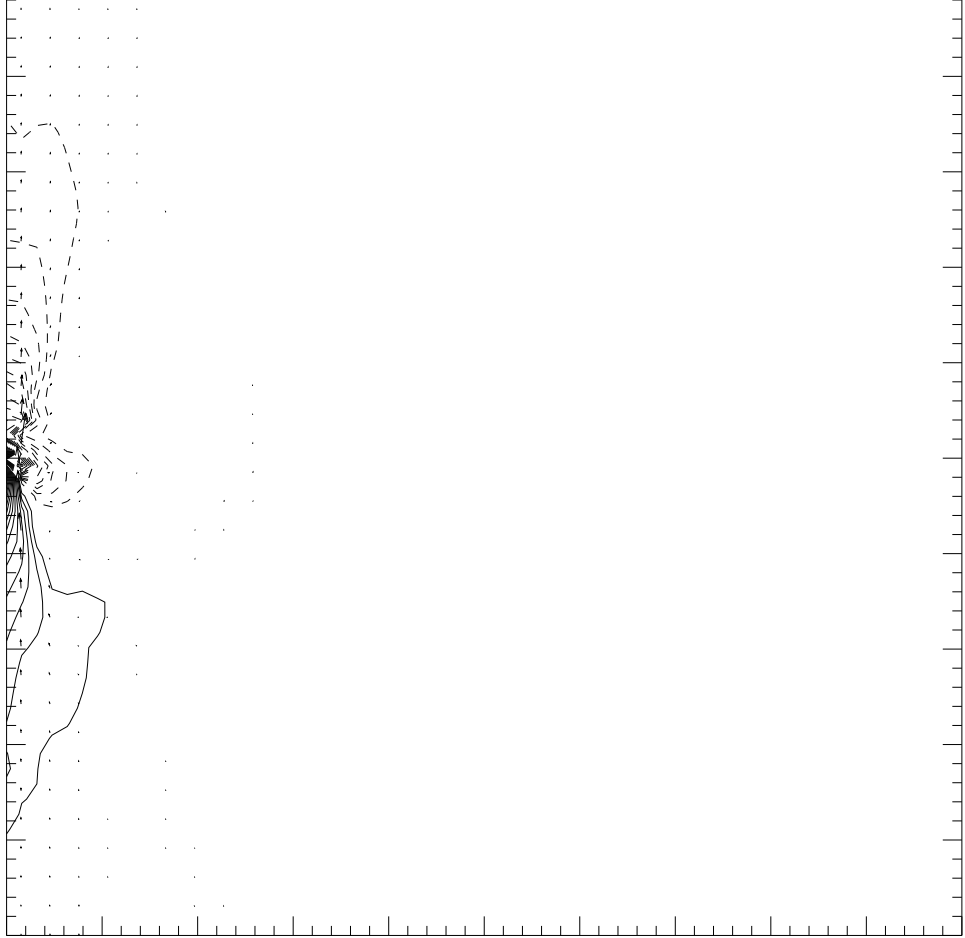
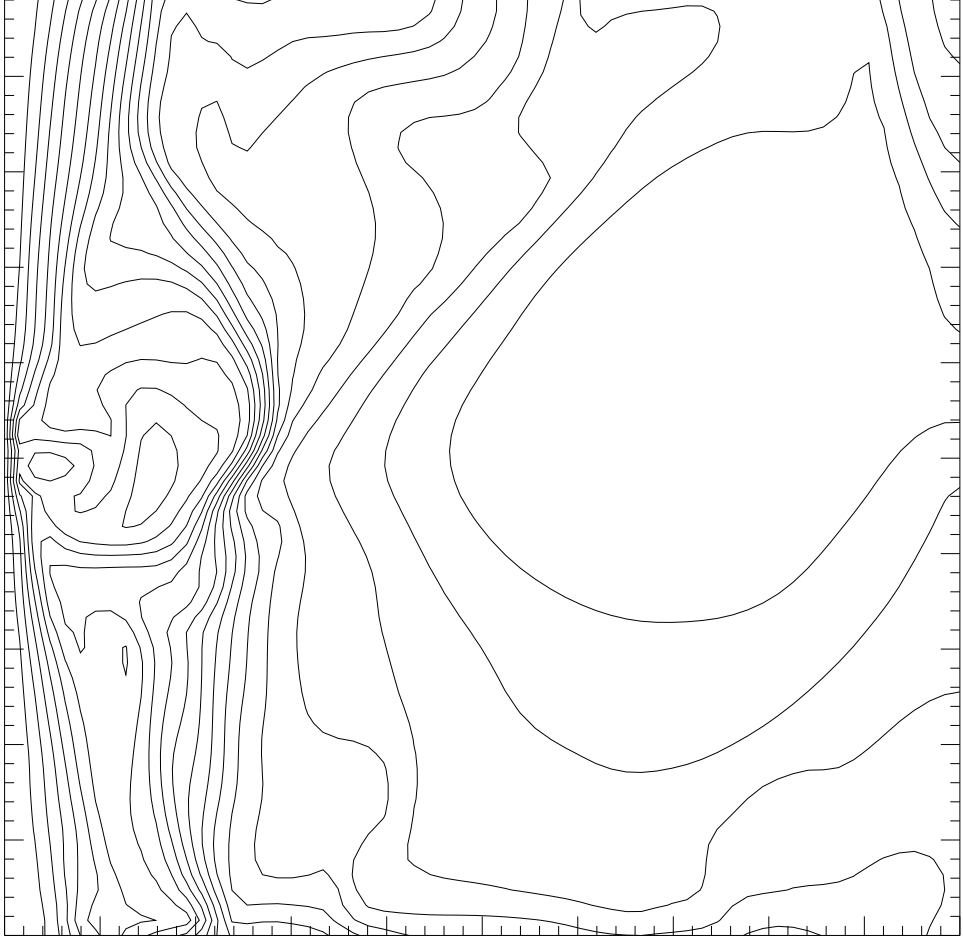
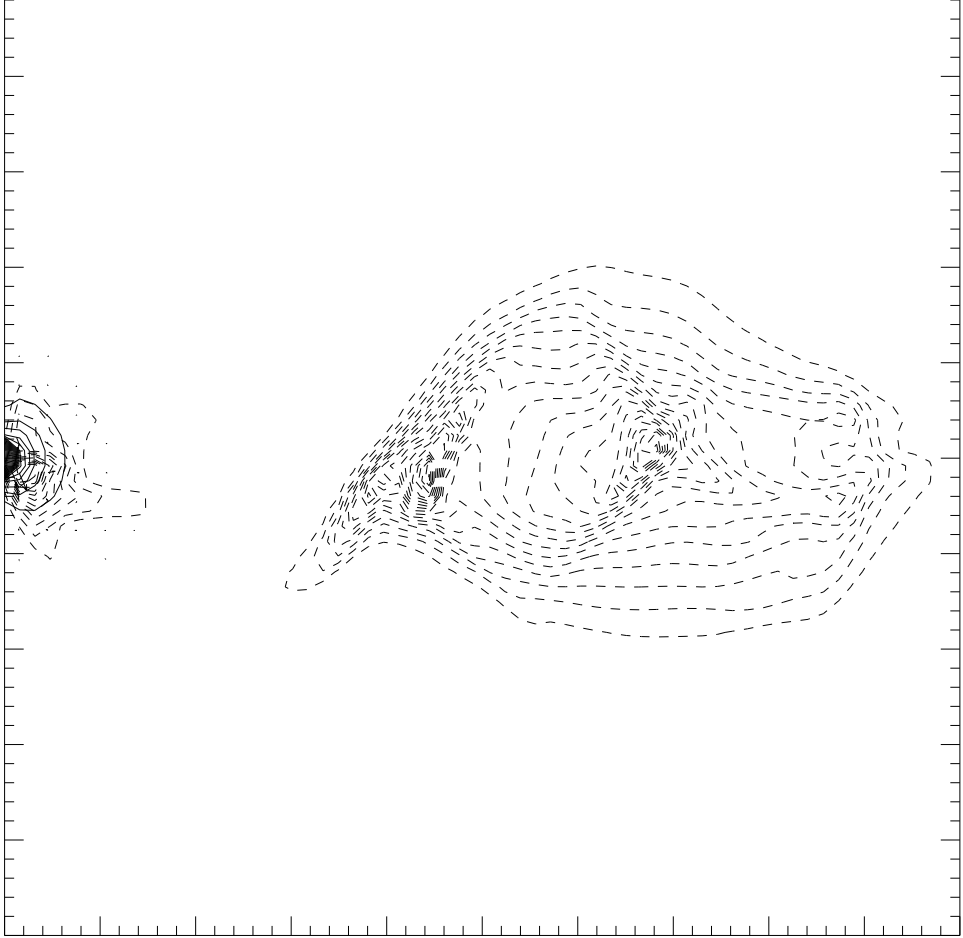
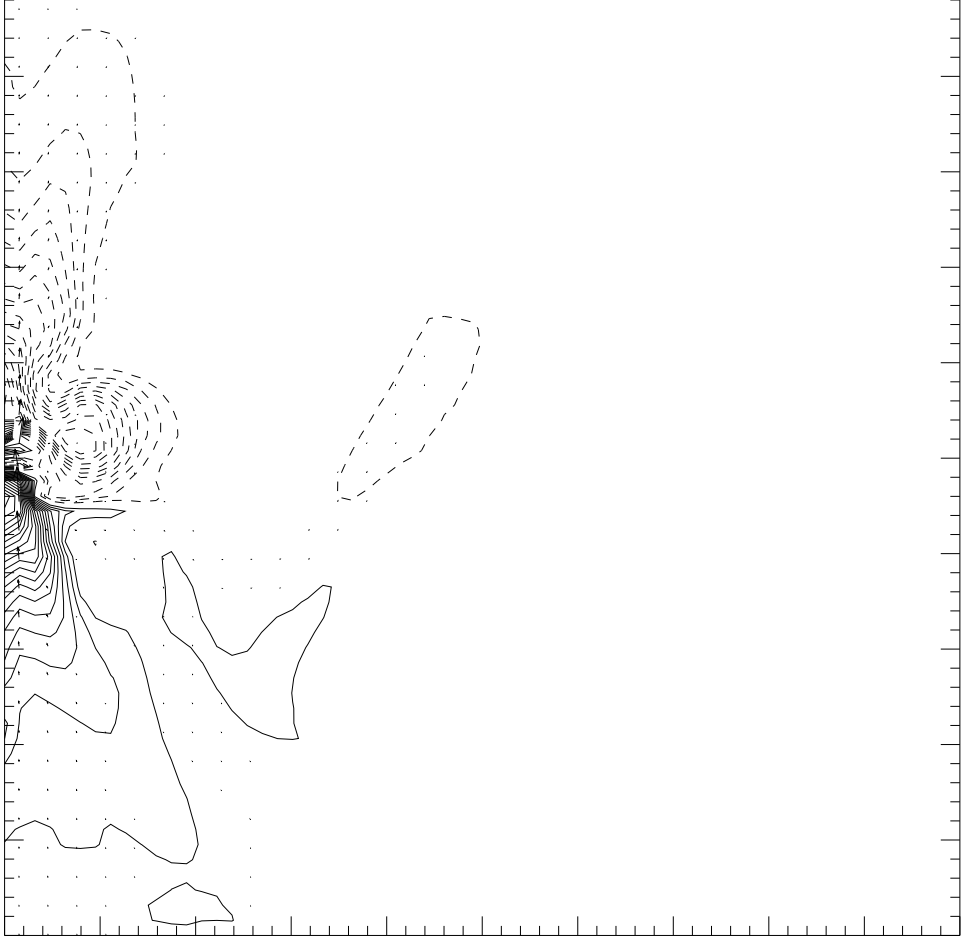
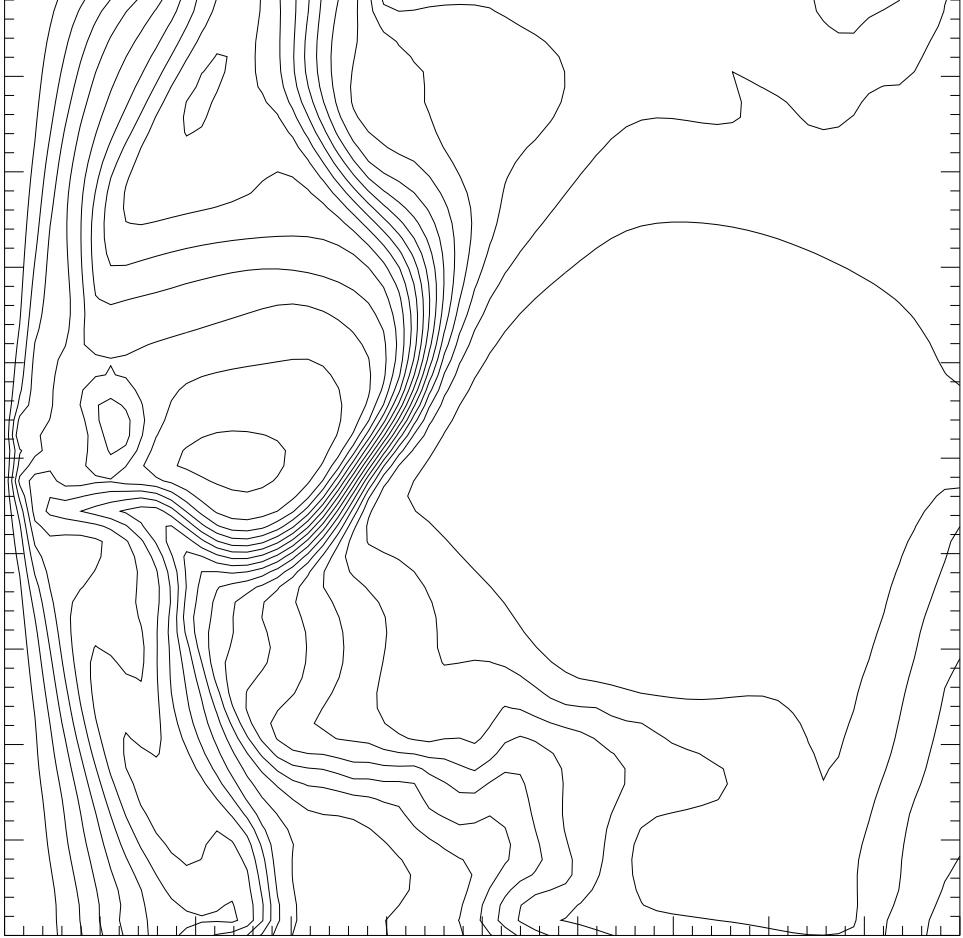
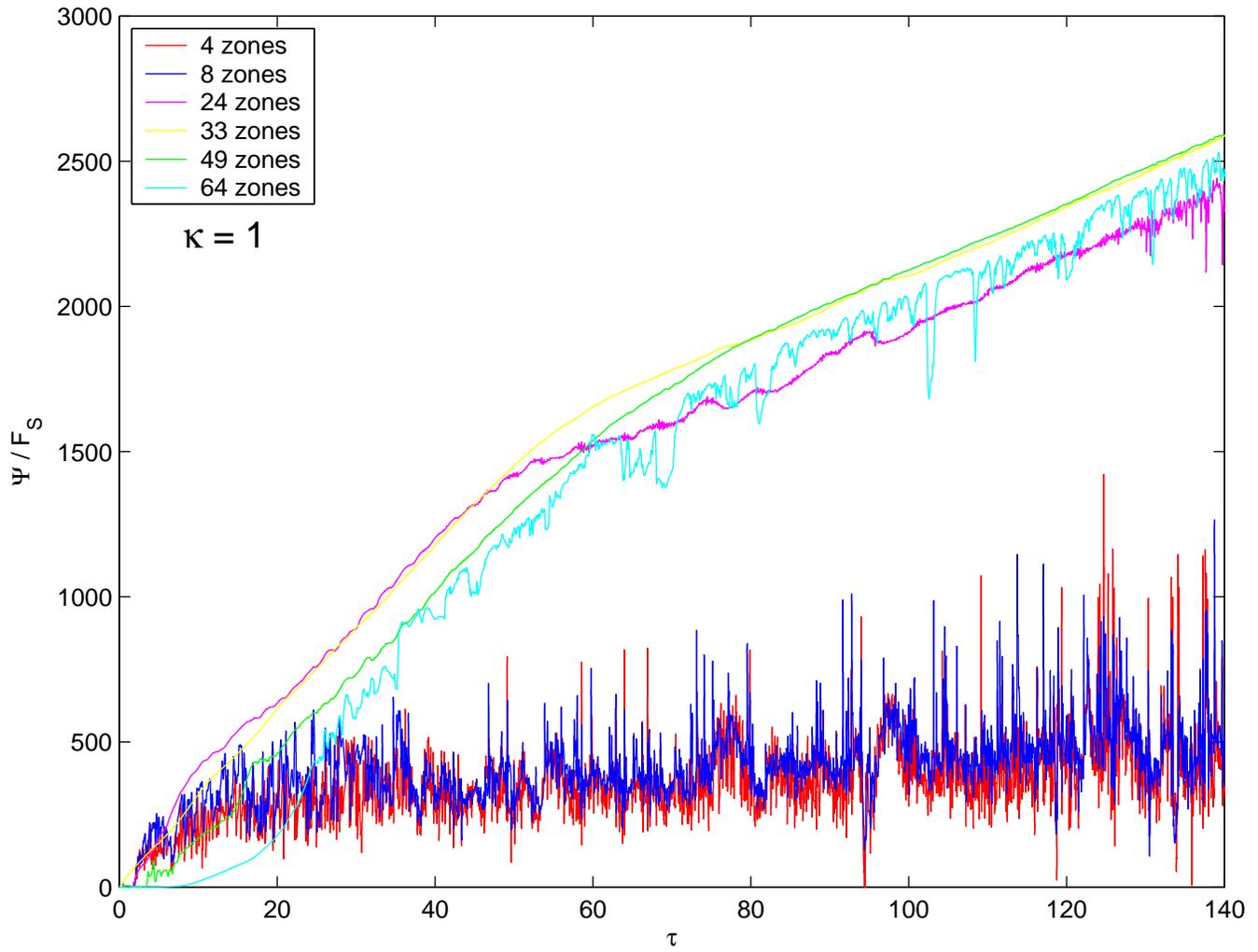
\end{document}